\documentclass[aps,prd,eqsecnum,notitlepage,
superscriptaddress,
nofootinbib,longbibliography]{revtex4-2}

\usepackage{amsmath,physics}
\usepackage{bm}
\usepackage{graphicx}
\usepackage{natbib}
\usepackage[caption=false]{subfig}
\usepackage{hyperref}
\usepackage{array}
\usepackage{float}
\usepackage{xcolor}

\global\arraycolsep=2pt
\DeclareMathOperator{\sgn}{sgn}
\DeclareMathOperator{\Li}{Li}

\usepackage{tikz,xcolor,hyperref}
\definecolor{lime}{HTML}{A6CE39}
\DeclareRobustCommand{\orcidicon}{%
	\begin{tikzpicture}
	\draw[lime, fill=lime] (0,0) 
	circle [radius=0.16] 
	node[white] {{\fontfamily{qag}\selectfont \tiny ID}};	\draw[white, fill=white] (-0.0625,0.095) 
	circle [radius=0.007];	\end{tikzpicture}
	\hspace{-2mm}}
\foreach \x in {A, ..., Z}{%
	\expandafter\xdef\csname orcid\x\endcsname{\noexpand\href{https://orcid.org/\csname orcidauthor\x\endcsname}{\noexpand\orcidicon}}
	}

\newcommand{\be}{\begin{equation}} 
\newcommand{\ee}{\end{equation}}

\begin{document}

%title%
\title{Energetics of quantum vacuum friction. II: Dipole fluctuations and field fluctuations}

%author information%
%\author{Xin Guo}
\author{Xin Guo\orcidA{}}
\email{guoxinmike@ou.edu}
%\homepage[]{Your web page}
%\thanks{}
%\altaffiliation{}
\affiliation{H. L. Dodge Department of Physics and Astronomy, University of Oklahoma, Norman, Oklahoma 73019, USA}

%\author{Kimball A. Milton}
\author{Kimball A. Milton\orcidB{}}
\email{kmilton@ou.edu}
%\homepage[]{Your web page}
%\thanks{}
%\altaffiliation{}
\affiliation{H. L. Dodge Department of Physics and Astronomy, University of Oklahoma, Norman, Oklahoma 73019, USA}

%\author{Gerard Kennedy}
\author{Gerard Kennedy\orcidC{}}
\email{g.kennedy@soton.ac.uk}
%\homepage[]{Your web page}
%\thanks{}
%\altaffiliation{}
\affiliation{School of Mathematical Sciences, University of Southampton, Southampton SO17 1BJ, United Kingdom}

\author{William P. McNulty\orcidD{}}
%\author{William P. McNulty}
\email{william.p.mcnulty-1@ou.edu}
%\homepage[]{Your web page}
%\thanks{}
%\altaffiliation{}
\affiliation{H. L. Dodge Department of Physics and Astronomy, University of Oklahoma, Norman, Oklahoma 73019, USA}

\author{Nima Pourtolami\orcidF{}}
%\author{Nima Pourtolami}
\email{nima.pourtolami@gmail.com}
%\homepage[]{Your web page}
%\thanks{}
%\altaffiliation{}
\affiliation{National Bank of Canada, Montreal, Quebec H3B 4S9, Canada}

\author{Yang Li\orcidE{}}
\email{leon@ncu.edu.cn}
%\homepage[]{Your web page}
%\thanks{}
%\altaffiliation{}
\affiliation{Department of Physics, Nanchang University, Nanchang 330031, China}

%date%
\date{\today}

%abstract%
\begin{abstract}
Quantum vacuum friction experienced by an atom, where the only dissipative mechanism is through its interaction with the radiation field, has been studied in our recent paper [Phys. Rev. D \textbf{104}, 116006 (2021)]. Quantum vacuum friction on an intrinsically dissipative particle is different in that the friction arises not only from the field fluctuations but also from the dipole fluctuations intrinsic to the particle. As a result, the dissipative particle can be out of the nonequilibrium steady state (NESS), and therefore loses or gains internal energy (rest mass). Only if the temperature of the particle equals a special NESS temperature will the particle be in NESS. In this paper, general NESS conditions are derived which give the NESS temperature of the particle as a function of the temperature of the radiation and the velocity of the particle. Imposing the NESS conditions, we also obtain an expression for the quantum vacuum friction in NESS. The NESS quantum vacuum friction is shown to be negative definite (opposing the motion of the particle) and equivalent to that found in our previous paper if the dissipation mechanism is restricted to radiation reaction. The NESS temperature and quantum vacuum friction are calculated numerically for various models. In particular, we show that, for a gold nanosphere, the deviation of its NESS temperature from the temperature of the radiation can be substantial and it is also possible to detect the NESS quantum vacuum friction directly at sufficiently high temperatures. Out of NESS, even though the quantum vacuum friction no longer has a definite sign in the rest frame of the radiation, the friction in the rest frame of the particle is still negative definite. Also, the external force needed to keep the particle moving must be in the same direction as the motion of the particle, therefore excluding the possibility of a perpetual motion machine, which could convert the vacuum energy into useful mechanical work. In addition, we find that the deviation of the temperature of the particle from its NESS temperature causes the particle to lose or gain internal energy in such a way that the particle would return to NESS after deviating from it. This enables experimental measurements of the NESS temperature of the particle to serve as a feasible signature for these quantum vacuum frictional effects.
\end{abstract}

%keywords%
%\keywords{}

%\maketitle must follow title, authors, abstract, and keywords%
\maketitle

%introduction
\section{introduction}
\label{introduction}
Quantum friction (also known as Casimir friction) \cite{Kim:reality} is usually associated with two typical configurations: two sliding plates \cite{Pendry:1997, VP:1999} or a particle moving parallel to a plate \cite{Dedkov:2003, Barton:friction, Intravaia:etal}, even though the notion can be applied more broadly to contexts like the expanding universe, rotating black holes, moving mirrors and even activities of sub-cellular biosystems \cite{Davies:QVF}. Although much effort has been invested over several decades to calculate quantum friction in both configurations, discrepancies in the theoretical results and even doubts concerning the existence of such friction remain \cite{Marty:thesis}. 
However, relative motion between two macroscopic bodies is not a necessity for quantum vacuum friction to occur. A neutral polarizable particle moving through free space will experience a force in the opposite/same direction as the particle's motion due to its interaction with surrounding blackbody radiation. This is what we refer to as quantum vacuum friction in Ref.~\cite{Xin:eqf1} and in the present paper.

The key ingredient of quantum vacuum friction is the presence of fluctuations. Due to the dissipative nature of the fluctuations, the electromagnetic vacuum behaves like a complex fluid, which modifies the motion of objects in it \cite{Kardar:friction}. Fluctuations are related to the imaginary part of the corresponding susceptibilities according to the fluctuation-dissipation theorem (FDT) \cite{Kubo:FDT}.  Typical examples are the Green's dyadic for electromagnetic field fluctuations and the polarizability of the particle for dipole fluctuations.
Both fluctuations can be responsible for quantum vacuum friction. In Ref.~\cite{Xin:eqf1}, where the intrinsic polarizability, $ \bm{\alpha} $, of the particle is considered to be purely real, it is the imaginary part of the electromagnetic Green's dyadic, $ \bm{\Gamma} $, that allows for the existence of fluctuations of the electromagnetic field and the corresponding induced dipole fluctuations. In this paper, we continue to study quantum vacuum friction, but for a neutral particle whose intrinsic electric polarizability is complex, $ \Im\bm{\alpha}(\omega)\ne 0 $.\footnote{For simplicity of the discussion, we assume the particle to possess no intrinsic magnetic polarizability, $ \bm{\beta}=0 $, and therefore we do not consider magnetic dipole fluctuations or those of higher multipoles.} This setting exhibits some new features beyond those already considered in Ref.~\cite{Xin:eqf1}.  First, the temperature of the particle, $ T' $, comes into play, through the intrinsic dipole fluctuations. Second, the particle now has the freedom to leave the nonequilibrium steady state (NESS) because it can absorb or emit net energy.

We still focus on the NESS situation from Sec.~\ref{NESS} to Sec.~\ref{nano}.
In Sec.~\ref{NESS}, we find the NESS conditions by requiring the particle to absorb and emit net energy at the same rate. The NESS conditions provide a relation between the temperature of the particle, $ T' $, and the temperature of the radiation, $ T $, which defines the NESS temperature of the particle, $ \tilde{T} $. Interestingly, we are able to prove that the NESS temperature of the particle is, quite generally, greater than the Planck-Einstein transformed temperature of the blackbody radiation, that is, $ \tilde{T}>T/\gamma $. Using the NESS conditions, the NESS temperature ratio, $\tilde{T}/T $, can be calculated as a function of the particle's velocity $ v $ and the radiation temperature $ T $ once $ \Im\bm{\alpha}(\omega) $ has been specified. For simplicity, we first work out the NESS conditions explicitly for the resonance model, $ \Im\bm{\alpha}(\omega)\propto \delta(\omega-\omega_{0}) $, and the monomial model, $ \Im\bm{\alpha}(\omega)\propto \omega^{n} $. Even though these models are simple, they provide insights into more realistic situations because more realistic models often reduce to these simple models in different temperature and velocity regimes. In Sec.~\ref{sign}, we impose the NESS conditions on the general formula for quantum vacuum frictional force previously derived in Ref.~\cite{Kim:dipole} and find that the resultant expression for the NESS quantum vacuum frictional force on the particle is negative definite and equivalent to that in Ref.~\cite{Xin:eqf1} if the dissipation mechanism is restricted to radiation reaction.  The nonrelativistic limit of the NESS friction reproduces the famous Einstein-Hopf drag~\cite{Einstein:Hopf}. The classical high-temperature limit of the NESS friction is found to be linear in $ T $. The NESS quantum vacuum friction for the resonance and monomial models are also explicitly calculated. In Sec.~\ref{nano}, we estimate the NESS temperature ratio and NESS quantum  vacuum friction for a gold nanosphere moving in vacuum. The Lorenz-Lorentz relation is used for the polarizability of the particle and the Drude model for the permittivity of gold. We assume a constant damping parameter for the gold nanosphere in \ref{nano1}. The model for the gold nanosphere turns out to reduce to the $ n=1 $ monomial model in the low-temperature limit and the $ n=-3 $ monomial model in the high-temperature limit. In the transition region between these two extreme limits, there also exists a temperature regime where the model reduces to the resonance model. In reality, the damping parameter is temperature dependent. Therefore, in \ref{nano2}, we employ the Bloch-Gr\"uneisen model to describe the temperature dependence of the damping parameter. The effect on the NESS temperature ratio and the NESS quantum vacuum friction of including this temperature dependence in the damping parameter is also discussed. 

We then extend our analysis to the out-of-NESS situation in Sec.~\ref{out}, and draw several interesting conclusions. Even though the frictional force, $ F $, in the rest frame of radiation ($ \mathcal{R} $), no longer has a definite sign, the frictional force, $ F' $, in the rest frame of particle ($ \mathcal{P} $), is equal to the NESS quantum vacuum friction, $ \tilde{F} $, and is therefore negative definite. Also, the external force, $ F_{\rm{ext}} $, required to maintain the configuration is the negative of $ F' $, and always pushes the particle forward. In addition, depending on whether the temperature of the particle is higher or lower than the NESS temperature, the sign of the total force, $ F_{\rm{tot}}=F+F_{\rm{ext}} $, on the particle is negative or positive, respectively. And the sign of the total force reflects the loss or gain of the internal energy (rest mass) of the particle. Finally, we numerically obtain the condition for the quantum vacuum friction, $F$, on a gold nanosphere (with constant damping) to be zero in frame $ \mathcal{R} $. The zeroes for $ F $ occur only in the high-temperature regime unless the velocities are ultrarelativistic.

In Appendix A, we provide a proof of the equivalence between the Lorentz force law and the principle of virtual work for friction on a moving dipole. In Appendix B, we supply high-temperature asymptotic expressions for the frictional power and force in the case of the Bloch-Gr\"{u}neisen model. In Appendix C, we extend the discussion about the zeroes of $ F $ to the low-temperature, high velocity regime.

In this paper we use Heaviside-Lorentz (rationalized) electromagnetic units and set $ k_{B}=c=\hbar=1 $ in the formulas. SI units are used in the numerical evaluations.

%mainbody

\section{The NESS conditions}\label{NESS}
In this paper, we explore again the energetics of the so-called quantum vacuum friction, the electromagnetic force that a neutral but polarizable particle experiences while it is moving through vacuum (free space) with constant velocity and interacting with the surrounding blackbody radiation. Unlike our previous investigation in Ref.~\cite{Xin:eqf1}, we study a particle that has intrinsic dissipation. This imparts to the particle the freedom of absorbing/releasing net energy and allows for the existence of independent dipole fluctuations.

There are three main players in the scene: the particle, the radiation, and an external agent which compensates for the energy lost into the particle and the radiation fields and maintains the uniform motion of the particle \cite{Intravaia:NTQF}. The external agent and the radiation both directly interact with the particle. Even though the particle itself is an open quantum system, the overall energy of all three players should be conserved, which can be summarized into a power balance relation,
\begin{equation}\label{eq01}
P_{\rm{par}}=P_{\rm{ext}}+P.
\end{equation}
Here, $ P_{\rm{ext}}=F_{\rm{ext}}v $ is the time rate of the external agent doing work on the particle. $ P $ is the time rate of the radiation field doing work on the particle. $ P_{\rm{par}} $ is the net power absorbed by the particle.\footnote{Of course, this results in a  change in the internal dynamics of the particle, which we will not discuss in detail in the current paper.}

If the particle does not absorb net energy,  $ P_{\rm{par}}=0 $, we say that it is in the nonequilibrium steady state (NESS). A neutral particle (e.g., a gold atom) that has no intrinsic dissipation is ensured to be in NESS. But for an intrinsically dissipative particle (e.g., a gold nanosphere that is made of millions of atoms), $ P_{\rm{par}}=0 $ defines a NESS condition that turns out to be a relation between the temperature of the particle, $ T' $, and the temperature of the radiation, $ T $. 

In frame $ \mathcal{R} $, NESS requires the external force to balance the electromagnetic force, $ F_{\rm{ext}}+F=0 $.\footnote{It will be clear in Sec.~\ref{out} that this is not the case out of NESS due to the change in the particle's mass.} As a result, the NESS condition also translates to a power-force relation for the radiation 
\begin{equation}
P=-P_{\rm{ext}}=-F_{\rm{ext}}v=Fv.
\end{equation}
%except that primes on the polarizabilities are all omitted as they are always evaluated in frame $ \mathcal{P} $ whenever they appear.
In frame $ \mathcal{P} $, the external agent cannot do any work on the particle, $ P'_{\rm{ext}}=0 $. And the NESS condition simplifies to
\begin{equation}
P'_{\rm{par}}=P'=0.
\end{equation}
Therefore, we could  calculate either $ P-Fv $ or $ P' $ in order to find the NESS conditions explicitly. 

\subsection{The frictional power}
We now calculate the electromagnetic power on the polarizable particle. Since the electromagnetic force that corresponds to this power is precisely the quantum vacuum friction, we may also call it the frictional power. 

Before proceeding to the calculation, let us define the problem more concretely. We will use $ \bm{\alpha}(\omega) $ for the electric polarizability of the neutral particle. To incorporate intrinsic dissipation, $ \bm{\alpha}(\omega) $ must be complex.  Without loss of generality, we assume that the particle is moving relative to the blackbody radiation in the $ x $ direction with constant velocity $ \vb{v}=v\hat{\vb{x}} $ and therefore lies on the trajectory $ \vb{r}(t)=\vb{v}t $. In the rest frame of the particle $ \mathcal{P} $, the particle sits at a fixed position, which we will assume to be the origin, $ \vb{r'}=\vb{0} $. Throughout the paper, we will use primes on quantities and coordinates in frame $ \mathcal{P} $, except for the polarizability $ \bm{\alpha} $, which is always defined in frame $ \mathcal{P} $. We will now calculate the rate at which the electromagnetic force does work (frictional power) on the particle.

In Ref.~\cite{Xin:eqf1}, the frictional power $ P $ in frame $ \mathcal{R} $ is derived using the  Joule heating law
\begin{equation}\label{eq2.1}
P(t)=\int d\vb{r} \, \vb{j}(t, \vb{r})\cdot \vb{E}(t, \vb{r}),
\end{equation}
while the frictional power $ P' $ in frame $ \mathcal{P} $ is computed by differentiating the free energy $ \mathcal{F'} $ in $ \mathcal{P} $ 
\begin{equation}\label{eq2.2}
P'(t')=\frac{\partial}{\partial t'}\mathcal{F'}=-\frac{\partial}{\partial t_{1}'}\left[\vb{d}(t_{0}')\cdot\vb{E}(t_{1}',\vb{0})\right]|_{t_{0}'=t_{1}'\to t'}.
\end{equation}
Here, we use the same notation as in Ref. \cite{Xin:eqf1}: $ t_{0}' $ for the time coordinate of the dipole operator and $ t_{1}' $ for that of the field operator and they are set equal after the differentiation. Effectively, this apparent separation of coordinates serves as a prescription that only the field coordinates are to be differentiated. 

One might be questioning the consistency between Eq.~\eqref{eq2.1} and Eq.~\eqref{eq2.2} and, in particular, the validity of the differentiation prescription of Eq.~\eqref{eq2.2}. Here we eliminate any such doubt by showing that the electromagnetic power on a moving dipole according to Eq.~\eqref{eq2.1} can always be written as a derivative of a free energy. And in frame $ \mathcal{P} $ where the velocity of the particle is zero, it reduces to the special form Eq.~\eqref{eq2.2}.

For a uniformly moving dipole $ \vb{d}(t) $ with velocity $ \vb{v} $, the corresponding classical current density is
\begin{equation}\label{eq2.3}
\vb{j}(t, \vb{r})=\left[\dot{\vb{d}}(t)-\vb{v} \grad\cdot\vb{d}(t)\right]\delta(\vb{r}-\vb{v}t).
\end{equation}
When this current is inserted into Eq.~\eqref{eq2.1}, we obtain the power
\begin{equation}\label{eq2.4}
P(t)=\dot{\vb{d}}(t)\cdot\vb{E}(t,\vb{v}t)+\left[\vb{d}(t)\cdot\grad\right]\left[ \vb{v}\cdot\vb{E}(t,\vb{v}t)\right],
\end{equation}
where the first term can be rewritten as 
\begin{equation}\label{eq2.5}
\dot{\vb{d}}(t)\cdot\vb{E}(t,\vb{v}t)=\frac{d}{dt}\left[\vb{d}(t)\cdot\vb{E}(t,\vb{v}t)\right]-\vb{d}(t)\cdot\frac{\partial}{\partial t}\vb{E}(t,\vb{v}t)-\left[\vb{v}\cdot\grad\right]\left[\vb{d}(t)\cdot\vb{E}(t,\vb{v}t)\right].
\end{equation}
Note the second term of Eq.~\eqref{eq2.4} and the last term of Eq.~\eqref{eq2.5} combine and give 
\begin{equation}\label{eq2.6}
\left[\vb{d}(t)\cdot\grad\right]\left[ \vb{v}\cdot\vb{E}(t,\vb{v}t)\right]-\left[\vb{v}\cdot\grad\right]\left[\vb{d}(t)\cdot\vb{E}(t,\vb{v}t)\right]=\left[\vb{d}(t)\cross \vb{v}\right]\cdot\left[\grad\cross\vb{E}(t,\vb{v}t)\right]=-\left[\vb{d}(t)\cross \vb{v}\right]\cdot\frac{\partial}{\partial t}\vb{B}(t,\vb{v}t),
\end{equation}
where we have used the Faraday's law in the last equality.
As a result, the power is written as
\begin{equation}\label{eq2.7}
P(t)=\frac{d}{dt}\left[\vb{d}(t)\cdot\vb{E}(t,\vb{v}t)\right]-\vb{d}(t)\cdot\frac{\partial}{\partial t}\vb{E}(t,\vb{v}t)-\left[\vb{d}(t)\cross \vb{v}\right]\cdot\frac{\partial}{\partial t}\vb{B}(t,\vb{v}t).
\end{equation}

The total derivative term in Eq.~\eqref{eq2.7} is generally present but it will not contribute to the quantum vacuum frictional power considered in this paper. The quantum vacuum frictional power is induced entirely by the quantum  fluctuations of the dipole operator $ \vb{d} $ or the field operator $ \vb{E} $. These fluctuations are determined by the imaginary part of the corresponding susceptibilities via the FDT. It follows from the linear response assumption (embedded in the FDT) and the time-translational invariance of the  susceptibilities that the dipole interaction energy $ \vb{d}\cdot \vb{E} $ is  time independent.
The next two terms in Eq.~\eqref{eq2.7} each contains a partial time derivative of the field operator. Indeed, in the above derivation, the prescription that only the time coordinates in the field operators should be differentiated follows straightforwardly from the Joule heating law. Similar to Eq.~\eqref{eq2.2}, we can now also write $ P $ as a time derivative of a free energy $ \mathcal{F} $,
\begin{equation}\label{eq2.8}
P(t)=\frac{\partial}{\partial t}\mathcal{F}=-\frac{\partial}{\partial t_{1}}\left[\vb{d}(t_{0})\cdot\vb{E}(t_{1},\vb{v}t)+\bm{\mu}_{\vb{v}}(t_{0})\cdot\vb{B}(t_{1},\vb{v}t)\right]|_{t_{0}=t_{1}\to t},
\end{equation}
where we have identified $ \vb{d}(t)\cross\vb{v}=\bm{\mu}_{v}(t) $ as the magnetic dipole moment induced by the movement of the electric dipole. The induced magnetic dipole term in $ \mathcal{F} $ vanishes if we set $ \vb{v}=\vb{0} $ in Eq.~\eqref{eq2.8} and the expression for $ P' $ in Eq.~\eqref{eq2.2} is reproduced.

If we apply the same reasoning for the frictional force, we obtain a proof for the equivalence of the Lorentz force law and the principle of virtual work. This is  detailed in Appendix \ref{apC}.

%we will consider the neutral particle to be dissipative intrinsically, that is, its intrinsic polarizability $ \bm{\alpha}(\omega) $ has an imaginary part before it is dressed by the radiation. A nonvanishing imaginary part of the intrinsic polarizability gives the particle freedom to absorb net energy, 

%The particle moves in vacuum (free space) and interacts with the surrounding blackbody radiation. The vacuum Green's dyadic of the electromagnetic field naturally develops imaginary part. We refer the readers to Appendix A in Ref.~\cite{Xin:eqf1} for details. According to FDT, both dipole fluctuations and field fluctuations exist in the combined system of particle and blackbody radiation.  
The NESS condition can be expressed as either $ P-Fv=0 $ in frame $ \mathcal{R} $ or $ P'=0 $ in frame $ \mathcal{P} $. Let us work in frame $ \mathcal{P} $, where the calculation is simpler. 

Recall the FDT for dipole fluctuations and field fluctuations in frame $ \mathcal{P} $:
\begin{subequations}\label{FDTs}
\begin{equation}\label{eq2.9}
\langle \vb{d}'(\omega')\vb{d}'(\nu') \rangle =2\pi \delta (\omega'+\nu') \Im \bm{\alpha}(\omega')\coth\left(\frac{\beta'\omega'}{2}\right),
\end{equation}
\begin{equation}
\label{eq2.10}
\langle\vb{E'}(\omega',\vb{k}'_{\perp};z')\vb{E'}(\nu',\bar{\vb{k}}'_{\perp};\bar{z}')\rangle=(2\pi)^{3}\delta(\omega'+\nu')\delta^{(2)}(\vb{k}'_{\perp}+\bar{\vb{k}}'_{\perp}) \Im \vb{g}'(\omega', \vb{k}'_{\perp};z',\bar{z}')\coth\left(\frac{\beta\gamma(\omega'+k_{x}'v)}{2}\right),
\end{equation}
\end{subequations}
where $ \beta'=1/T' $ is the inverse temperature of the particle while $ \beta=1/T $ is the inverse temperature of the radiation, each in their respective frame and $ \gamma $ is the relativistic dilation factor. Here, $ \vb{g}' $ is the reduced Green's dyadic, which is the Fourier transform of the retarded Green's dyadic,
\begin{equation}\label{g}
\vb{\Gamma}(\omega; \vb{r}, \vb{r'})=\int \frac{d^{2}\vb{k}_{\perp}}{(2\pi)^{2}} e^{i\vb{k}_{\perp}\cdot (\vb{r}_{\perp}-\vb{r'}_{\perp})} \vb{g}(\omega,\vb{k}_{\perp}; z, z'),
\end{equation}
where $\vb{\Gamma} $ is only transformed in the transverse spatial directions so that our analysis can be readily extended to a geometry with a planar symmetry. The specific form of the Green's dyadic is detailed in our previous paper, Ref.~\cite{Xin:eqf1}.

In general, the symbol $ \Im $ in Eq.~\eqref{FDTs} denotes the generalized imaginary part of the corresponding matrices, i.e., 
\begin{equation}
\Im \bm{\chi}=\frac{\bm{\chi}-\bm{\chi}^{\dagger}}{2i},
\end{equation}
so that $ \Im\bm{\chi} $ is Hermitian, which is required because the product of operators in Eq.~\eqref{FDTs} are symmetrized. In the calculation for the quantum vacuum friction, however, $ \Im \vb{g}' $ reduces to the ordinary imaginary part, since the vacuum Green's dyadic is symmetric. And as we will see, only the diagonal components of the polarization tensor $ \bm{\alpha} $ contributes to the quantum vacuum friction due to the additional symmetry of the vacuum problem.

%Here, we only transform the the coordinates in the two transverse directions so that our formulation could be easily extended to more general backgrounds than vacuum, which are not translational invariant in $ z $ direction.
%If the atom is dissipative, $ \Im\alpha(\omega)\neq 0 $, the dipole fluctuation is no longer solely induced by the field fluctuation but has to be treated explicitly. In this more general situation, the energy of the atom is not necessarily conserved, which means the atom can be out of NESS. Therefore, the power-force relation $ P=Fv $ is not always guaranteed but becomes a condition to characterize NESS.
%The excess power $ \tilde{P}=P-Fv $ can be viewed as a measure of the deviation of the system from NESS and the NESS condition can be restated as
%\begin{equation}\label{eq5-0.4}
%\tilde{P}=0.
%\end{equation}
%It turns out to be a relation between the inverse temperature of the atom $ \beta' $ and the inverse temperature of the blackbody radiation $ \beta $. The specific form of the NESS condition is dependent on the specific model for the atom's polarizability and the polarization state of the atom.
%
%The analysis outlined in Eq.~\eqref{eq4-1} is general and can be used even out of NESS.
%It follows from Eq.~\eqref{eq4-1} that
%\begin{equation}\label{eq5-1}
%\tilde{P}=\frac{1}{\gamma^{2}}P'=\frac{1}{\gamma^{2}}\frac{\partial}{\partial t'}\mathcal{F'}.
%\end{equation}

The interaction free energy in $ \mathcal{P} $ is
\begin{equation}\label{eq5-3}
 \mathcal{F}'(t')=-\vb{d}'(t')\cdot \vb{E}'(t',\vb{0})=-\int\frac{d\omega}{2\pi} e^{-i\omega t'} \vb{d}'(\omega) \cdot \int \frac{d\nu}{2\pi} e^{-i\nu t'} \vb{E}'(\nu;\vb{0}).
\end{equation}
To leading order in the intrinsic polarizability $ \bm{\alpha}(\omega) $, the free energy is split into a dipole-fluctuation-induced part
\begin{subequations}\label{eq5-4}
\begin{equation}\label{eq5-4a}
\mathcal{F}'_{dd}(t')=-\int\frac{d\omega}{2\pi} e^{-i\omega t'} \left\langle\vb{d}'(\omega) \cdot \int \frac{d\nu}{2\pi} e^{-i\nu t'}\int\frac{d^{2}\vb{k}_{\perp}}{(2\pi)^{2}} \vb{g}'(\nu,\vb{k}_{\perp})\cdot\vb{d}'(\nu) \right\rangle,
\end{equation}
and a field-fluctuation-induced part
\begin{equation}\label{eq5-4b}
 \mathcal{F}'_{EE}(t')=-\int\frac{d\omega}{2\pi} e^{-i\omega t'}\left\langle \left(\bm{\alpha}(\omega)\cdot\int\frac{d^{2}\vb{k}_{\perp}}{(2\pi)^{2}}\vb{E}'(\omega,\vb{k}_{\perp})\right) \cdot \int \frac{d\nu}{2\pi} e^{-i\nu t'} \int\frac{d^{2}\vb{k}'_{\perp}}{(2\pi)^{2}} \vb{E}'(\nu,\vb{k}'_{\perp})\right\rangle.
\end{equation}
\end{subequations}
%The correlation of the field operators can be evaluated by the FDT in Eq.~\eqref{eq4-8}. The FDT for the correlation of the dipole operators is
%\begin{equation}\label{eq5-6}
%\langle d_{i}'(\omega)d_{j}'(\nu) \rangle =2\pi \delta (\omega+\nu) \Im \alpha_{ij}(\omega)\coth\left(\frac{\beta'\omega}{2}\right).
%\end{equation}

Now we proceed to differentiate $ \mathcal{F'} $ for the power $ P' $ using Eq.~\eqref{eq2.2}. With the understanding that the derivative is only taken with respect to the time dependence of the original field operator $ \vb{E}'(t',\vb{0}) $ in Eq.~\eqref{eq5-3}, we find
\begin{subequations}
\begin{equation}\label{eq5-7}
P'_{dd}=\frac{\partial}{\partial t'}\mathcal{F'}_{dd}=-\int\frac{d\omega}{2\pi}\frac{d^{2}\vb{k}_{\perp}}{(2\pi)^{2}} \,\omega \tr \Im \bm{\alpha}(\omega) \cdot \Im \vb{g}'(\omega,\vb{k}_{\perp}) \coth\frac{\beta'\omega}{2},
\end{equation} 
\begin{equation}\label{eq5-8}
P'_{EE}=\frac{\partial}{\partial t'}\mathcal{F'}_{EE}=\int\frac{d\omega}{2\pi}\frac{d^{2}\vb{k}_{\perp}}{(2\pi)^{2}} \,\omega \tr \Im \bm{\alpha}(\omega)\cdot \Im \vb{g}'(\omega,\vb{k}_{\perp}) \coth\left[\frac{\beta}{2}\gamma(\omega+k_{x}v)\right].
\end{equation}
\end{subequations}
The $ dd $ and $ EE $ contributions are summed to yield $ P' $
\begin{equation}\label{eq5-9}
P'=\int\frac{d\omega}{2\pi}\frac{d^{2}\vb{k}_{\perp}}{(2\pi)^{2}}\, \omega \tr \Im \bm{\alpha}(\omega)\cdot \Im \vb{g}'(\omega,\vb{k}_{\perp}) \left\lbrace\coth\left[\frac{\beta}{2}\gamma(\omega+k_{x}v)\right]-\coth(\frac{\beta'\omega}{2})\right\rbrace.
\end{equation} 
For the vacuum background, the Green's dyadic is invariant in different frames,
%The derivation above uses the rest frame method introduced in Sec.~\ref{rest}. If we were to compute $ \tilde{P}=P-Fv $ in the rest frame of blackbody radiation, we would find exactly the same result as in Eq.~\eqref{eq5-9}. The advantage of quantization in the rest frame of atom is that it makes finding the contributions from different polarization states of the atom very convenient. For example, the diagonal contributions for the vacuum frictional power are
\begin{align}\label{eqdyadic}
\vb{g}'(z,z';\omega,\vb{k}_{\perp})=\vb{g}(z,z';\omega,\vb{k}_{\perp})=\frac{1}{2\kappa}e^{-\kappa|z-z'|}
&\mqty(\omega^{2}-k_{x}^{2}
&
-k_{x}k_{y}
&
-ik_{x}\kappa\sgn(z-z')
\\
-k_{x}k_{y}
&
\omega^{2}-k_{y}^{2}
&
-ik_{y}\kappa\sgn(z-z')
\\
ik_{x}\kappa\sgn(z'-z)
&
\quad ik_{y}\kappa\sgn(z'-z)
&
k^{2}).
\end{align}
When the Green's functions in Eq.~\eqref{eqdyadic} are inserted into Eq.~\eqref{eq5-9}, it is immediately seen that the off-diagonal polarizations ($ i\ne j$)  do not contribute to $ P' $ because of the oddness of the integrand in $ k_{y} $ or the vanishing of the signum function in the coincident spatial coordinate limit. The diagonal contributions are found to be, respectively,
\begin{subequations}\label{eq5-10}
\begin{equation}\label{eq5-10a}
P'^{\rm{X}}=\frac{1}{4\pi^{2}\gamma v}\int_{0}^{\infty} d\omega \Im \alpha_{xx}(\omega)\, \omega^{4} \int_{y_{-}}^{y_{+}} dy \left[1-\frac{1}{\gamma^{2}v^{2}}(y-\gamma)^{2}\right]\left[\frac{1}{e^{\beta\omega y}-1}-\frac{1}{e^{\beta'\omega}-1}\right],
\end{equation}
\begin{equation}\label{eq5-10b}
P'^{\rm{Y}}=\frac{1}{8\pi^{2}\gamma v}\int_{0}^{\infty} d\omega \Im \alpha_{yy}(\omega)\, \omega^{4} \int_{y_{-}}^{y_{+}} dy \left[1+\frac{1}{\gamma^{2}v^{2}}(y-\gamma)^{2} \right]\left[\frac{1}{e^{\beta\omega y}-1}-\frac{1}{e^{\beta'\omega}-1}\right],
\end{equation}
\end{subequations}
where the integral limits on $ y $ are $ y_{-}=\gamma (1-v)$ and $ y_{+}=\gamma (1+v) $. Note we have taken advantage of the integrand's evenness in $ \omega $ and the cancellation of the divergent piece in Eq.~\eqref{eq5-10}.
If the particle is isotropic, the contributions to $ P' $ from all diagonal polarization states sum to
\begin{equation}\label{eq5-11}
P'^{\rm{ISO}}=\frac{1}{2\pi^{2}\gamma v}\int_{0}^{\infty} d\omega \Im \alpha(\omega)\, \omega^{4} \int_{y_{-}}^{y_{+}} dy \left[\frac{1}{e^{\beta\omega y}-1}-\frac{1}{e^{\beta'\omega}-1}\right].
\end{equation}
Using the momentum distribution functions introduced in Ref.~\cite{Xin:eqf1},
\begin{equation}\label{eqfP}
\addtolength{\arraycolsep}{-3pt}
f^{\rm{P}}(y)=
\left\{%
 \begin{array}{lcrcl}
 \frac{3}{2\gamma v},& \qquad \rm{P}=\rm{ISO}\\\\
\frac{3}{4\gamma v} \left[1-\frac{1}{\gamma^{2}v^{2}}(y-\gamma)^{2}\right],& \qquad \rm{P}=\rm{X}\\\\
\frac{3}{8\gamma v}\left[1+\frac{1}{\gamma^{2}v^{2}}(y-\gamma)^{2}\right],& \qquad \rm{P}=\rm{Y}, \rm{Z}\\
 \end{array}
 \right.
\end{equation}
these formulas can be summarized as
\begin{equation}\label{eqPprime}
P'^{\rm{P}}=\frac{1}{3\pi^{2}}\int_{0}^{\infty} d\omega \Im \alpha_{\rm{P}}(\omega) \,\omega^{4}\int_{y_{-}}^{y_{+}} dy f^{\rm{P}}(y)\left(\frac{1}{e^{\beta\omega y}-1}-\frac{1}{e^{\beta'\omega}-1}\right).
\end{equation}

%The NESS conditions $ P'=0 $ for different polarization states can therefore be summarized as
%\begin{equation}\label{eq5-11.5}
%\int_{0}^{\infty} d\omega \Im\alpha(\omega)\, \omega^{4} \int_{y_{-}}^{y_{+}} dy \frac{f^{\rm{P}}(y)}{e^{\beta'\omega}-1}=\int_{0}^{\infty} d\omega \Im\alpha(\omega)\, \omega^{4} \int_{y_{-}}^{y_{+}} dy \frac{f^{\rm{P}}(y)}{e^{\beta\omega y}-1},
%\end{equation}
%where $ \rm{P} $ denotes different polarization states with corresponding characteristic functions
%\begin{equation}\label{eqP}
%\addtolength{\arraycolsep}{-3pt}
%f^{\rm{P}}(y)=
%\left\{%
% \begin{array}{lcrcl}
% 2,& \qquad \rm{P}=\rm{ISO}\\\\
% y^{2}-\left(y-\frac{1}{\gamma}\right)^{2}\frac{1}{v^{2}},& \qquad \rm{P}=\rm{X}\\\\
% 1-\frac{1}{2}\left[y^{2}-\left(y-\frac{1}{\gamma}\right)^{2} \frac{1}{v^{2}}\right],& \qquad \rm{P}=\rm{Y}.\\
% \end{array}
% \right.
%\end{equation} 

\subsection{The NESS temperature of the neutral particle and its lower bound}
Given a particular model for the particle's polarizability, the NESS condition, $ P'^{P}=0 $, 
defines a special NESS temperature of the particle $ \tilde{T} $ (the corresponding inverse temperature is denoted as $ \tilde{\beta} $) for a fixed radiation temperature $ T $ and polarization state $ \rm{P} $.

Interestingly, we find that, independent of the model for $ \Im\alpha(\omega) $ and the polarization state of the particle, the NESS temperature of the particle $ \tilde{T} $ must be greater than the Planck-Einstein transformed temperature of the blackbody radiation $ T/\gamma $ \cite{Farias:temperature}. 
We need the following assumptions to prove this theorem:
\begin{equation}\label{assumption}
\Im\alpha(\omega)\ge 0 \quad \rm{and} \quad  \lim_{\omega\to 0}\omega^{4}\Im\alpha(\omega)=0.
\end{equation}
The first assumption is usually satisfied by a realistic particle made of ordinary lossy materials~\cite{Jung:nanowire}. The second assumption is to avoid an infrared divergence in the power formula Eq.~\eqref{eqPprime}.

Let us define a function
\begin{equation}\label{eq5-21}
I(\xi)=\int_{0}^{\infty} d\omega \, \Im\alpha(\omega)\,\omega^{4}\frac{1}{e^{\xi\omega}-1}.
\end{equation}
The general NESS condition $ P'^{P}=0 $ for different polarization states can then be written as
\begin{equation}\label{eq5-22}
\int_{y_{-}}^{y_{+}} dy \, f^{\rm{P}}(y)I(\tilde{\beta})=\int_{y_{-}}^{y_{+}} dy \,f^{\rm{P}}(y)I(\beta y).
\end{equation}
Noting $ \gamma $ is the midpoint of the interval $ [y_{-},y_{+}] $ and $ I(\xi) $ is a decreasing and convex function, that is, $ I'(\xi)<0 $ and $ I''(\xi)>0 $, the following inequality follows,
\begin{equation}\label{eq5-23}
\int_{y_{-}}^{y_{+}} dy \, I(\beta y)>\int_{y_{-}}^{y_{+}} dy \, I(\beta \gamma).
\end{equation}
Because the momentum distribution functions $ f^{\rm{P}}(y) $ are all even with respect to $ y=\gamma $, we also have
\begin{equation}\label{eq5-23.1}
\int_{y_{-}}^{y_{+}} dy \, f^{\rm{P}}(y)I(\beta y)>\int_{y_{-}}^{y_{+}} dy \, f^{\rm{P}}(y)I(\beta \gamma).
\end{equation}
Combining Eq.~\eqref{eq5-23.1} and Eq.~\eqref{eq5-22}, we find
\begin{equation}\label{eq5-23.5}
\int_{y_{-}}^{y_{+}} dy \, f^{\rm{P}}(y)I(\tilde{\beta})>\int_{y_{-}}^{y_{+}} dy \, f^{\rm{P}}(y)I(\beta \gamma).
\end{equation}
Since both $ I(\tilde{\beta}) $ and $ I(\beta \gamma) $ are independent of $ y $, they can be taken out of the integral. In addition, the remaining integral $\int_{y_{-}}^{y_{+}} dy \, f^{\rm{P}}(y)$ is positive definite. It then follows that $  I(\tilde{\beta})>I(\beta\gamma)$. Recalling $ I(\xi) $ is a decreasing function, we conclude
\begin{equation}\label{eq5-23.6}
\tilde{T}>\frac{T}{\gamma}.
\end{equation}
The theorem therefore predicts a lower bound for the NESS temperature of the particle.

The nonrelativistic limit of the NESS temperature is also independent of the model for $ \Im\alpha (\omega) $ and the polarization state. Expanding the NESS condition Eq.~\eqref{eq5-22} in $ v $ and keeping only the terms first order in $ v $, we find the nonrelativistic limit of the NESS temperature is precisely the temperature of the surrounding blackbody radiation, 
\begin{equation}\label{eqNR}
\tilde{T}\sim T, \qquad v\to 0.
\end{equation}
%For the isotropic case, the $ y $ integral in the NESS condition Eq.~\eqref{eq5-11.5} can in fact be readily carried out, leading to the formula
% \begin{equation}\label{eq5-26}
% \int_{0}^{\infty} d\omega \Im\alpha(\omega)\,\omega^{4}\, 2\gamma v\frac{1}{e^{\beta'\omega}-1}=\int_{0}^{\infty} d\omega \Im\alpha(\omega)\,\omega^{4}\frac{1}{\beta\omega}\ln\left[\frac{1-e^{-\beta\omega y_{+}}}{1-e^{-\beta\omega y_{-}}}\right].
% \end{equation}
%If we expand both sides of Eq.~\eqref{eq5-26} in $ v $ and keep only the linear terms, we find Eq.~\eqref{eq5-26} reduces to $ \beta=\beta' $. In the nonrelativistic limit, the NESS temperature of the atom in its rest frame equals the temperature of the blackbody radiation in its rest frame, independent of the model for the polarizability.
Therefore, the deviation of the NESS temperature of the particle from the temperature of the radiation is a relativistic effect. 

\subsection{The NESS temperature ratio for resonance and monomial models}
Let us define the NESS temperature ratio to be 
\begin{equation}\label{eqrtilde}
\tilde{r}=\frac{\tilde{T}}{T}.
\end{equation}
In order to study the behavior of $ \tilde{r} $ concretely, we will discuss two ideal models here, namely the resonance model and the monomial model. The more realistic situation that we consider later reduces to these ideal models in various limits. We will also work with isotropic particles, the momentum distribution functions for which are much simpler.

The resonance model is characterized by a sharp resonance at $ \omega_{0} $,
\begin{equation}\label{resonance}
\Im\alpha(\omega)\propto\delta(\omega-\omega_{0}),
\end{equation}
where the numerical coefficient of the delta function is irrelevant for studying the NESS temperature ratio $ \tilde{r} $.
Inserting Eq.~\eqref{resonance} into Eq.~\eqref{eq5-22} for the isotropic polarization,  $ \tilde{r}$ can be determined in terms of the dimensionless frequency $ x_{0}=\beta\omega_{0}/2 $,
\begin{equation}\label{resratio}
\tilde{r}=x_{0}\left\lbrace\coth^{-1}\left[\frac{1}{2\gamma v x_{0}}\ln\frac{\sinh(x_{0}y_{+})}{\sinh(x_{0}y_{-})}\right]\right\rbrace^{-1}.
\end{equation}

\begin{figure*}[h!]
\subfloat[]{\label{figsub1}%
\includegraphics[width=0.48\linewidth]{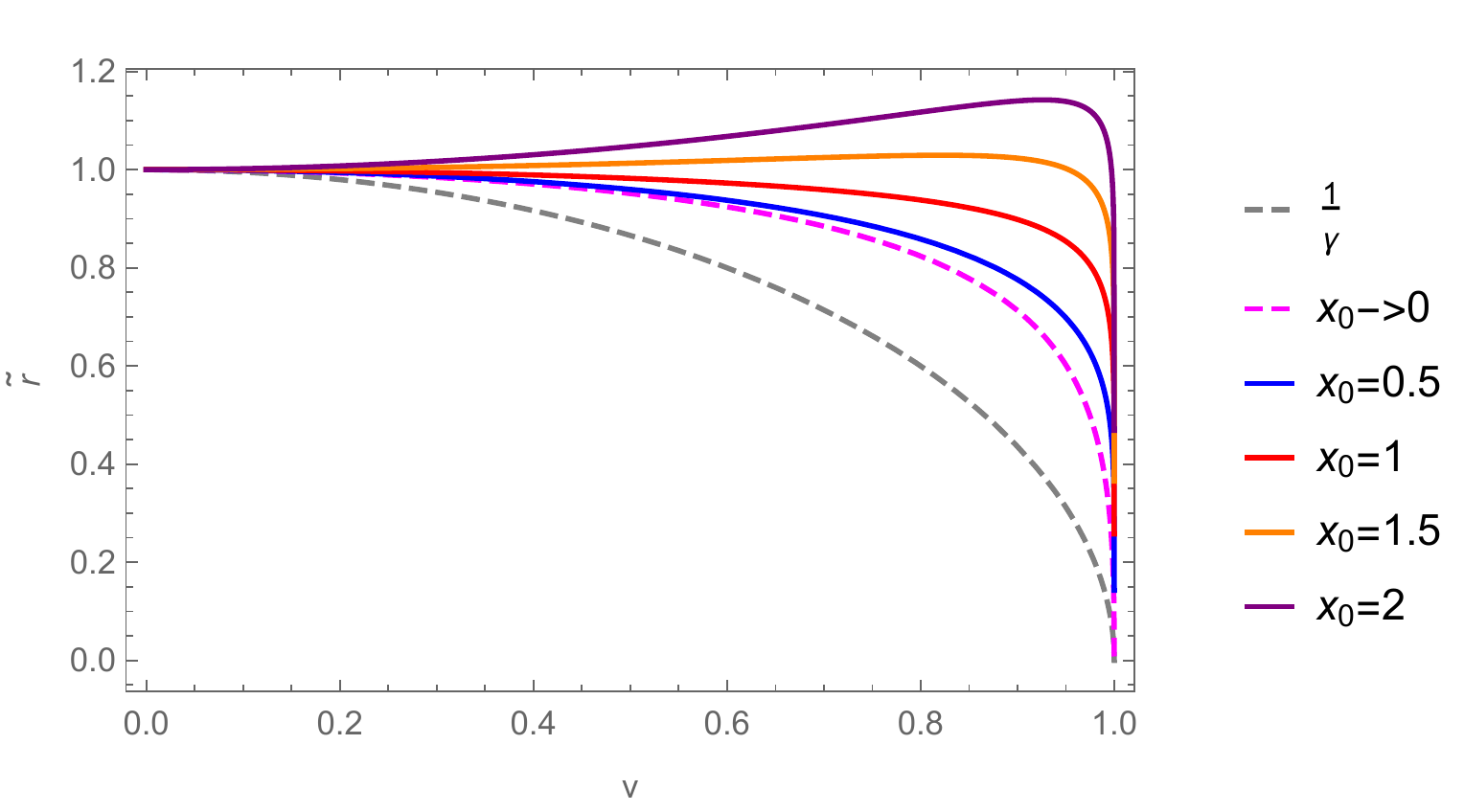}%
}
\subfloat[]{\label{figsub2}%
\includegraphics[width=0.48\linewidth]{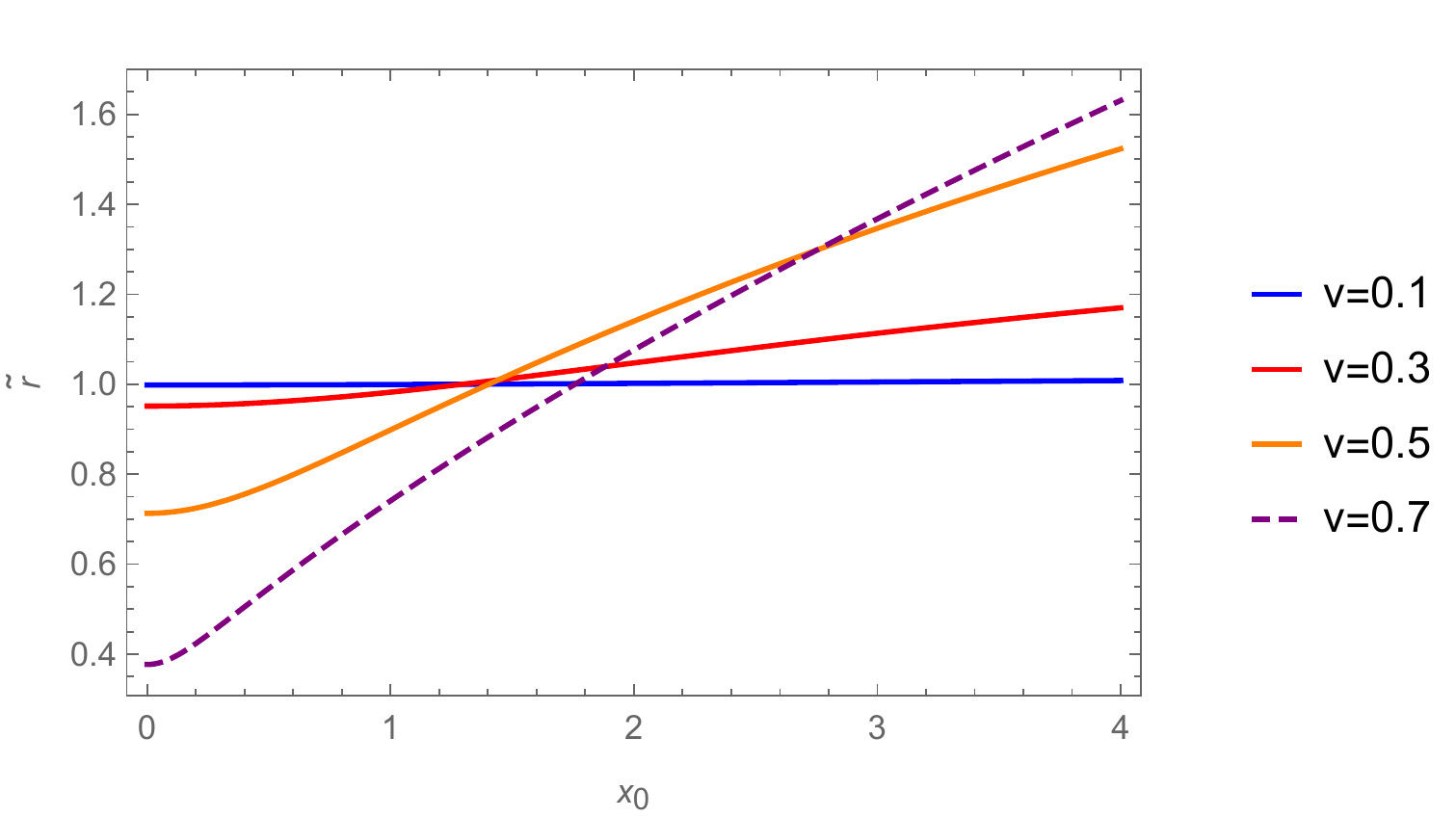}%
}
\caption{For the resonance model, the NESS temperature ratio $ \tilde{r}$ in Eq.~\eqref{resratio} is illustrated. (a) $\tilde{r}$ is plotted as a function of the velocity $ v $ for various values of  $ x_{0} $. All of the curves with different $ x_{0} $ drop to $ 0 $ in the limit of $ v\to 1 $. (b) $\tilde{r}$ is plotted as a function of $x_{0}$ for different values of $v$.}
\label{FIGresNESS}
\end{figure*}

The NESS temperature ratio $ \tilde{r} $ determined by Eq.~\eqref{resratio} is plotted as a function of velocity $ v $ for different $ x_{0} $ in Fig.~\ref{figsub1}. As seen in Fig.~\ref{figsub1}, the temperature ratio increases with $ x_{0} $ for a given velocity $ v $. Therefore, the ratio is bounded below by its zero frequency/high-temperature limit ($ \omega_{0}\to 0 $ or $ \beta\to 0 $), 
\begin{equation}
\tilde{r}\sim\frac{1}{\gamma v}\ln y_{+}, \qquad x_{0}\to 0,
\end{equation}
which is illustrated by the dashed magenta curve in Fig.~\ref{figsub1}. Note that this is still well above the dashed grey curve, which is the lower bound of the NESS temperature ratio given by  Eq.~\eqref{eq5-23.6}, $ 1/\gamma $.
In addition, independent of the resonance frequency or radiation temperature, the NESS temperature ratio $ \tilde{r} $ drops to zero as the velocity of the particle approaches the speed of light.
In fact, the ultrarelativistic limit of $ \tilde{r} $ can be shown to be
\begin{equation}\label{reslimit}
\tilde{r}\sim\frac{2x_{0}}{\ln(4\gamma x_{0})-\ln\ln (\gamma/x_{0})}, \qquad \gamma\to\infty,
\end{equation}
so the decay of NESS temperature of the particle is logarithmic in $ \gamma $. 

In Fig.~\ref{figsub2}, the NESS temperature ratio $ \tilde{r} $ is plotted as a function of $ x_{0}$ for various velocities $ v $. It is confirmed that $ \tilde{r} $ generally grows with $ x_{0} $. The figure also shows that the separatrix between $\tilde{T}/T>1$ and $\tilde{T}/T<1$ is close to $x_{0}=1.3$ for most velocities unless the particle becomes quite relativistic. To the order of $ v^{2} $, the position of the separatrix is found to be $ x_{0}=1.288 $ by setting $ \tilde{r}=1 $ in Eq.~\eqref{resratio}.  

We now turn to the monomial model,
\begin{equation}\label{monomial}
\Im\alpha(\omega)\propto \omega^{n}.
\end{equation}
Of course, physically, $ n $ must be an odd integer to make sure $ \alpha(t-t') $ is real. We will ease this restriction in the following discussion because the NESS temperature ratio $ \tilde{r} $ is mathematically well defined for monomial models with any real power $ n $. However, we will see in the next section that the frictional force would only be convergent for monomial models with $ n > -4 $.
%Physically, $ n $ must be an odd integer to respect the Kramers-Kronig relation. But the following arguments are independent of this physical requirement. 

Inserting Eq.~\eqref{monomial} into the NESS condition Eq.~\eqref{eq5-22} for the isotropic polarization, it's easier to first take the integral on $ \omega $ and then that on $ y $ with the result
%\begin{equation}\label{eq5-insert}
% \int_{y_{-}}^{y_{+}} dy \left(\frac{1}{\beta'}\right)^{5+n}= \int_{y_{-}}^{y_{+}} dy \left(\frac{1}{\beta y}\right)^{5+n}.
%\end{equation}
%After taking the $ y $ integral in Eq.~\eqref{eq5-insert}, the NESS temperature ratio is found to be
\begin{equation}\label{eq5-19}
\tilde{r}=\left[\frac{1}{2\gamma v(4+n)}\left(y_{+}^{4+n}-y_{-}^{4+n}\right)\right]^{\frac{1}{5+n}}.
\end{equation}
In particular, $ n=3 $ corresponds to the low-frequency radiation-reaction model \cite{Xin:eqf1}, for which the NESS temperature ratio is 
\begin{equation}
\tilde{r}=\left[\gamma^{6}\left(1+5v^{2}+3v^{4}+\frac{1}{7}v^{6}\right)\right]^{\frac{1}{8}},
\end{equation}
which agrees with the NESS temperature ratio obtained from $ P'=0$ given in Eq.~(8.150) of Ref.~\cite{Volokitin:book}.

The low velocity approximation of the NESS temperature ratio for a general power $ n $ is readily obtained by expanding Eq.~\eqref{eq5-19} in $ v $,
\begin{equation}
v\ll 1 : \qquad \tilde{r}\to 1+\frac{1}{6}(n+3)v^{2}+\cdots.
\end{equation}
Just as expected, the NESS temperature of the particle and the radiation temperature coincide in the nonrelativistic limit. The high velocity behavior of the NESS temperature ratio, however, is rather different for different values of $ n $,
\begin{align}\label{eq5-20}
\gamma\gg 1 &: \qquad \tilde{r}\to\left\{%
 \begin{array}{lcrcl}
\left[\frac{(2\gamma)^{3+n}}{4+n}\right]^{\frac{1}{5+n}}, &  \qquad n>-4\\\\
\left[\frac{1}{-(4+n)}\right]^{\frac{1}{5+n}}\frac{1}{2\gamma},& \qquad n<-4\\\\
\frac{\ln 2\gamma}{\gamma}, & \qquad n=-4.
 \end{array}
 \right.
\end{align}
We note $ n=-5 $ is included in the second situation, where the ratio evaluates to $ e/2\gamma $. In the limit of $ \gamma\to\infty $, the NESS temperature ratio diverges for $ n>-3 $ but vanishes for $ n<-3 $. Only for $ n=-3 $, as we can see from the exact expression Eq.~\eqref{eq5-19}, the temperature ratio for the $ n=-3 $ model equals $ 1 $ at all velocities.

\begin{figure}[h!]
 \includegraphics[width=0.7\linewidth]{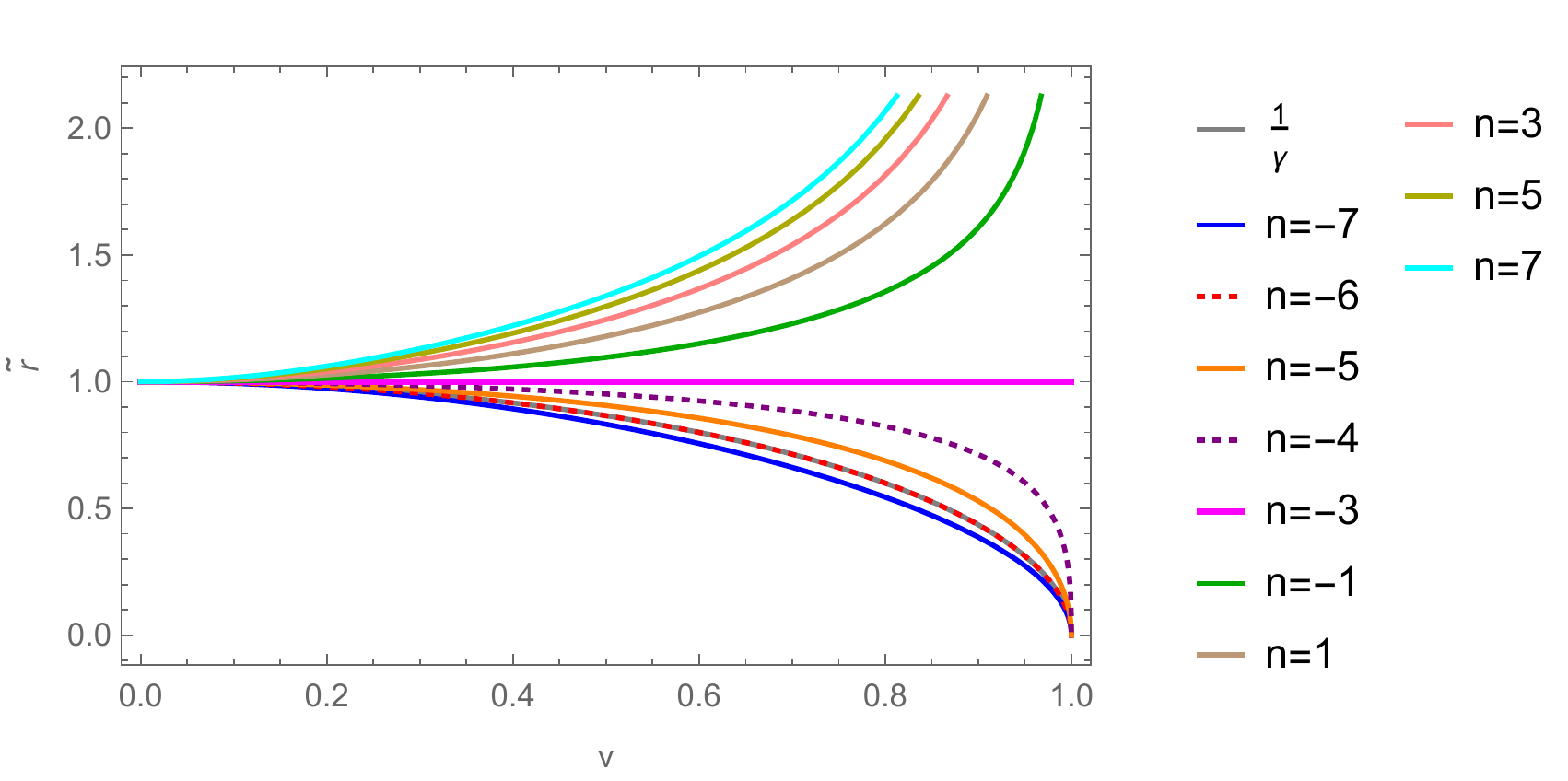}
 \caption{For the monomial model, the NESS temperature ratio  $\tilde{r}$ in Eq.~\eqref{eq5-19} is plotted as a function of the velocity $ v $ for various powers $ n $. The separatrix $ n=-3 $ is shown by the solid magenta curve. The curve for $ n=-4 $ is $ \ln y_{+}/\gamma v $ and the curve for $ n=-5 $ is $ e y_{-}^{1/v} $, obtained by taking the corresponding limits of Eq.~\eqref{eq5-19}. The red dotted curve for $ n=-6 $ exactly coincides with $ 1/\gamma $.
 \label{FIG1}}
 \end{figure}

These limits are more clearly seen and confirmed in Fig.~\ref{FIG1} where we plot the NESS temperature ratio $ \tilde{r} $ in Eq.~\eqref{eq5-19} as a function of the velocity $ v $ for various powers $ n $. At the same velocity, the temperature ratio is generally greater for a bigger power $ n $. Apart from $ n=-3 $, another special power is $ n=-6 $, for which the NESS temperature ratio precisely equals $ 1/\gamma $. As a result, the inequality Eq.~\eqref{eq5-23.6} actually holds beyond the assumption of the theorem Eq.~\eqref{assumption} ($ n>-4 $ for these monomial models). 

\section{Quantum vacuum friction in NESS}\label{sign}

\subsection{General features of quantum vacuum friction}\label{sec6a}
A formula for quantum friction in a general background, including both the $ \vb{dd} $ and $ \vb{EE} $ contributions has been been worked out in Ref.~\cite{Kim:dipole},
\begin{equation}\label{eq6-0}
F=\int_{-\infty}^{\infty} \frac{d\omega}{2\pi}\frac{d^{2}\vb{k}_{\perp}}{(2\pi)^{2}}(k_{x}+\omega v) \tr \Im \bm{\alpha}(\omega)\cdot \Im \vb{g}'(\omega, \vb{k}_{\perp})\left[\coth\left(\frac{\beta\gamma(\omega+k_{x}v)}{2}\right)-\coth\left(\frac{\beta'\omega}{2}\right)\right].
\end{equation}
Of course, one can derive the above formula by calculating the Lorentz force directly. Or, one can find $ F $ by applying the principle of virtual work in frame $\mathcal{R} $, $ F=-\partial_{x} \mathcal{F} $. Finally, one can also first apply the principle of virtual work in frame $ \mathcal{P} $, $ F'=-\partial_{x'} \mathcal{F'} $ and then find $ F $ using the relation $ F=F'+P'v $ \cite{Xin:eqf1}. The equivalence of the Lorentz force law and the principle of virtual work is established in Appendix \ref{apC} for a moving electric dipole.

For the vacuum background, we insert the Green's dyadic in Eq.~\eqref{eqdyadic} into Eq.~\eqref{eq6-0}. Again, due to the symmetry of the Green's dyadic, only the diagonal polarizations contributes to the quantum vacuum friction. The vacuum frictional force on a particle only polarizable in direction $ \rm{P} $ is found to be
\begin{equation}\label{eq6-1}
F^{\rm{P}}=\frac{1}{6\pi^{2}\gamma v}\int_{0}^{\infty} d\omega \,\omega^{4} \Im \alpha_{\rm{P}}(\omega)\int_{y_{-}}^{y_{+}} dy \left(y-\frac{1}{\gamma}\right)f^{\rm{P}}(y) \left[\coth(\frac{\beta\omega y}{2})-\coth(\frac{\beta'\omega}{2})\right],
\end{equation}
where the momentum distribution functions $ f^{\rm{P}}(y) $ are illustrated in Eq.~\eqref{eqfP}. In Eq.~\eqref{eq6-1}, the first term in the bracket originates from the field fluctuations ($ \mathbf{EE} $) and the second term from the dipole fluctuations ($ \mathbf{dd} $). Both the $ \mathbf{EE} $ and $ \mathbf{dd} $ contributions have ultraviolet divergent pieces unless $ \Im\alpha(\omega) $ falls off faster than $ 1/\omega^{5} $ for high frequencies. This, however, is not an issue for the total frictional force because of the exact cancellation of the divergent pieces between the two contributions. After the cancellation, the frictional force becomes
\begin{equation}\label{FISO}
F^{\rm{P}}=\frac{1}{3\pi^{2}\gamma v}\int_{0}^{\infty} d\omega \,\omega^{4} \Im \alpha_{\rm{P}}(\omega)\int_{y_{-}}^{y_{+}} dy \left(y-\frac{1}{\gamma}\right)f^{\rm{P}}(y) \left[\frac{1}{e^{\beta\omega y}-1}-\frac{1}{e^{\beta'\omega}-1}\right].
\end{equation}
Therefore, it is crucial to keep both $ \mathbf{dd} $ and $ \mathbf{EE} $ contributions. Otherwise, one would encounter an unphysical divergence in the quantum vacuum friction. The quantum vacuum friction for the case of an isotropic particle has been worked out by various authors. Volokitin and Persson omitted the $ \mathbf{dd} $ contribution for the blackbody friction in one of their early papers on quantum friction \cite{Volokitin:2008}.\footnote{Even though the frictional force they got in Eq.~(49) of  Ref.~\cite{Volokitin:2008} agrees with our NESS quantum vacuum friction formula Eq.~\eqref{eq6-2} for the isotropic case, it is clear that they had not imposed the NESS condition and failed to consider an independent temperature of the particle because of  their omission of the $ \mathbf{dd} $ contribution.} This error was pointed out by Dedkov and Kyasov in Ref.~\cite{Dedkov:tangential} and it was corrected in later works of Volokitin and Persson, such as Ref.~\cite{Volokitin:book}. In addition, Pieplow and Henkel \cite{P&H:covariant} also consider both contributions. The quantum vacuum frictional force we obtain in our Eq.~\eqref{FISO} is in agreement with those found in Refs.~\cite{Dedkov:tangential, Volokitin:book,
P&H:covariant} where both $ \mathbf{dd} $ and $ \mathbf{EE} $ contributions to the force are correctly included.

In Ref.~\cite{Kim:dipole}, we have also pointed out that the quantum vacuum frictional force is not of a definite sign when both dipole  and field fluctuations coexist. Only when some particular model for polarizability is selected, for example, the monomial model with power $ n\ge -2 $, is the frictional force on an isotropic particle shown to be negative definite.\footnote{\label{ft4}For these monomial models with power $ n $, a simple calculation shows that the field fluctuation ($ \vb{EE} $) contribution to the quantum vacuum frictional force changes sign when $ n=-2 $. To be more specific, the $ \vb{EE} $ contribution is positive for $ n<-2 $, zero for $ n=-2 $ and negative for $ n>-2 $. The dipole fluctuation ($ \vb{dd} $) contribution is always negative, independent of the power $ n $.} 

In order to find out the quantum vacuum friction that the particle feels in NESS, we impose the NESS condition Eq.~\eqref{eq5-22} on Eq.~\eqref{FISO} so that the independent temperature of the particle could be eliminated in the expression for the force in favor of the the radiation temperature,
\begin{equation}\label{eq6-2}
\tilde{F}^{\rm{P}}=\frac{1}{3\pi^{2}\gamma v}\int_{0}^{\infty} d\omega \,\omega^{4} \Im \alpha(\omega)\int_{y_{-}}^{y_{+}} dy  \left(y-\gamma\right)f^{\rm{P}}(y) \frac{1}{e^{\beta\omega y}-1},
\end{equation}
where the tilde on $ F $ is to emphasize that it is the NESS quantum vacuum friction.
We note that the NESS frictional force is negative definite as long as the neutral particle is made of ordinary lossy material, $ \Im\alpha(\omega)\geq 0 $, independent of the particular model for the polarizability.

%A change of variable $ y-\gamma=\gamma z $ perhaps makes the claim more evident,
%\begin{equation}\label{eq6-2.2}
%\tilde{F}^{\rm{ISO}}_{dd+EE}=\frac{1}{2\pi^{2}v^{2}}\int_{0}^{\infty} d\omega\, \omega^{4} \Im \alpha(\omega)\int_{-v}^{v} dz  \frac{z}{e^{\beta\omega\gamma(z+1)}-1}.
%\end{equation}
%Therefore, the NESS frictional force must be a drag opposing the particle's motion, independent of the model for polarizability of the atom.

%Only the diagonal polarization states contribute to quantum vacuum friction  because the off-diagonal vacuum Green's functions are either odd in $ k_{y} $ or involve a factor of $ \sgn(z-z') $ which evaluates to zero in the coincident spatial coordinates limit. The NESS quantum vacuum friction for all the diagonal polarization states can be summarized into the following formula:
%\begin{equation}\label{eqFP}
%\addtolength{\arraycolsep}{-3pt}
%\tilde{F}^{\rm{P}}=\frac{1}{3\pi^{2}\gamma v}\int_{0}^{\infty} d\omega\,\omega^{4}\Im\alpha_{\rm{P}}(\omega)
%\int_{y_{-}}^{y_{+}} dy\frac{(y-\gamma)}{e^{\beta\omega y}-1}\,f^{\rm{P}}(y),
%\end{equation}
%where $ f^{\rm{P}}(y) $ is defined in Eq.~\eqref{eqfP}.

The polarizability in Eq.~\eqref{eq6-2} should be understood as the effective polarizability, including all possible sources of dissipation in its imaginary part. Let us temporarily resume our notation in Ref.~\cite{Xin:eqf1} for the effective polarizability, $ \hat{\bm{\alpha}} $, for clarity. Suppose the particle is dissipative only due to its interaction with the surrounding blackbody radiation, then the imaginary part of the effective polarizability is related to a real, intrinsic polarizability to the second order as
\begin{equation}\label{alphahat}
\Im\hat{\bm{\alpha}}(\omega)=\frac{\omega^{3}}{6\pi}\bm{\alpha}^{2}(\omega).
\end{equation}
If we substitute $ \omega^{3}\alpha_{\rm{P}}^{2}(\omega)/6\pi $ for $\Im \alpha_{\rm{P}}  $ in Eq.~\eqref{eq6-2}, the second order friction formulas Eqs.~(3.17), (3.20), (3.21), (3.22) in Ref.~\cite{Xin:eqf1} are recovered precisely.

\subsection{The limits of quantum vacuum friction in NESS}
From Eq.~\eqref{eq6-2}, it is not hard to see  that the NESS quantum vacuum friction vanishes both in the zero temperature limit $ \beta\to\infty $ and in the zero velocity limit $ v\to 0 $. 

Let us examine more closely, for the isotropic polarization, the velocity dependence of the NESS quantum vacuum friction Eq.~\eqref{eq6-2}, which is only contained in the following function
\begin{equation}\label{J}
J(x,v)=\frac{1}{\gamma^{2}v^{2}}\int_{y_{-}}^{y_{+}} dy \frac{y-\gamma}{e^{2xy}-1}, \quad x=\frac{\beta\omega}{2}.
\end{equation}
The function $ J(x,v) $ can be explicitly expressed as
\begin{equation}\label{Jexplicit}
J(x,v)=\frac{1}{2\gamma v x}\left[\ln(1-e^{-2xy_{-}})+\ln(1-e^{-2xy_{+}})\right]+\frac{1}{4\gamma^{2}v^{2}x^{2}}\left[\Li_{2}(e^{-2xy_{-}})-\Li_{2}(e^{-2xy_{+}})\right].
\end{equation}
Though this form for $ J(x,v) $ is exact, it is not particularly illuminating. See Eq.~\eqref{Jsum2} for a useful series representation of it.

In the low velocity limit, $ J(x,v) $ can be expanded in series of $ v $,
\begin{equation}\label{Jlowv}
J(x,v)\sim -\frac{v}{3}\frac{x}{\sinh^{2}(x)} -\frac{v^{3}}{60}\frac{x}{\sinh^{4}(x)}\left[-5+8x^{2}+(5+4x^{2})\cosh(2x)-10x\sinh(2x)\right],
\end{equation}
where we have kept the first two terms in the series. If we insert only the linear term in Eq.~\eqref{Jlowv} into Eq.~\eqref{eq6-2}, we obtain
\begin{equation}\label{eq6-2.4}
\tilde{F}^{\rm{ISO}}\sim -\frac{v}{6\pi^{2}}\int_{0}^{\infty} d\omega\, \omega^{4}\Im\alpha(\omega) \frac{\beta\omega/2}{\sinh^{2}(\beta\omega/2)},
\end{equation}
which is a generalization of the  Einstein-Hopf drag \cite{Einstein:Hopf, Mkrtchian:universal, Lach:EH}. In Ref.~\citep{Kim:dipole}, we obtained this formula by taking the nonrelativistic limit of Eq.~\eqref{eq6-1} and mandating the particle to have the same temperature as the blackbody radiation. But now we understand the physics better: the radiation temperature $ T $ is indeed the nonrelativistic limit of the NESS temperature $ \tilde{T} $ and the Einstein-Hopf drag is therefore simply the nonrelativistic NESS quantum vacuum friction!

%The contribution of the next term in Eq.~\eqref{Jlowv} to the quantum friction is
%\begin{equation}\label{eq6-2.6}
%\tilde{F}^{\rm{ISO} (3)}_{dd+EE}=-\frac{v^{3}}{120\pi^{2}}\int_{0}^{\infty} d\omega\, \omega^{4}\Im\alpha(\omega) \frac{\beta\omega/2}{\sinh^{4}(\beta\omega/2)} \left[-5+2\beta^{2}\omega^{2}+(5+\beta^{2}\omega^{2})\cosh(\beta\omega)-5\beta\omega\sinh(\beta\omega)\right].
%\end{equation}

Let us now turn to the high velocity limit of $ J(x,v) $. In Eq.~\eqref{J}, $ J(x, v) $ can be recast into a sum,
\begin{equation}\label{Jsum}
J(x,v)=\frac{1}{\gamma^{2}v^{2}}\sum_{k=1}^{\infty} \int_{y_{-}}^{y_{+}} dy \, (y-\gamma)e^{-2xyk}.
\end{equation}
The integral on $ y $ for each term in the sum is easily done, yielding
\begin{equation}\label{Jsum2}
J(x,v)=-\frac{1}{\gamma^{2}v^{2}}\sum_{k=1}^{\infty}\left\lbrace \frac{\gamma v}{2xk}\left[e^{-2xy_{+}k}+e^{-2xy_{-}k}\right]+\frac{1}{(2xk)^{2}}\left[e^{-2xy_{+}k}-e^{-2xy_{-}k}\right]\right\rbrace.
\end{equation}
This expression is so far exact and equivalent to Eq.~\eqref{Jexplicit}. It can be used to get an approximation for $ J(x,v) $ to arbitrary precision by truncating the series. In the high velocity limit with $ \gamma\gg 1 $ and $y_{+}\to 2\gamma$, we can drop the exponentials
involving $y_{+}$ in Eq.~\eqref{Jsum2}. If we further assume $ \gamma x \gg 1 $ (no assumption on the relative magnitude $ \gamma/x $ needs to be made), the expression in the braces can be approximated by only keeping the second term of the first bracket, after which the sum is easily written in closed form,
\begin{equation}\label{Jvhigh}
J(x,v)\sim\frac{1}{2x\gamma v}\ln\left(1-e^{-2xy_{-}}\right).
\end{equation}

\begin{figure*}[h!]
\subfloat[]{\label{figJa}%
\includegraphics[width=0.48\linewidth]{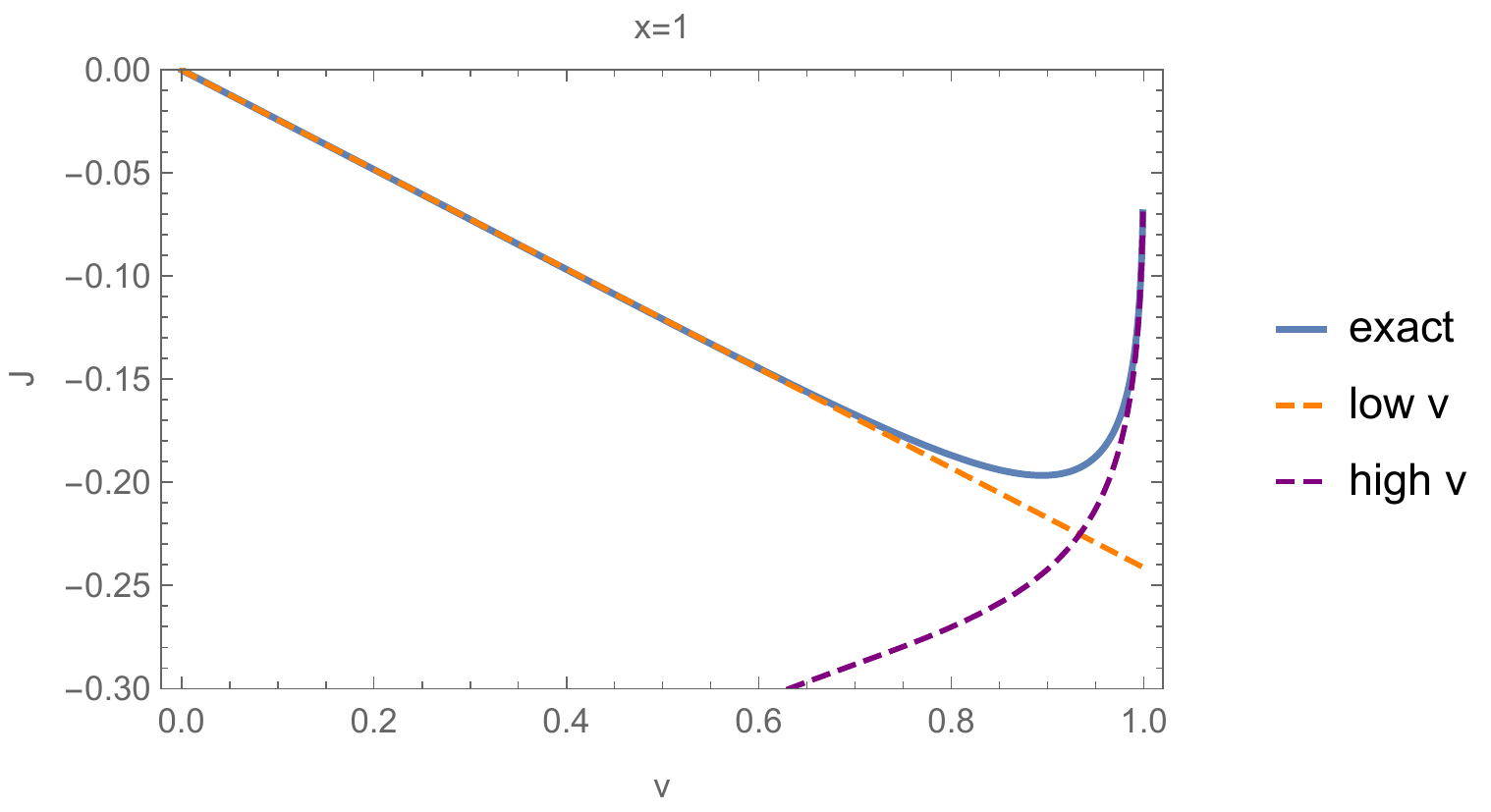}%
}
\subfloat[]{\label{figJb}%
\includegraphics[width=0.48\linewidth]{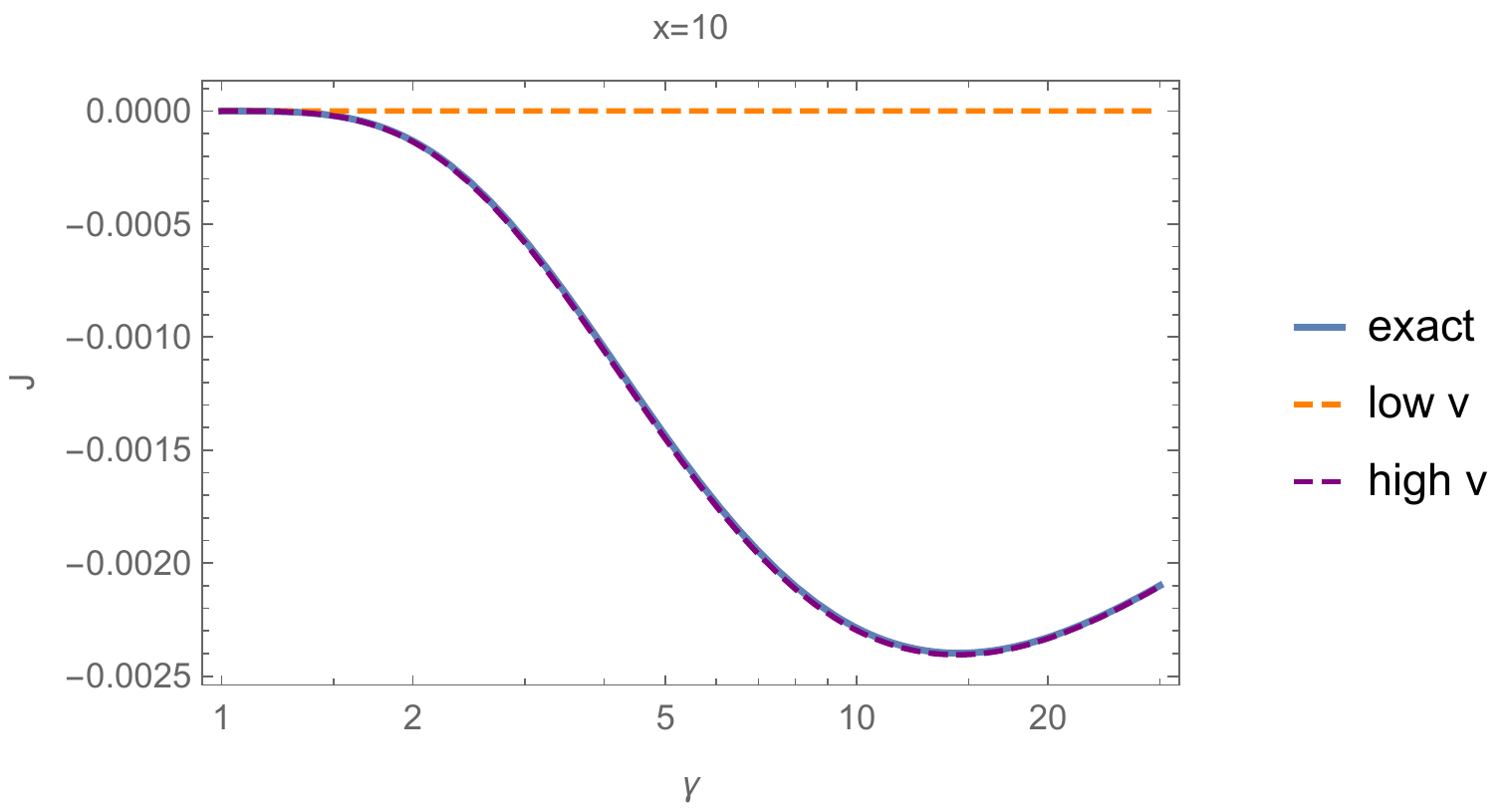}%
}
\caption{The velocity dependence of $ J $ in Eq.~\eqref{J} is illustrated. The exact expression in Eq.~\eqref{Jexplicit} is shown by the solid blue curves. The dashed orange curves show the linear term in Eq.~\eqref{Jlowv}. The dashed purple curves show the high velocity approximation of $ J $ in Eq.~\eqref{Jvhigh}. (a) $ J $ is plotted as a function of $ v $ for $ x=1 $. (b) $ J $ is plotted as a function of $ \gamma $ for $ x=10 $.} 
\label{FigJ}
\end{figure*}

In Fig.~\ref{FigJ}, we illustrate the velocity dependence of the exact function $ J $ together with its low velocity approximation [the linear term in Eq.~\eqref{Jlowv}] and high velocity approximation [Eq.~\eqref{Jvhigh}]. The exact function $ J $ clearly exhibits a minimum. That is to say, the magnitude of the contribution from a particular frequency $ x=\beta\omega/2 $ to the NESS quantum vacuum friction is maximized at an intermediate velocity. For smaller $ x $ (e.g., $x=1$), the minimum of $ J $ lies in the transition region between the low velocity regime and the high velocity regime. For larger $ x $ (e.g., $ x=10 $), the minimum is well captured by the high velocity approximation in Eq.~\eqref{Jvhigh}. In the approximation of $ v\to 1 $ and $ y_{-}\to1/2\gamma $, the location of the minimum is found analytically from Eq.~\eqref{Jvhigh} to be $ \gamma=x/\ln 2 $.

% \begin{figure}
% \includegraphics[width=0.7\linewidth]{eqfII-3_2021-11-05_Jv}
% \caption{The function $ J $ is illustrated for fixed $ x=1 $. The exact expression in Eq.~\eqref{Jexplicit} is plotted with the solid blue curve. The dashed orange curve shows the linear term in Eq.~\eqref{Jlowv}. The dashed purple curve shows the high velocity approximation of $ J $ in Eq.~\eqref{Jvhigh}.
% \label{FIGJ}}
% \end{figure}

Finally, we also want to examine the classical limit of vacuum quantum friction Eq.~\eqref{eq6-2}. In order to render the thermal contributions more manifest, we temporarily reinsert $ \hbar $ in Eq.~\eqref{eq6-2} with change of variable $ y=\gamma z $ and obtain
\begin{equation}\label{frictionhbar}
\tilde{F}^{\rm{ISO}}=\frac{\hbar}{2\pi^{2}v^{2}}\int_{0}^{\infty} d\omega \,\omega^{4}\Im\alpha(\omega)\int_{1-v}^{1+v} dz \frac{z-1}{e^{\hbar\beta\omega\gamma z}-1}.
\end{equation}
In the limit of $ \beta\gamma\to 0 $, that is the high-temperature and modest-velocity regime, the exponential factor can be expanded in $ \beta\gamma $. Keeping only the first term in the expansion, we find $ \hbar $ disappears explicitly and the integration on $ z $ can be readily carried out, leading to
\begin{equation}\label{thermal}
\tilde{F}^{\rm{ISO}}_{\rm{CL}}=\frac{v-\ln y_{+}}{\pi^{2}\beta\gamma v^{2}}\int_{0}^{\infty} d\omega \Im\alpha(\omega)\,\omega^{3},
\end{equation}
which can also be derived by using the classical FDT from the outset.
In its nonrelativistic limit, the friction in Eq.~\eqref{thermal} further reduces to 
\begin{equation}\label{thermalnr}
\tilde{F}^{\rm{ISO}}_{\rm{CL},\rm{NR}}=-\frac{v}{3\pi^{2}\beta}\int_{0}^{\infty} d\omega \Im\alpha(\omega)\,\omega^{3},
\end{equation}
linear in temperature and in velocity.\footnote{Obviously, the above arguments for the high-temperature limit of quantum vacuum friction do not apply to the situation when the imaginary part of the polarizability is temperature dependent, e.g., in the Bloch-Gr\"{u}neisen model discussed in  Sec.~\ref{nano2}. In addition, $ \Im\alpha(\omega) $ must be properly behaved in both infrared and ultravilolet regimes so that the $ \omega $ integral in Eq.~\eqref{thermalnr} is convergent.} 

\subsection{The NESS quantum vacuum friction for resonance model and monomial model}\label{sec6b}
As simple examples, let us work out the NESS quantum vacuum friction for the resonance and monomial models.

In order to evaluate the NESS quantum vacuum friction for the resonance model in Eq.~\eqref{resonance}, the coefficient of the delta function is needed. So, let us be specific and consider a simple model of an harmonically bound electron with vanishing damping for the polarizability. See, for example, Ref.~\cite{ Schwinger:ce, Jackson} for details. The imaginary part of the polarizability is therefore
\begin{equation}\label{resonance2}
\Im\alpha(\omega)=\frac{\pi e^{2}}{2m\omega_{0}}\delta(\omega-\omega_{0}),
\end{equation}
where $ e $ and $ m $ are the charge and mass of the electron and $ \omega_{0}>0 $ is the resonance frequency.
We will assume the polarizability to be isotropic for simplicity here. Inserting Eq.~\eqref{resonance2} into Eq.~\eqref{eq6-2} with the dimensionless variables $ x_{0}=\beta\omega_{0}/2 $, we obtain
\begin{equation}\label{h}
\tilde{F}^{\rm{ISO}}=\frac{2e^{2}}{\pi m\beta^{3}}x_{0}^{3}J(x_{0},v),
\end{equation}
where $ J $ is defined in Eq.~\eqref{J}.

\begin{figure}[h!]
 \includegraphics[width=0.7\linewidth]{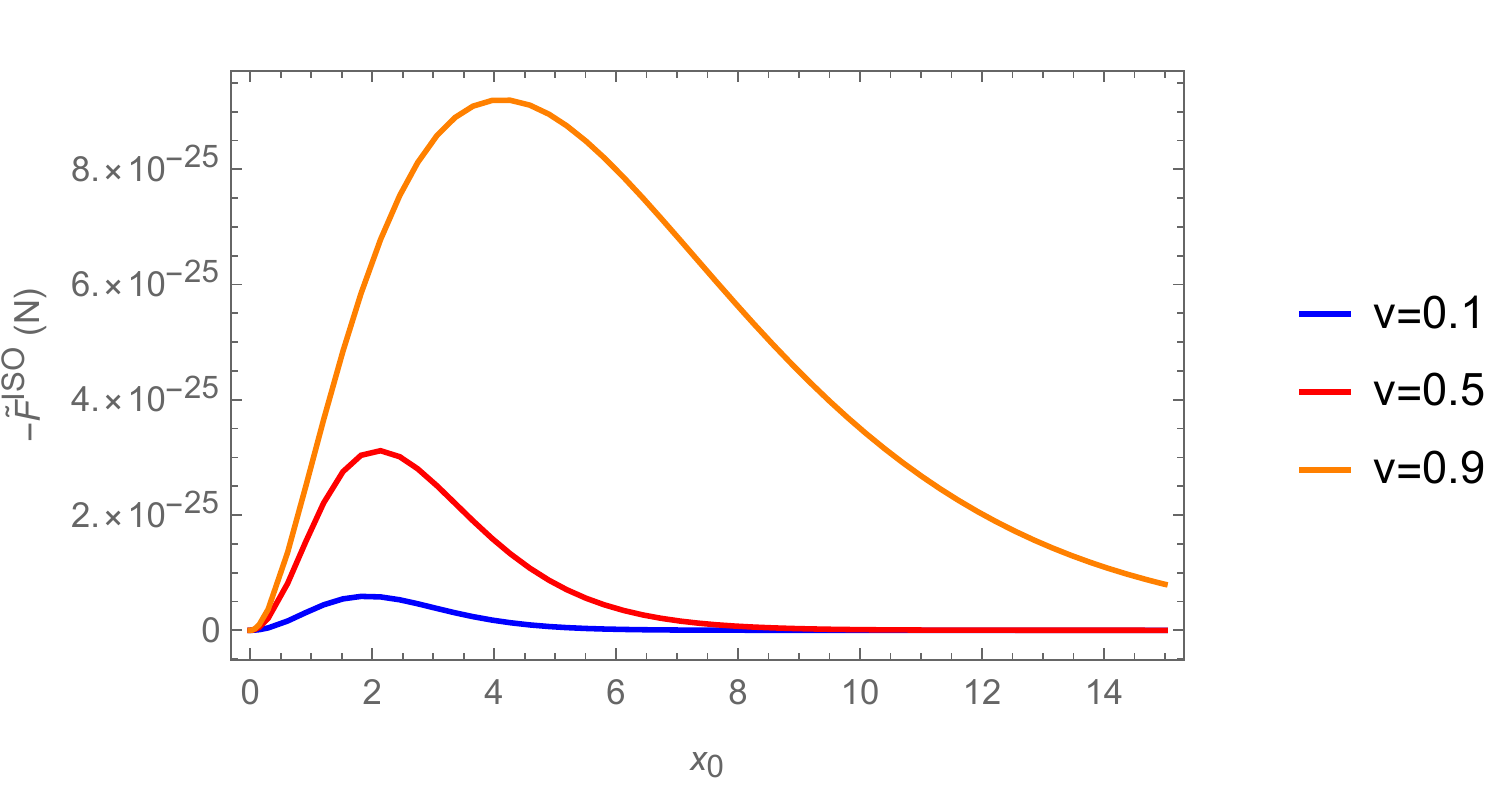}
 \caption{For the resonance model in Eq.~\eqref{resonance2}, the magnitude of the NESS quantum vacuum friciton $ -\tilde{F}^{\rm{ISO}} $ in Eq.~\eqref{h} at room temperature $ T=300\, \rm{K} $ is plotted as a function of $ x_{0} $ for different velocities.
 \label{FIG4}}
 \end{figure}

At the room temperature $ T=300\, \rm{K} $,  the dimensional coefficient in Eq.~\eqref{h} is evaluated as $ 1.60\cross10^{-24}\rm{N} $, \footnote{The conversion factors used in the estimate are $k_{B}=8.62\cross 10^{-5} \rm{eV/K}  $ and $ \hbar c=1.97\cross 10^{-5}\, \rm{eV}\cdot\rm{cm} $.} after converting to SI units. The velocity dependence of the friction is completely controlled by the function $ J $ illustrated in Fig.~\ref{FigJ}. Here, we illustrate the $ x_{0} $ dependence of the magnitude of the friction in Eq.~\eqref{h} for different velocities in Fig.~\ref{FIG4}. 
As is seen in the figure, the magnitude of the friction exibits a peak at a certain frequency ($ x_{0} $) for each fixed velocity. And for greater velocities, the peak is higher with the position of the peak blueshifted. The resultant quantum friction for the typical velocities shown in the figure is on the order of $ 10^{-25} \,\rm{N} $, which is still far from the reach of current experiments.

Next, we turn to the monomial model in Eq.~\eqref{monomial}. To be specific, let us consider the low-frequency radiation reaction model with $ n=3 $,
\begin{equation}\label{RRmodel}
\Im\alpha(\omega)=\frac{\omega^{3}}{6\pi}\alpha_{0}^{2}.
\end{equation}
After inserting this model into Eq.~\eqref{eq6-2}, again it's easier to first work out the integral on $ \omega $ followed by the integral on $ y $. The quantum vacuum friction for various polarization states is found to be 
\begin{equation}\label{eq6-3.5}
 \addtolength{\arraycolsep}{-3pt}
\tilde{F}^{\rm{P}}=-\frac{32\alpha_{0}^{2}\pi^{5}}{4725\beta^{8}} \gamma^{6}v\left\{%
 \begin{array}{lcrcl}
15v^{4}+70v^{2}+35,& \qquad \rm{P}=\rm{ISO},\\\\
-3v^{4}-4v^{2}+7,& \qquad \rm{P}=\rm{X},\\\\
9v^{4}+37v^{2}+14,& \qquad \rm{P}=\rm{Y,Z}.\\
 \end{array}
 \right.
\end{equation} 
The expected identity $\tilde{F}^{\rm{ISO}}= \tilde{F}^{\rm{X}} +\tilde{F}^{\rm{Y}}+\tilde{F}^{\rm{Z}}$ can be immediately confirmed. 

\begin{figure}[h!]
 \includegraphics[width=0.7\linewidth]{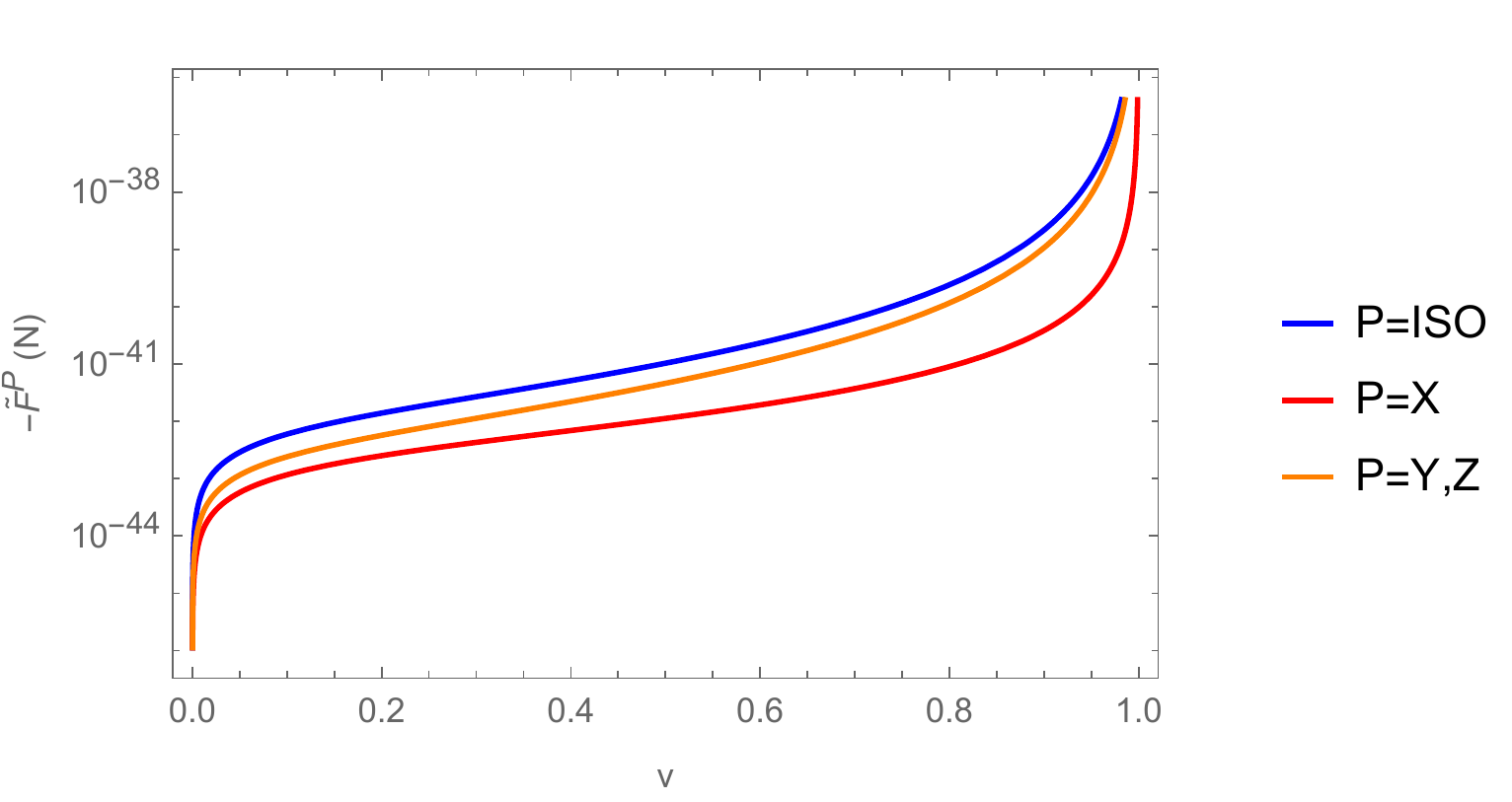}
 \caption{For the low-frequency radiation reaction model in Eq.~\eqref{RRmodel}, the magnitude of the NESS quantum vacuum friction $ -\tilde{F}^{\rm{P}} $ for different polarization states in Eq.~\eqref{eq6-3.5} are plotted as functions of $ v $ at room temperature $ T=300\, \rm{K} $. The static polarizability of a gold atom is used for $ \alpha_{0} $ in the numerical evaluation. \label{FIG3}}
 \end{figure}

For a neutral gold atom, the recommmended value of its static scalar polarizability found in Ref.~\cite{polarizability-table} is $ 36 \,\rm{a.u.}=5.33\cross 10^{-24} \rm{cm^{3}} $. Using this value for $ \alpha_{0} $, the dimensional factor in Eq.~\eqref{eq6-3.5} is estimated to be $- 1.63\cross 10^{-43}\, \rm{N} $ at $ T=300 \,\rm{K} $. We plot the magnitude of the resultant NESS quantum vacuum friction on a gold atom as a function of $ v $ for different polarizations in Fig.~\ref{FIG3}. From the figure, it is seen that the friction increases with velocity and the friction for the parallel ($ x $) polarization is always smaller than that for the transverse ($ y $ and $ z $) polarizations.

If the gold atom is replaced by a perfectly conducting sphere, the polarizability becomes directly proportional to the volume of the sphere, $ \alpha_{0}=4\pi a^{3} $. The magnitude of the friction can be enhanced for a sphere bigger than an atom. For example, the dimensional prefactor in Eq.~\eqref{eq6-3.5} becomes $ 9.09\cross 10^{-25}\,\rm{N} $ for a perfectly conducting sphere of radius $a=100\,\rm{nm}  $, much greater than that for the gold atom.

Of course, the radiation reaction model in Eq.~\eqref{RRmodel} is only a low-frequency approximation of the more complete model already discussed in Sec.~V of Ref.~\cite{Xin:eqf1}. Considering the exponential factor in Eq.~\eqref{eq6-2}, the low-frequency contributions dominate so long as the temperature is low.\footnote{For a gold atom moving at a velocity $ v=0.5 $, the low frequency approximation is accurate up to a temperature $ T=10^{6}\,\rm{K} $. See Fig. 2b of Ref.~\cite{Xin:eqf1}} Indeed, the friction for isotropic polarization in Eq.~\eqref{eq6-3.5} agrees with the low-temperature friction obtained in Eq.~(5.4a) of Ref.~\cite{Xin:eqf1}.

%Our relativistic result in Eq.~\eqref{eq6-3.5} reproduces the nonrelativistic blackbody friction coefficient given by Ref.~\cite{Lach:Imalpha} in the low velocity limit. Using our Eq.~\eqref{eq6-3.5}, we can extract the quantum friction to any order in $ v $. The ratio of the leading relativistic correction cubic in velocity to the nonrelativistic approximation of the friction for this low frequency radiation reaction model is found to be
%\begin{equation}\label{forceratio}
% \addtolength{\arraycolsep}{-3pt}
%\frac{\tilde{F}^{\rm{P} (3)}_{dd+EE}}{\tilde{F}^{\rm{P} (1)}_{dd+EE}}=v^{2}
%\left\{%
% \begin{array}{lcrcl}
%5,& \qquad \rm{P}=\rm{ISO}\\\\
%\frac{17}{7},& \qquad \rm{P}=\rm{X}\\\\
%\frac{79}{14},& \qquad \rm{P}=\rm{Y}.\\
% \end{array}
% \right.
%\end{equation} 
%The ratio can as well be calulated directly from Eq.~\eqref{eq6-2.4} and Eq.~\eqref{eq6-2.6}.
%For the isotropic case, the relativistic correction becomes comparable to the Einstein-Hopf drag at $ v=1/\sqrt{5} $. Such analysis can be readily extended to the monomial model with an arbitrary power $ n $, $ \Im\alpha(\omega)\propto \omega^{n} $, and the same ratio is found to be
%\begin{equation}\label{eqCn}
%\frac{\tilde{F}^{\rm{ISO} (3)}_{dd+EE}}{\tilde{F}^{\rm{ISO} (1)}_{dd+EE}}=\frac{1}{10}(n^{2}+8n+17)v^{2}.
%\end{equation}

Contrary to the resonance model, quantum friction for the model in Eq.~\eqref{RRmodel} ($ n=3 $ monomial model) diverges as the velocity approaches the speed of light. This turns out to be true for all monomial models with a power $ n> -3 $. For a monomial model with an arbitrary power $ n $, $ \Im\alpha(\omega)=\alpha_{n}\omega^{n} $ ($ \alpha_{n} $ is a constant in the model which has the proper dimension according to the power $ n $), the NESS quantum friction for the isotropic polarization can be calculated explicitly from Eq.~\eqref{eq6-2},
\begin{equation}\label{fn}
\tilde{F}^{\rm{ISO}}=\frac{\alpha_{n}\Gamma(5+n)\zeta(5+n)}{2\pi^{2}\beta^{5+n}}M_{n}(v),\quad M_{n}(v)\equiv\frac{1}{\gamma^{2}v^{2}}\left[\frac{1}{3+n}\left(y_{+}^{3+n}-y_{-}^{3+n}\right)-\frac{\gamma}{4+n}\left(y_{+}^{4+n}-y_{-}^{4+n}\right)\right].
\end{equation}
%For the monomial model with power $ n $, the NESS quantum vacuum friction can be determined up to a constant in the model for polarizability. Considering again the isotropic polarization for simplicity, the friction according to Eq.~\eqref{eq6-2} is
The velocity dependence of the friction in Eq.~\eqref{fn} is all contained in the function $ M_{n}(v) $. The high velocity behavior ($ \gamma\to\infty $) of $ M_{n} $ depends on $ n $,
\begin{equation}\label{gn}
\addtolength{\arraycolsep}{-3pt}
\lim_{\gamma\to \infty}M_{n}=\left\{%
 \begin{array}{lcrcl}
-\frac{2^{4+n}}{4+n}\gamma^{(3+n)},& \qquad n>-4\\\\
\left(\frac{2^{-(4+n)}}{4+n}-\frac{2^{-(3+n)}}{3+n}\right)\gamma^{-(5+n)},& \qquad n<-4\\\\
-\frac{2}{\gamma}\ln \gamma,& \qquad n=-4.\\
 \end{array}
 \right.
\end{equation}
Even though $ M_{n} $ can be defined for an arbitrary value of $ n $, $ \tilde{F}^{\rm{ISO}} $ in Eq.~\eqref{fn} is not well defined for negative integers $ n\le -4 $ due to the singularities of the prefactor. As a result, for integer powers $ n $, the friction is almost always divergent in the high velocity limit, except for $ n=-3 $,
\begin{equation}\label{-3}
\tilde{F}^{\rm{ISO}}=\frac{\alpha_{-3}}{12\beta^{2}}\frac{1}{\gamma^{2}v^{2}}\left[\ln(\frac{1+v}{1-v})-2\gamma^{2}v\right].
\end{equation}
It is not surprising that $ n=-3 $ is special, recalling that the corresponding  NESS temperature ratio $ \tilde{r}=1 $ for all velocities. In addition, the full radiation reaction model in Ref.~\cite{Xin:eqf1} reduces to $ n=-3 $ monomial model in the high-frequency limit. Indeed, the friction in Eq.~\eqref{-3} precisely equals the high-temperature quantum vacuum friction recorded in Eq.~(5.4b) of Ref.~\cite{Xin:eqf1} if we set $ \alpha_{-3}=6\pi $ in accordance with Eq.~(5.1) of Ref.~\cite{Xin:eqf1}.

\section{NESS temperature ratio and NESS friction for a gold nanosphere}\label{nano}
%As a more realistic example, in this section, we will propose models for the dissipation of a gold nanosphere and evaluate the NESS temperature ratio and friction for it. Assuming the nanosphere is isotropic and using the Lorenz-Lorentz relation~\footnote{See, for example, Ref.~\cite{Schwinger:ce} for a reference.}, the polarizability of the nanosphere $ \alpha(\omega) $ can be written in terms of the permittivity of gold $ \varepsilon(\omega) $ and radius of the nanosphere $ a $ as,
%\begin{equation}\label{LL}
%\alpha(\omega)=4\pi a^{3} \,\frac{\varepsilon(\omega)-1}{\varepsilon(\omega)+2}.
%\end{equation}
%The permittivity of gold is well described by the Drude model
%\begin{equation}\label{Lorentzian}
%\varepsilon (\omega)=1-\frac{\omega_{p}^{2}}{\omega^{2}+i\omega\nu}.
%\end{equation}
%Here, $ \omega_{p} $ is the plasma frequency of gold and we will take its value to be 
As a realistic example, we will in this section evaluate the NESS temperature ratio and friction for a nanosphere made of gold. For simplicity, we assume the nanosphere is isotropic and ignore any surface effect of the nanosphere. The polarizability of the nanosphere $ \alpha(\omega) $ can be expressed in terms of its radius $ a $ and the permittivity of gold $ \varepsilon(\omega) $ through the Lorenz-Lorentz relation 
\begin{equation}\label{LL}
\alpha(\omega)=4\pi a^{3} \,\frac{\varepsilon(\omega)-1}{\varepsilon(\omega)+2}.
\end{equation}
An introduction of the Lorenz-Lorentz relation can be found in Ref.~\cite{Schwinger:ce}. The permittivity of gold $\varepsilon(\omega) $ is often described by the Drude model
\begin{equation}\label{Lorentzian}
\varepsilon (\omega)=1-\frac{\omega_{p}^{2}}{\omega^{2}+i\omega\nu},
\end{equation}
where $ \omega_{p} $ is the plasma frequency and $ \nu $ the damping parameter of gold. In \ref{nano1}, the damping parameter will be treated as a temperature-independent constant. In \ref{nano2}, we will instead consider the damping parameter to be temperature dependent and describe it using the
Bloch-Gr\"{u}neisen model.

\subsection{Constant damping model}\label{nano1}
Combining Eq.~\eqref{LL} and Eq.~\eqref{Lorentzian}, we find
\begin{equation}\label{model3}
\Im\alpha(\omega)=V \frac{\omega_{p}^{2}\,\omega\nu}{(\omega_{1}^{2}-\omega^{2})^{2}+\omega^{2}\nu^{2}}.
\end{equation}
where $ V=\frac{4}{3} \pi a^{3}$ denotes the volume of the nanosphere and we have introduced the rescaled plasma frequency $ \omega_{1}=\omega_{p}/\sqrt{3} $. The imaginary part of the polarizability is temperature independent if all the parameters in Eq.~\eqref{model3} are constant. In the following numerical evaluation, we use the nominal room temperature value for the plasma frequency and damping parameter given in Ref.~\cite{Hoye:Does}: $ \omega_{p}=9.00\, \rm{eV} $ and $ \nu=0.0350\, \rm{eV} $.

When the gold nanosphere moves through vacuum in a constant velocity $ v $, the NESS temperature $\tilde{T} $ deviates from the temperature of the blackbody radiation $ T $ determined by the NESS condition Eq.~\eqref{eq5-22}. And in NESS, the nanosphere experiences a frictional drag given by Eq.~\eqref{eq6-2}.

In order to evaluate the NESS temperature ratio $ \tilde{r}=\tilde{T}/T $, we insert Eq.~\eqref{model3} into Eq.~\eqref{eq5-22} and introduce a new set of dimensionless variables
\begin{equation}\label{dimensionless}
u=\frac{\omega}{\omega_{1}},\qquad
\epsilon=\frac{\nu}{\omega_{1}},
\qquad
x_{1}=\frac{\beta\omega_{1}}{2},
\qquad
\tilde{r}=\frac{\beta}{\tilde{\beta}}.
\end{equation}
We therefore obtain
\begin{equation}\label{3NESS}
\int_{0}^{\infty} du\frac{u^{5}}{(1-u^{2})^{2}+u^{2}\epsilon^{2}}\frac{1}{e^{2x_{1}u/\tilde{r}}-1}=\int_{0}^{\infty} du\frac{u^{5}}{(1-u^{2})^{2}+u^{2}\epsilon^{2}}\frac{1}{4\gamma v x_{1}u}\ln\left(\frac{1-e^{-2x_{1}y_{+}u}}{1-e^{-2x_{1}y_{-}u}}\right),
\end{equation}
which can be solved to determine the NESS temperature ratio $\tilde{r}$ numerically. Due to the cancellation of the volume factor on both sides of the equation, the temperature ratio is independent of the size of the nanosphere $ a $. The dimensionless damping for gold is $ \epsilon=0.00673 $; the temperature ratio $ \tilde{r} $ then depends only on the velocity of the nanosphere $ v $ and the temperature of the blackbody radiation  $ T $. 

\begin{figure*}
\subfloat[]{\label{figr1a}%
\includegraphics[width=0.48\linewidth]{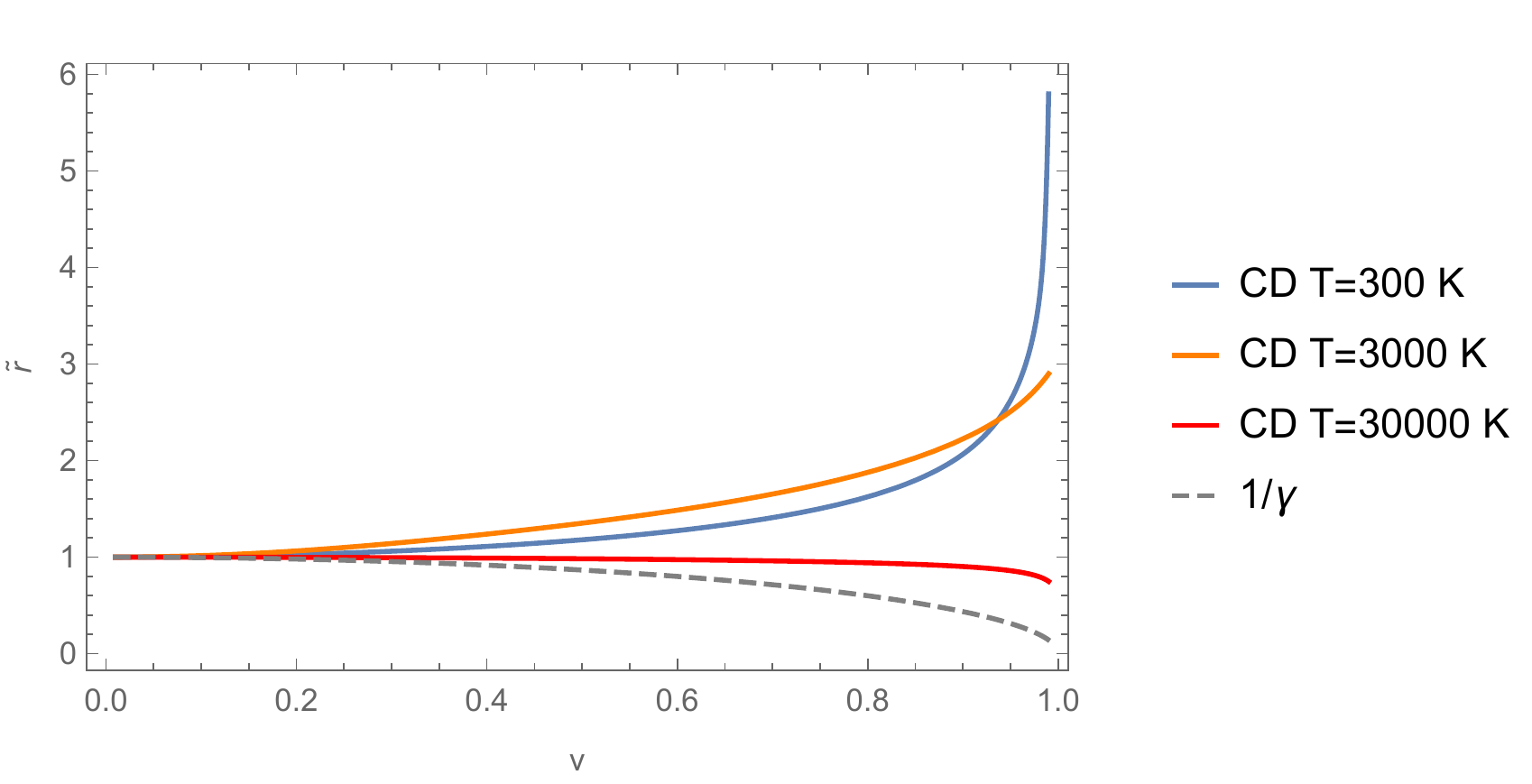}%
}
\subfloat[]{\label{figr1b}%
\includegraphics[width=0.45\linewidth]{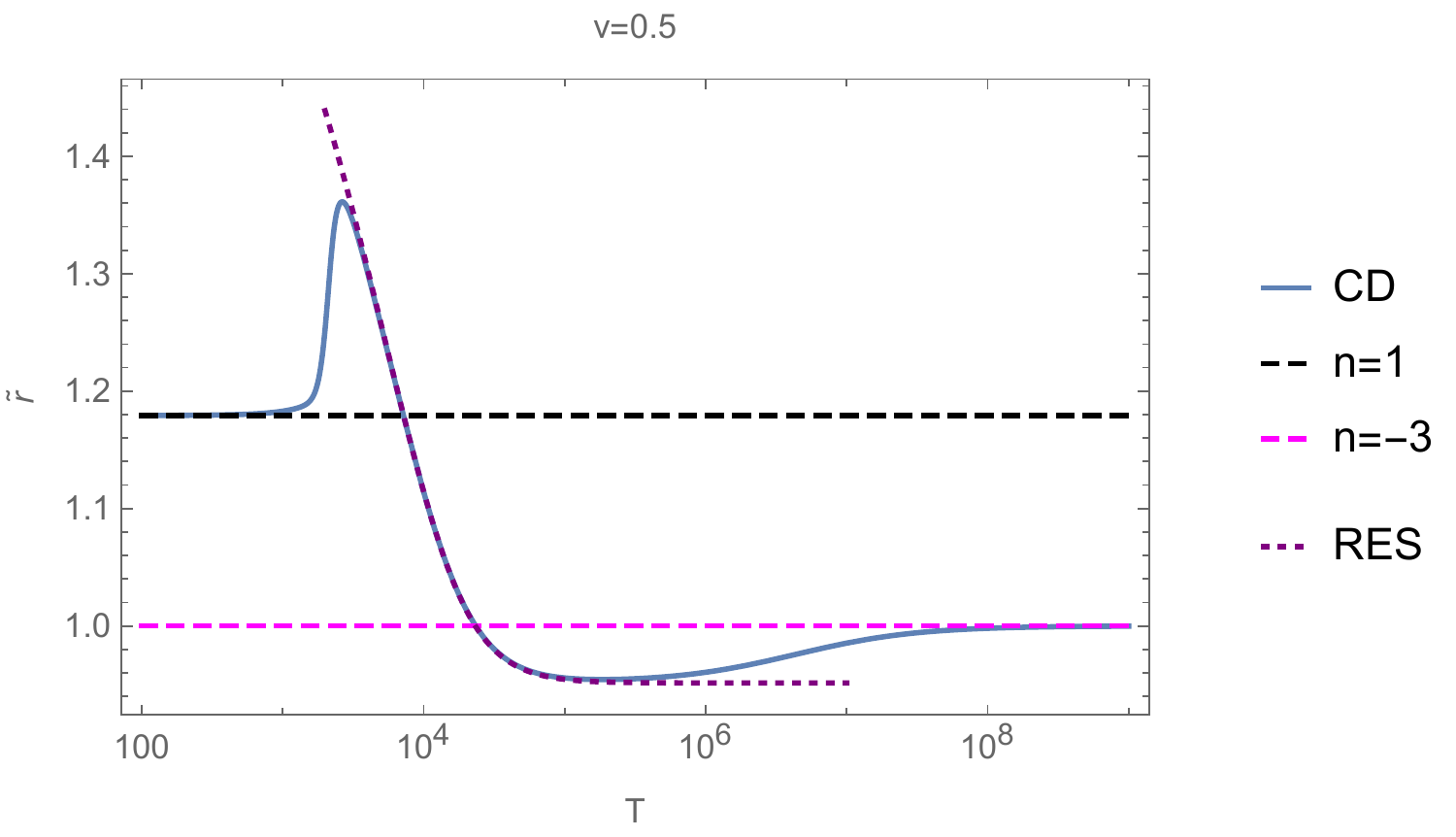}%
}
\caption{The velocity dependence and temperature dependence of the NESS temperature ratio $ \tilde{r} $ for a gold nanosphere with constant damping parameter (CD) is illustrated. (a)\, At three different temperatures of the blackbody radiation $ T=300\, \rm{K}$ $(x_{1}=101)$, $ T=3000\, \rm{K}$ $(x_{1}=10.1)$ and $ T=30\,000\, \rm{K}$ $(x_{1}=1.01)$, $ \tilde{r} $ computed numerically from Eq.~\eqref{3NESS} is plotted as a function of velocity for $ 0.01\le v\le 0.99 $. The dashed gray curve shows the lower bound of the temperature ratio $ 1/\gamma $ given by the theorem in Eq.~\eqref{eq5-23.6}. (b)\, At fixed velocity $ v=0.5 $, $ \tilde{r} $ is plotted as a function of the radiation temperature for the constant damping model (solid, blue), the $ n=1 $ monomial model (dashed, black), the $ n=-3 $ monomial model (dashed, magenta) and the resonance model (dotted, purple).} 
\label{Figr1}
\end{figure*}

In Fig.~\ref{figr1a}, we plot the NESS temperature ratio $ \tilde{r} $ as a function of velocity  at different temperatures of the blackbody radiation. For $ T=300\,\rm{K} $ and $ T=3000\,\rm{K} $, $ \tilde{r} $ is greater than $ 1 $ and increasing with velocity while it is less than $ 1 $ and decreasing with velocity for $ T=30\,000\,\rm{K} $. Increasing the background temperature from $ T=300 \,\rm{K} $ to $ T=3000\, \rm{K} $ enhances the temperature ratio for modest velocities but suppresses it for sufficiently high velocities. 

Figure \ref{figr1b} shows the NESS temperature ratio $\tilde{r} $ as a function of radiation temperature $ T $ for a fixed velocity $ v=0.5 $ for various models. The NESS temperature ratio of the gold nanosphere with constant damping parameter is shown by the blue curve, which has a peak at $ T=2650\,\rm{K} $. And at the peak, the deviation of the NESS temperature of the nanosphere from the temperature of the environment reaches $ 36\% $, being quite noticeable.

It is also seen from Fig.~\ref{figr1b} that the constant damping model for the nanosphere in various temperature regimes can be mimicked by the simple models discussed earlier.  In the low-temperature regime (lower than $ 10^{3}\,\rm{K} $), $ \tilde{r} $ for the nanosphere (CD model) almost remains constant $ \tilde{r}=1.18 $, the same as the NESS temperature ratio for the $ n=1 $ monomial model with $ v=0.5 $. In the intermediate temperature regime (roughly from the peak to the valley of the blue curve), the behavior of $ \tilde{r} $ for the nanosphere is well captured by the resonance model. In the high-temperature regime (higher than $ 10^{8}\,\rm{K} $), $ \tilde{r} $ for the nanosphere reaches and stablizes at $ \tilde{r}=1 $, which coincides with the NESS temperature ratio for the very special $ n=-3 $ monomial model.

Apart from the numerical result, the above behavior of $ \tilde{r} $ as a solution to Eq.~\eqref{3NESS} can be understood qualitatively. Due to the exponential factor on both sides of Eq.~\eqref{3NESS}, at a modest velocity (i.e., $v=0.5$), the small $ u $ (frequency) contribution dominates the integral at low temperatures while the large $ u $ (frequency) contribution dominates the integral at high temperatures. In Eq.~\eqref{model3}, $ \Im\alpha(\omega) $ reduces to the $ n=1 $ monomial model and the $ n=-3 $ monomial model in the low and high-frequency limits, respectively. In addition, due to the smallness of the damping $\epsilon$, there ought to be a region where the integral is dominated by the resonance contributions around $ u=1 $, corresponding to a resonance model with the resonance frequency $ \omega_{1} $.

Let's now calculate the NESS quantum vacuum friction on the moving gold nanosphere with a constant damping parameter. After inserting Eq.~\eqref{model3} into Eq.~\eqref{eq6-2} and introducing the dimensionless variables in  Eq.~\eqref{dimensionless}, we find the quantum vacuum friction to be
%\begin{equation}\label{force3}
%\tilde{F}^{\rm{ISO}}_{dd+EE}=-Af_{1}(x_{1},\epsilon,v),\quad A= \frac{V\omega_{p}^{2}\omega_{1}^{2}\nu}{2\pi^{2}}, \quad f_{1}(x_{1},\epsilon,v)=-\int_{0}^{\infty} du \frac{u^{5}}{(1-u^{2})^{2}+u^{2}\epsilon^{2}} \,J(x_{1}u,v).
%\end{equation}
\begin{equation}\label{force3}
\tilde{F}^{\rm{ISO}}=\frac{V\omega_{p}^{2}\omega_{1}^{3}}{2\pi^{2}\gamma^2 v^2}\int_{0}^{\infty} du \frac{u^{5}\epsilon}{(1-u^{2})^{2}+u^{2}\epsilon^{2}}\int_{y_{-}}^{y_{+}} dy \,(y-\gamma)\frac{1}{e^{2x_{1}uy}-1}.
\end{equation}
For a gold nanosphere of radius $ a=100\,\rm{nm} $, the dimensional coefficient independent of $ v $ and $ T $ is evaluated to be $ 2.57\cross 10^{-10} \rm{N} $. 

\begin{figure*}
\subfloat[]{\label{figf1a}%
\includegraphics[width=0.45\linewidth]{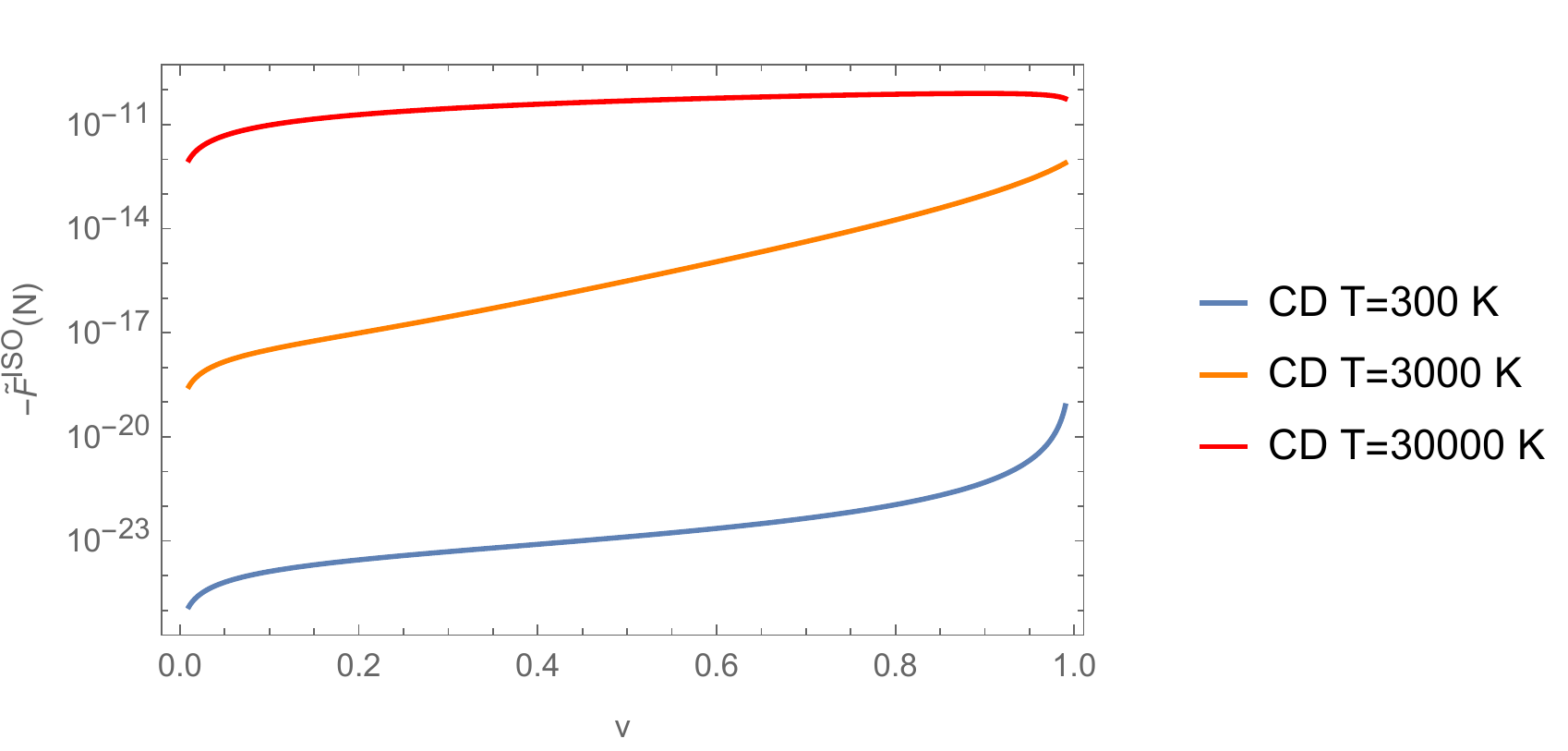}%
}
\subfloat[]{\label{figf1b}%
\includegraphics[width=0.42\linewidth]{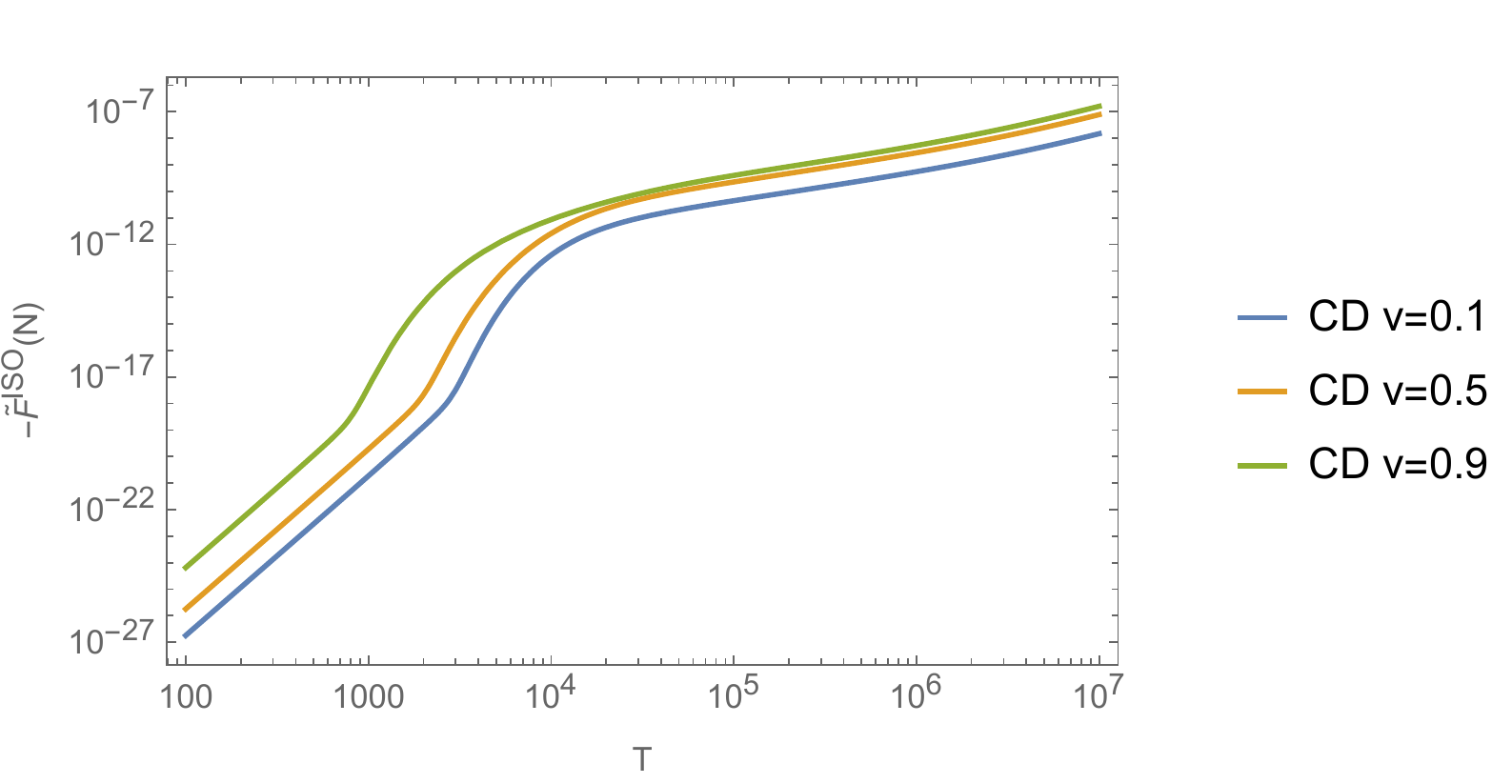}%
}\\
\subfloat[]{\label{figf1c}%
\includegraphics[width=0.45\linewidth]{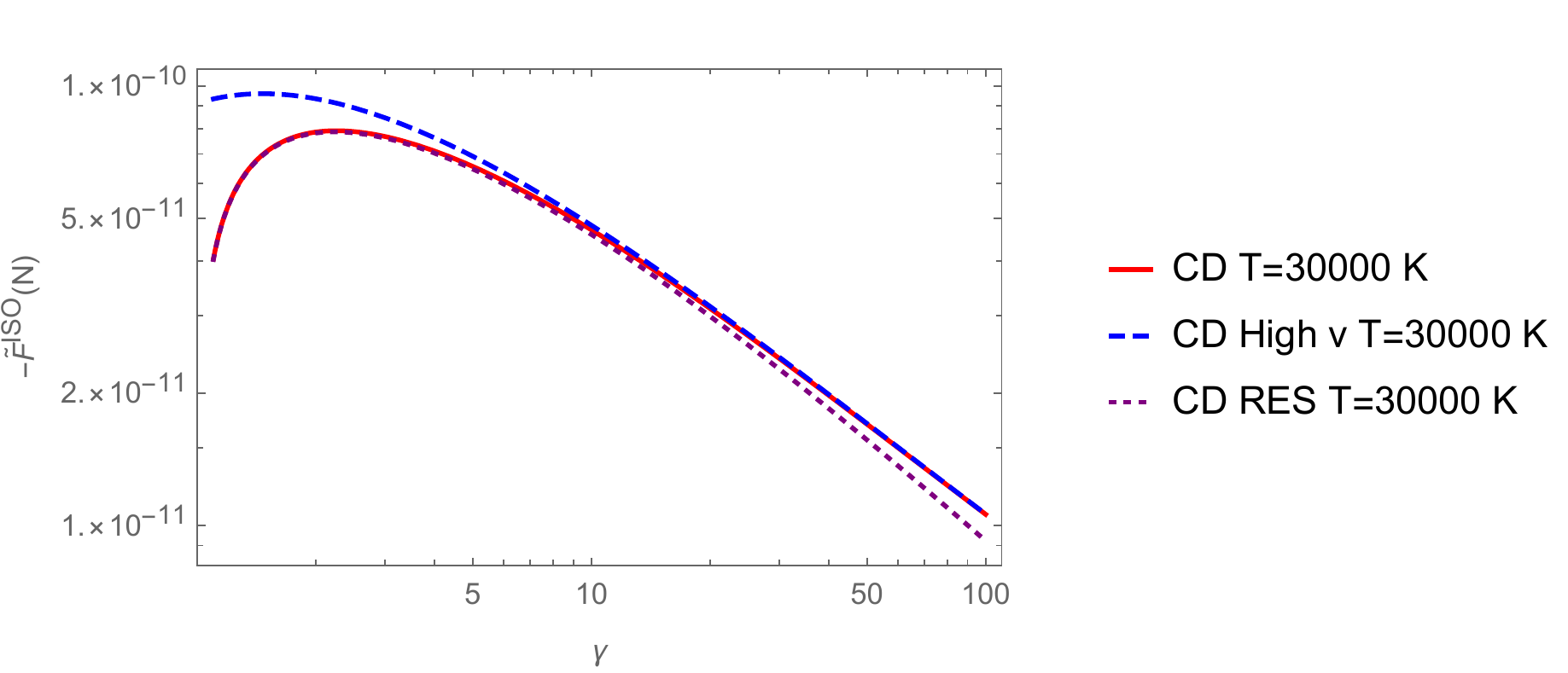}%
}
\subfloat[]{\label{figf1d}%
\includegraphics[width=0.42\linewidth]{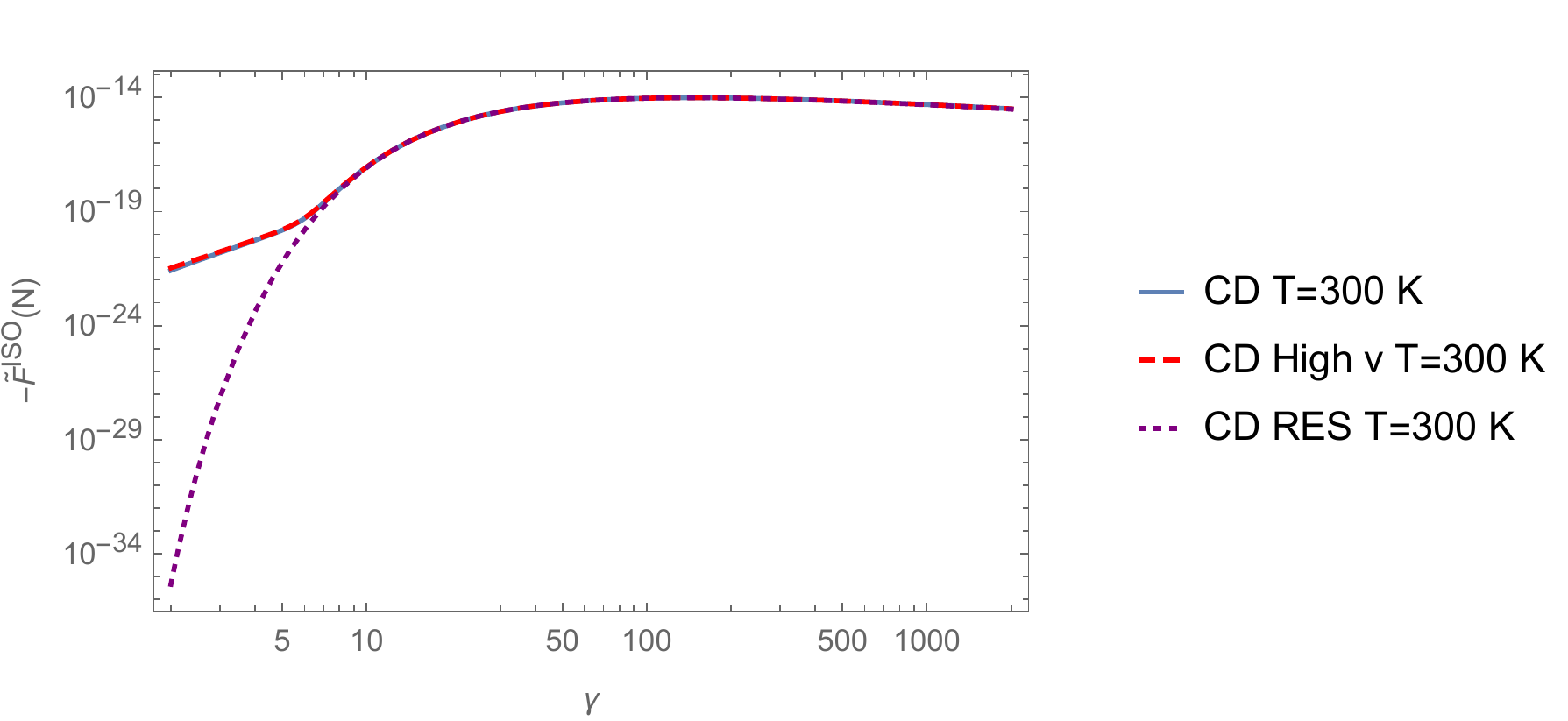}%
}
\caption{The magnitude of the NESS quantum vacuum friction in Eq.~\eqref{force3}, $-\tilde{F}^{\rm{ISO}}$, for a gold nanosphere with constant damping parameter is illustrated. (a)\, At three different temperatures of the blackbody radiation $ T=300\, \rm{K}$ $(x_{1}=101)$, $ T=3000\, \rm{K}$ $(x_{1}=10.1)$ and $ T=30\,000\, \rm{K}$~$(x_{1}=1.01)$, $ -\tilde{F}^{\rm{ISO}} $ is plotted as a function of velocity for $ 0.01\le v \le 0.99 $.  (b)\, For three different velocities $ v=0.1 $, $ v=0.5 $ and $ v=0.9 $, $ -\tilde{F}^{\rm{ISO}} $ is plotted as a function of the temperature of the blackbody radiation $ T $. (c) For $ T=30\,000\,\rm{K} $, $ -\tilde{F}^{\rm{ISO}} $ in Eq.~\eqref{force3} is shown again as a function of $ \gamma $ by the solid red curve. The dashed blue curve shows its high velocity approximation using Eq.~\eqref{Jvhigh}. The dotted purple curve plots only the contributions coming from the resonance at $ u=1 $. (d) For $ T=300\,\rm{K} $, the exact friction is shown again as a function of $ \gamma $ by the solid blue curve. The dashed red curve plots the high velocity approximation of the friction using Eq.~\eqref{Jvhigh}. The dotted purple curve plots the resonance contributions.} 
\label{Figf1}
\end{figure*}

In Fig.~\ref{figf1a}, the magnitude of the NESS quantum vacuum friction $ -\tilde{F}^{\rm{ISO}} $ is plotted as a function of velocity $ v $ for the three different radiation temperatures. Even though $ -\tilde{F}^{\rm{ISO}} $ increases with velocity for $ T=300\,\rm{K} $ and $ T=3000\,\rm{K} $ in the velocity range shown ($ 0.01\le v \le 0.99 $), it does not monotonically increase with velocity for $ T=30\,000\,\rm{K} $. 

Figure \ref{figf1b} plots $ -\tilde{F}^{\rm{ISO}} $ as a function of the radiation temperature for fixed velocities. It is seen that the magnitude of the NESS quantum vacuum friction generally increases with temperature and the temperature effects are more prominent in the low-temperature regime than the high-temperature regime. Again, this can be understood as a result of the different limiting behavior of the model for the nanosphere in Eq.~\eqref{model3}. It reduces to the $ n=1 $ monomial model in the low-temperature regime and the $ n=-3 $ monomial model in the high-temperature regime. In Eq.~\eqref{fn}, the NESS quantum vacuum friction for a monomial model with power $ n $ is found to be proportional to $ T^{5+n} $. Therefore, the temperature dependence of NESS friction on the nanosphere is $ T^{6} $ for low temperatures and then weakens to $ T^{2} $ for high temperatures.

In order to illustrate the high velocity behavior of the friction more clearly, we plot the magnitude of the NESS quantum vacuum friction as a function of $ \gamma $ in Figs.~\ref{figf1c} and \ref{figf1d}. Figure \ref{figf1c} illustrates that, for $ T=30\,000\,\rm{K} $, the peak of the magnitude of the friction occurs at $ \gamma=2.25$ and the behavior of the friction around the peak can be captured by the resonance contribution due to the smallness of the damping parameter. This reminds us of the  nonmonotonic behavior of the classical friction on a charged particle passing above a conducting plate \cite{Kim:charged}, also due to the small damping of the plate. For even higher $ \gamma $, the resonance approximation starts to show discrepancy and the actual friction is better approximated using the high velocity approximation Eq.~\eqref{Jvhigh}. Figure \ref{figf1d} further reveals that the nonmonotonic behavior of the friction also occurs for lower temperatures, only with the peak shifted towards more relativistic velocities. For low temperatures, contributions to the friction from frequencies lower than the resonance frequency tend to dominate. Therefore, the resonance approximation becomes less accurate unless the velocity gets really large.

\subsection{Bloch-Gr\"{u}neisen model}\label{nano2}
\begin{figure}[h!]
 \includegraphics[width=0.7\linewidth]{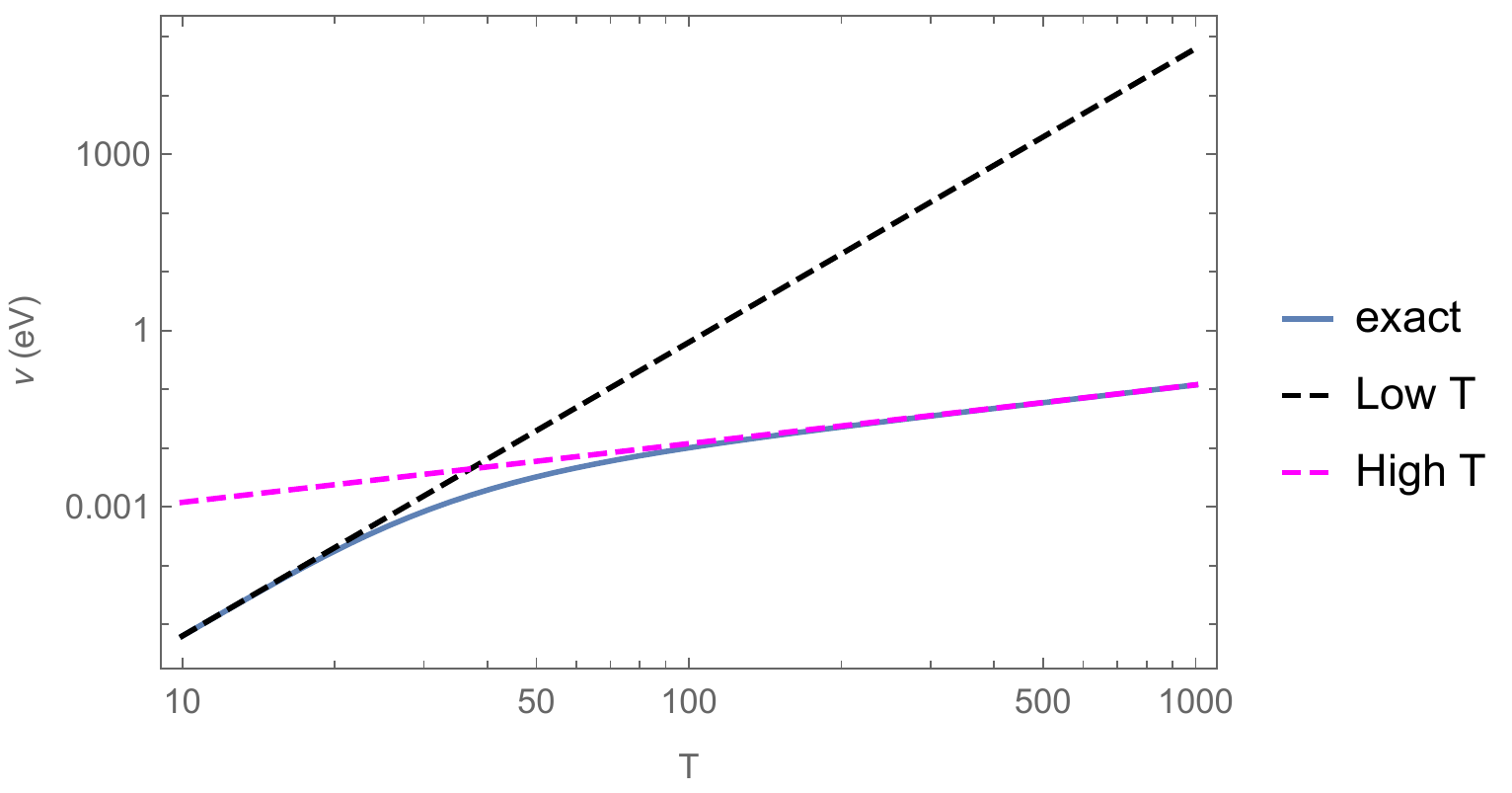}
 \caption{The temperature-dependent damping parameter $ \nu $ in Eq.~\eqref{BG1} is illustrated for gold with $ \theta=175\,\rm{K} $ and $ \nu_{0}=0.0832\,\rm{eV} $.\label{figBG1}}
 \end{figure} 
Even though the plasma frequency $ \omega_{p} $ has very weak temperature dependence, it is more realistic to adopt a temperature-dependent damping parameter $ \nu $ in Eq.~\eqref{model3}. Damping of a simple metal is mainly due to the scattering of electrons by phonons and can be well described by the Bloch-Gr\"{u}neisen (BG) model \cite{Bloch, Gruineisen},
\begin{equation}\label{BG1}
\nu(T)=\nu_{0}\left(\frac{T}{\theta}\right)^{5}\int_{0}^{\frac{\theta}{T}} dx \frac{x^{5}e^{x}}{(e^{x}-1)^{2}}.
\end{equation}
For gold, the Bloch-Gr\"{u}neisen temperature $ \theta $ is $ 175\,\rm{K} $. And the constant $ \nu_{0}$ in Eq.~\eqref{BG1} is determined to be $ 0.0832\,\rm{eV} $ by the room temperature ($ 300\,\rm{K} $) value of the damping parameter $ \nu=0.0350\,\rm{eV} $ \cite{Hoye:Does, Lambrecht:mirror, Palik}.\footnote{The value for $ \nu_{0} $ we use is slightly different than that in Appendix D of \cite{Hoye:Does} where the room temperature is taken to be $ 295\,\rm{K} $. There is, of course, no definite consensus on the meaning of the room temperature. Nonetheless, taking it to be $ 300\,\rm{K} $ is more consistent with the source of the raw data in \cite{Palik}.} The low and high temperature limits of the Bloch-Gr\"{u}neisen damping can be easily worked out,
\begin{equation}\label{BG2}
\addtolength{\arraycolsep}{-3pt}
\nu(T)\to\left\{%
 \begin{array}{lcrcl}
5\Gamma(5)\zeta(5)\nu_{0}\left(\frac{T}{\theta}\right)^{5},& \qquad T\ll \theta,\\\\
\frac{\nu_{0}}{4}\left(\frac
{T}{\theta}\right),& \qquad T\gg \theta.\\
 \end{array}
 \right.
\end{equation}
In Fig.~\ref{figBG1}, we plot the Bloch-Gr\"{u}neisen damping parameter for gold as a function of temperature. The transition between the low and high temperature behavior of $ \nu $ is seen to occur at a rather low temperature around $ T=40\,\rm{K} $.

\begin{figure*}[h!]
\subfloat[]{\label{figBG2a}%
\includegraphics[width=0.48\linewidth]{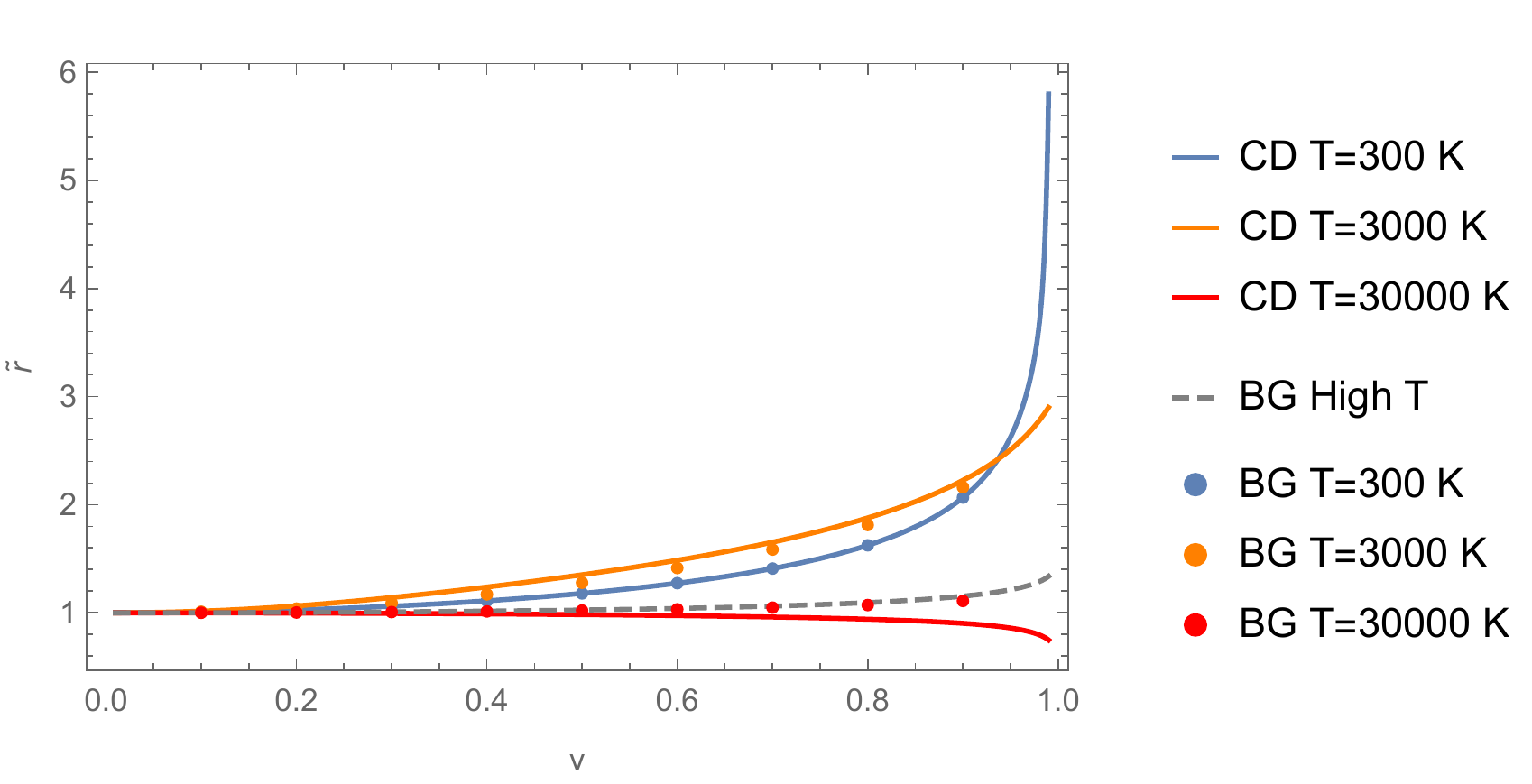}%
}
\subfloat[]{\label{figBG2b}%
\includegraphics[width=0.45\linewidth]{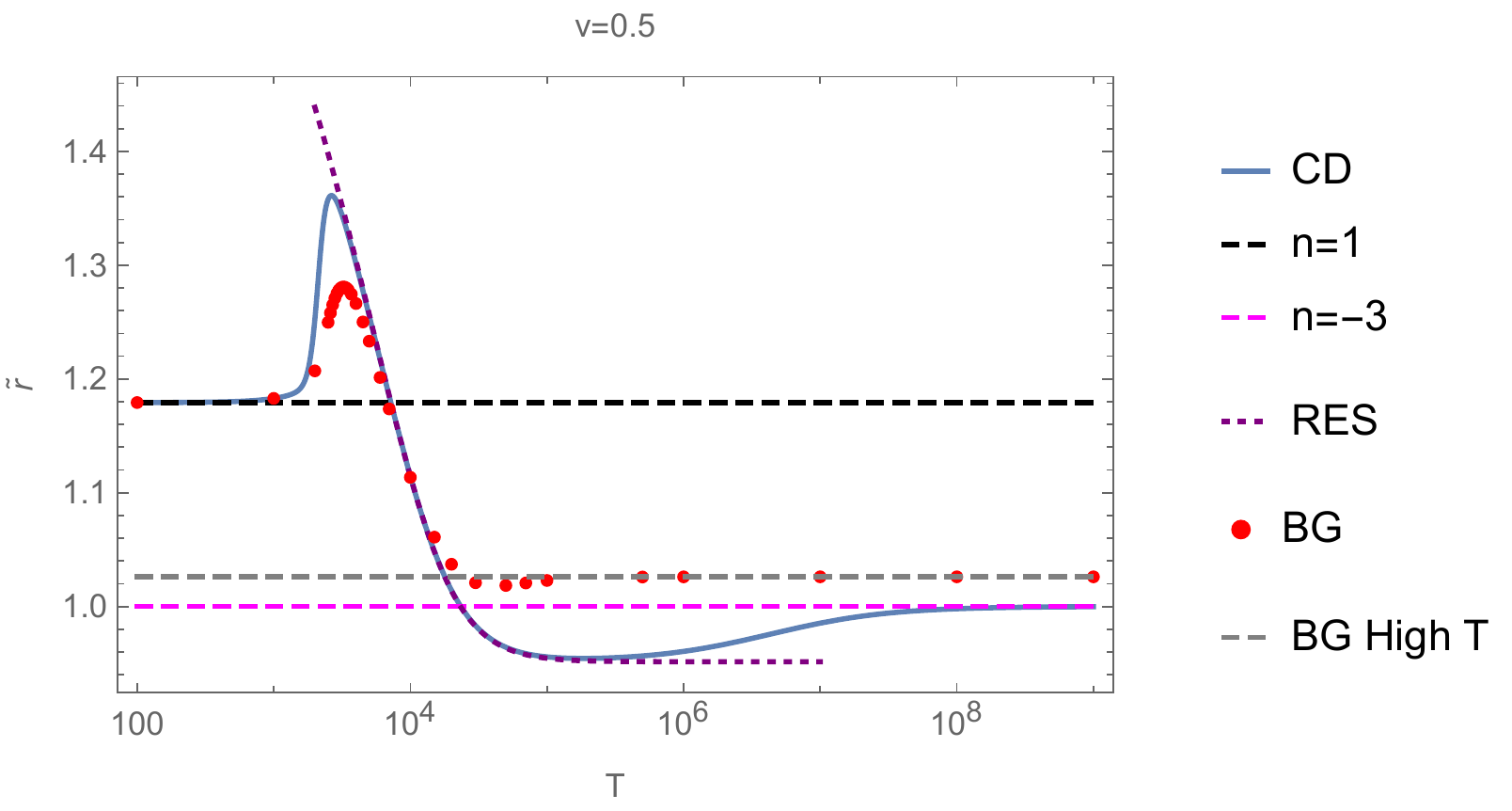}%
}
\caption{The NESS temperature ratio $ \tilde{r} $ for the BG model is  compared with that for the CD model. (a)\,  At three different temperatures of the blackbody radiation, $ \tilde{r} $ is plotted as a function of $ v $ for the BG model and the CD model by the dotted curves and the solid curves, respectively. The high-temperature prediction of the NESS temperature ratio for the BG model is plotted by the dashed, gray curve. (b)\, At fixed velocity $ v=0.5 $, $ \tilde{r} $ is plotted as a function of the radiation temperature for the CD model (solid, blue), the $ n=1 $ monomial model (dashed, black), the $ n=-3 $ monomial model (dashed, magenta), the resonance model (dotted, purple) and the BG model (dotted, red). The high-temperature prediction of $ \tilde{r} $ for the BG model at $ v=0.5 $ is $ 1.03 $, plotted by a dashed grey curve.} 
\label{figBG2}
\end{figure*}

Incorporating the temperature dependence of the damping parameter, $ \Im\alpha(\omega) $ in Eq.~\eqref{model3} becomes dependent on the temperature of the gold nanosphere $ T' $,
\begin{equation}\label{BG3}
\Im\alpha(\omega,T')=V \frac{\omega_{p}^{2}\,\omega\nu(T')}{(\omega_{1}^{2}-\omega^{2})^{2}+\omega^{2}\nu^{2}(T')}.
\end{equation}

Equation \eqref{3NESS} can still be used to find the NESS temperature ratio with the only modification that the dimensionless damping $ \epsilon $ now also depends on the temperature of the nanosphere, $\epsilon\to \epsilon(\tilde{T})=\epsilon(\tilde{r}T)$. The modified Eq.~\eqref{3NESS} is numerically solved for the NESS temperature ratio of the gold nanosphere with the temperature-dependent damping described by the BG model and the results are compared with that for the CD model in Fig.~\ref{figBG2}.
%\begin{equation}\label{BG4}
%\int_{0}^{\infty} du\frac{u^{5}}{(1-u^{2})^{2}+u^{2}\epsilon^{2}(r_{T}T)}\frac{1}{e^{2x_{1}u/r_{T}}-1}=\int_{0}^{\infty} du\frac{u^{5}}{(1-u^{2})^{2}+u^{2}\epsilon^{2}(r_{T}T)}\frac{1}{4\gamma v x_{1}u}\ln\left(\frac{1-e^{-2x_{1}y_{+}u}}{1-e^{-2x_{1}y_{-}u}}\right),
%\end{equation}
%where $ \epsilon $ now depends on the temperature of the gold nanosphere $ T'=r_{T}T $ and $ r_{T} $ is the NESS temperature ratio in this case with temperature dependent damping.

Figure \ref{figBG2a} shows that, at $T= 300\,\rm{K} $, $ \tilde{r} $ is almost the same for the two models even though the temperature-dependent damping $ \epsilon $ is evaluated at $ \tilde{r} T$ and will be different from the constant value used for the CD model. This indicates that, at room temperatures, the NESS temperature ratio $\tilde{r}  $ is rather insensitive to the actual value of the damping. The NESS temperature ratio for the BG model is seen to be generally smaller than that for the CD model at $ T=3000\,\rm{K} $ but larger at $ T=30\,000\,\rm{K} $. 

The phenomenon is more obvious in Fig.~\ref{figBG2b} where we see both the peak ($ \sim 3200\,\rm{K} $) and the valley ($ \sim 42000\,\rm{K} $) of the NESS temperature ratio are softened for the BG model in comparison to the CD model. In addition, the behavior of the BG and CD models in the two extreme temperature limits are also clearly shown in the figure. In the low-temperature limit, the NESS temperature ratio shows no difference between the two models and both models can be well approximated by the $ n=1 $ monomial model. In the high-temperature limit, however, the NESS temperature ratio for the BG model is raised to a constant above 1, while $ \tilde{r} $ for the CD model approaches $ 1 $, which is the NESS temperature ratio for the $ n=-3 $ monomial model. 

We can understand all these phenomena qualitatively as well. For temperatures higher than room temperature,  the damping $ \epsilon $ in Eq.~\eqref{3NESS} must be bigger for the BG model than the CD model, which will weaken the effect of the resonance. As to why adding the temperature dependence in damping would alter the high-temperature limit of $ \tilde{r} $ but keep its low-temperature limit unchanged, let us re-examine what happens to $ \Im\alpha(\omega) $ in Eq.~\eqref{BG3} in both limits. Recall the damping $ \nu(T') $ grows as $ T'^{5} $ in the low-temperature limit and grows linearly in the high-temperature limit. Therefore, the low-frequency (temperature) behavior for $ \Im\alpha(\omega) $ is still proportional to $ \omega $. Yet, the high-frequency behavior is no longer just proportional to $ \omega^{-3} $, because in the denominator, $ \omega^{2}\nu^{2} $ can be comparable to $ \omega^{4} $ in the high-temperature limit. Indeed, keeping both terms enables us to provide an analytical prediction of the correct high-temperature limit of $ \tilde{r} $ for the BG model, as is shown by the dashed, gray curves in both figures. The analysis is detailed in Appendix \ref{ApC} where $ \tilde{r} $ is given as a solution of an algebraic equation \eqref{C6}.

Finally, let us also calculate the NESS quantum vacuum friction for the BG model. In this regard, we can still use Eq.~\eqref{force3}, replacing the temperature-independent damping $ \epsilon $ with the temperature-dependent one evaluated at the NESS temperature, $ \epsilon(\tilde{T})$. The NESS temperature can be found through $ \tilde{T}=\tilde{r}T $, where $ \tilde{r} $ has already been found numerically as a function of velocity $ v $ and radiation temperature $ T $.
%\begin{equation}\label{BG5}
%F=-Af(v,T), \quad A=\frac{V\omega_{p}^{2}\omega_{1}^{3}}{2\pi^{2}}, \quad f(v,T)=-\frac{1}{\gamma^{2}v^{2}}\int_{0}^{\infty} du \frac{u^{5}\epsilon(r_{T}T)}{(1-u^{2})^{2}+u^{2}\epsilon^{2}(r_{T}T)}\int_{y_{-}}^{y_{+}} dy \,(y-\gamma)\frac{1}{e^{2x_{1}uy}-1}.
%\end{equation}
%For the gold nanosphere with a radius $ a=100\,\rm{nm} $, the force coefficient evaluates to $ A=2.57\cross 10^{-10} \,\rm{N} $.

\begin{figure*}
\subfloat[]{\label{figBG3a}%
\includegraphics[width=0.48\linewidth]{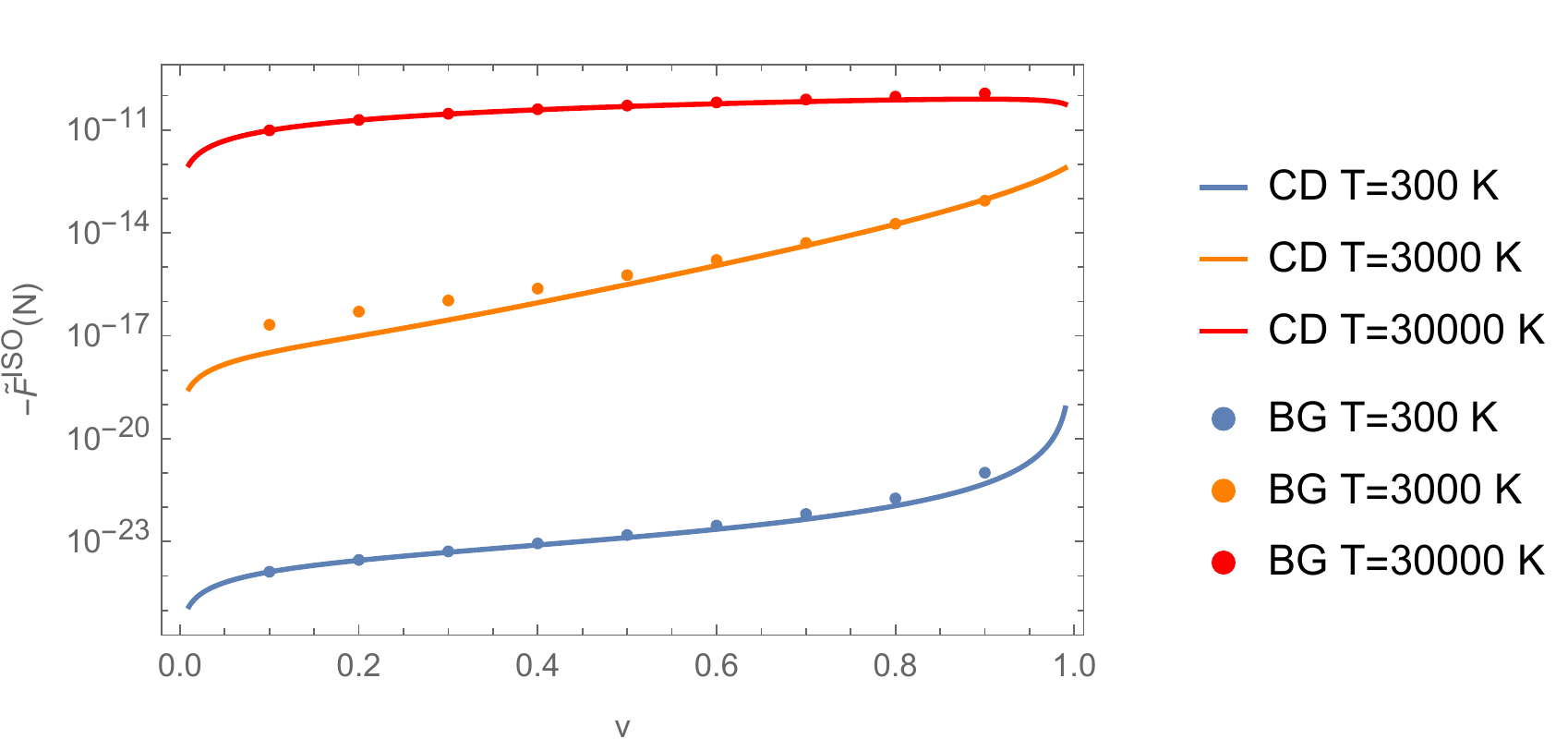}%
}
\subfloat[]{\label{figBG3b}%
\includegraphics[width=0.44\linewidth]{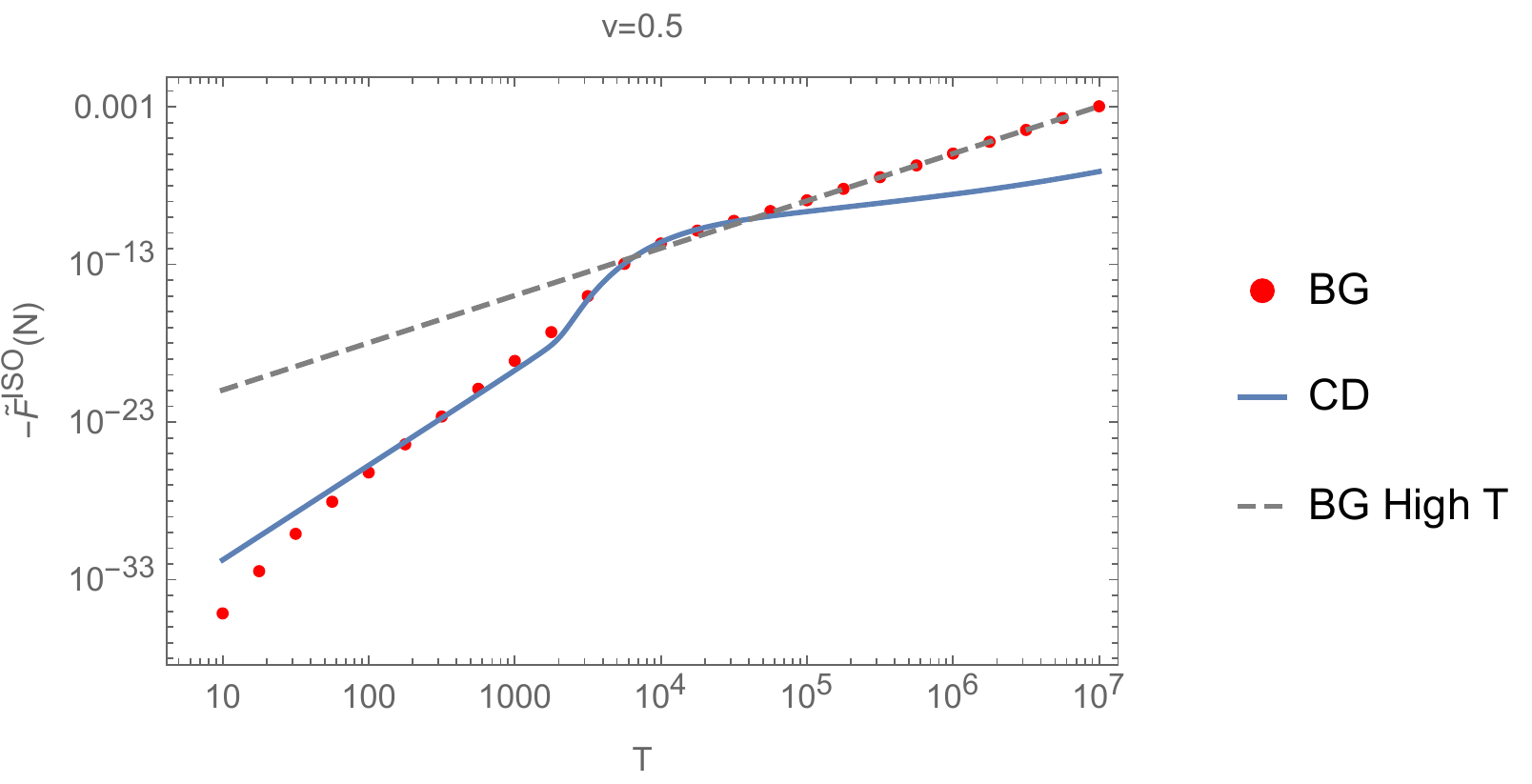}%
}
\caption{The magnitude of the NESS quantum vacuum friction, $ -\tilde{F}^{\rm{ISO}} $, for the BG model is compared with that for the CD model. (a)\, At the three different temperatures, $ -\tilde{F}^{\rm{ISO}} $ is plotted as a function of velocity $ v $ for the BG model and the CD model by the dotted curves and the solid curves, respectively. (b)\, For a fixed velocity $ v=0.5 $, $ -\tilde{F}^{\rm{ISO}} $ is plotted as a function of the radiation temperature $ T $ for the BG model and the CD model by the dotted curves and the solid curves, respectively. The high-temperature prediction of $ -\tilde{F}^{\rm{ISO}} $ for the BG model is plotted by the dashed, gray curve.} 
\label{figBG3}
\end{figure*}

In Fig.~\ref{figBG3a}, the magnitude of the NESS quantum friction is plotted as a function of velocity for both the BG model and the CD model. The behavior of the discrepancy between the two models seems intricate. At $ T=300\,\rm{K} $, the discrepancy is larger for higher velocities. At $ T=3000\,\rm{K} $, the discrepancy is larger for smaller velocities instead. At $ T=30\,000\, \rm{K} $, the data for the two models pretty much agrees until the velocity becomes very close to the speed of light. All of this can be understood as a result of different limiting behaviors of the nanosphere model in different temperature and velocity regimes. As the temperature increases, the behavior of the model goes from the $ n=1 $ monomial model through the resonance model and finally to the $ n=3 $ monomial model. Since the friction as seen in Eq.~\eqref{force3} is independent of damping for the resonance model, the discrepancies between the BG model and the CD model disappear when the resonant frequency dominates. And whenever the model is better approximated by the monomial models, the friction is proportional to the damping parameter. The damping for the  BG model is certainly bigger than that for the CD model at temperatures higher than $ 300\,\rm{K} $. This is even the case when the radiation temperature is exactly $ 300\,\rm{K} $ because, for the BG model, the damping is evaluated at the temperature of the particle, which is still greater than $ 300\,\rm{K} $.

In Fig.~\ref{figBG3b} where the magnitude of quantum friction over a wide range of temperature is plotted, it is seen that the temperature dependence in damping lowers the friction for very low temperatures ($ <100 \,\rm{K} $) because of a smaller damping parameter for the BG model. In the intermediate temperature regime, the temperature dependence of the damping has a relatively smaller effect on the magnitude of the friction. Whether we use the CD model or the BG model for the gold nanosphere, the magnitude of the quantum vacuum friction on the nanosphere reaches the order of picoNewtons\footnote{It may be possible to detect a fluctuation-induced force of such magnitude, since the precision measurement for the static Casimir force can easily reach the order of picoNewtons nowadays. See, for example, Ref.~\cite{Munday:repulsive}. The main challenge for experiments might then be keeping the particle moving at constant velocity in a rather hot background.} if it can be kept moving in a background with a temperature as high as $ 10^{4}\,\rm{K} $, which is unambiguously much greater than that for a gold atom calculated in Ref.~\cite{Xin:eqf1}.  Adding the temperature dependence in the damping, the high-temperature behavior of the force is enhanced from $ T^{2} $ to $ T^{3} $, which is clearly illustrated in the high-temperature prediction of the NESS friction force given in Eq.~\eqref{C8}.

\section{energetics of quantum vacuum friction out of NESS}\label{out}
The transformations of the frictional power and force between frame $ \mathcal{R} $ and frame $ \mathcal{P} $ are \cite{Xin:eqf1}
\begin{subequations}\label{5.1}
\begin{equation}\label{5.1b}
P=\frac{\partial}{\partial t}\mathcal{F}=\gamma \left(\frac{\partial}{\partial t'}-v\frac{\partial}{\partial x'}\right) \frac{1}{\gamma} \mathcal{F'}=\left(\frac{\partial}{\partial t'}-v\frac{\partial}{\partial x'}\right) \mathcal{F'}=P'+vF',
\end{equation}
\begin{equation}\label{5.1a}
F=-\frac{\partial}{\partial x}\mathcal{F}=-\gamma \left(\frac{\partial}{\partial x'}-v\frac{\partial}{\partial t'}\right) \frac{1}{\gamma} \mathcal{F'} =-\left(\frac{\partial}{\partial x'}-v\frac{\partial}{\partial t'}\right)\mathcal{F'}=F'+vP'.
\end{equation}
\end{subequations}
These relations are general and hold whether the particle is in or out of NESS. They can be easily inverted as
\begin{equation}\label{5.2}
P'=\gamma^{2}(P-vF), \qquad F'=\gamma^{2}(F-vP).
\end{equation}
From Eq.~\eqref{5.2}, together with Eq.~\eqref{eqPprime} for $P'^{\rm{P}}$ and Eq.~\eqref{FISO} for $F^{\rm{P}}$, we find the frictional power in frame $ \mathcal{R} $ for a particle in diagonal polarization state $ \rm{P} $ is
\begin{equation}\label{5.25}
P^{\rm{P}}=\frac{1}{3\pi^{2}\gamma}\int_{0}^{\infty} d\omega \Im\alpha_{\rm{P}}(\omega) \,\omega^{4}\int_{y_{-}}^{y_{+}} dy \,y f^{\rm{P}}(y) \left(\frac{1}{e^{\beta\omega y}-1}-\frac{1}{e^{\beta'\omega}-1}\right).
\end{equation}
In the case of an isotropic particle, we confirm that this expression for frictional power agrees with the time component of the four-force that Pieplow and Henkel derive for blackbody friction in Ref.~\cite{P&H:covariant} using a fully covariant formulation, and with the thermal radiation power (a different sign convention is used there, though) obtained by Dedkov and Kyasov in Ref.~\cite{Dedkov:review}.
%\footnote{\textcolor{red}{We note a typo in Eq.~(30) of Ref.~\cite{Dedkov:tangential} for the heating rate $ \dot{Q} $: the $ x $ factor must be omitted in order to be consistent with the following equation $ (31) $ and earlier paper \cite{Dedkov:fluctuation} by the authors. Even after correcting this error, their formula for $ \dot{Q} $ does not agree with our formula for $ P $ but coincides with our formula for $ P' $ if multiplying by an additional $ \gamma^{2} $. However, even if the connection can be made, it contradicts with their claim that $ \dot{Q} $ is the heating rate in the frame of resting background in Ref.~\cite{Dedkov:fluctuation}.}}

When the particle is in NESS, the friction must be balanced by an external force, in either frame $ \mathcal{R} $ or frame $ \mathcal{P} $,
\begin{subequations}\label{5.4}
\begin{equation}\label{5.4a}
\tilde{F}_{\rm{tot}}=\tilde{F}+\tilde{F}_{\rm{ext}}=0,
\end{equation}
\begin{equation}\label{5.4b}
\tilde{F}_{\rm{tot}}'=\tilde{F}'+\tilde{F}'_{\rm{ext}}=0.
\end{equation}
\end{subequations}
Recall that in NESS, the friction is the same in frame $ \mathcal{P} $ and frame $ \mathcal{R} $, $ \tilde{F}'=\tilde{F} $ \cite{Xin:eqf1}. As a result, the external force needed is the same in frame $ \mathcal{P} $ and frame $ \mathcal{R} $, $ \tilde{F}_{\rm{ext}}'=\tilde{F}_{\rm{ext}} $. Since we have demonstrated that $ \tilde{F} $ is negative definite, the external force is positive definite in both frames, 
 $ \tilde{F}_{\rm{ext}}'=\tilde{F}_{\rm{ext}}>0 $.

Now, let us discuss the different situation out of NESS. On the one hand, the internal energy of the particle, or equivalently, the rest mass of the particle, $ m $, is allowed to change.\footnote{The varying mass for moving dipoles has also been discussed pedagogically in Ref.~\cite{Mansuripur}. A quantum mechanical illustration can be found in Ref.~\cite{Barnett:decayatom}.} Therefore, the relativistic momentum of the particle $\gamma m v  $ in $ \mathcal{R} $ can change, while the velocity $ v $ is kept constant by the external force. The varying mass results in a net force on the particle in frame $ \mathcal{R} $, so that Eq.~\eqref{5.4a} must be modified out of NESS,
\begin{equation}\label{5.5}
F_{\rm{tot}}=F+F_{\rm{ext}}=\gamma v\frac{dm}{dt}=v \frac{dm}{dt'}=vP'\,\Rightarrow\, F_{\rm{ext}}=-F+vP',
\end{equation}
where we have used the time dilation relation $ dt=\gamma dt' $ and that the rate of rest mass change, $dm/dt'$ is the same as the rate of change in the particle's internal energy, $ P' $. The rate $ P' $, being precisely the frictional power in the rest frame of the particle, is already derived in Eq.~\eqref{eqPprime} for the different polarizations.

On the other hand, Eq.~\eqref{5.4b} must still hold out of NESS because the particle's momentum in frame $ \mathcal{P} $ remains zero, 
\begin{equation}\label{5.5.2}
F_{\rm{tot}}'=F'+F_{\rm{ext}}'=0
\,\Rightarrow\, F_{\rm{ext}}'=-F'=-F+vP',
\end{equation}
where we have used Eq.~\eqref{5.1a} in the last equality. Comparing Eq.~\eqref{5.5} with Eq.~\eqref{5.5.2}, we find the external force needed in frame $ \mathcal{R} $ and frame $ \mathcal{P} $ is still the same, $ F_{\rm{ext}} =F_{\rm{ext}}'=-F'$. 

%From Eq.~\eqref{5.2}, we are able to express $ F' $ in terms of the known quantities $ F $ and $ P' $,
%\begin{equation}\label{5.6}
%F'=F-vP'.
%\end{equation}
The quantum vacuum friction in frame $ \mathcal{P} $, $ F' $, for a particle in diagonal polarization state $ \rm{P} $, can again be obtained from Eq.~\eqref{FISO} and Eq.~\eqref{eqPprime}, 
\begin{equation}\label{5.7}
F'^{\rm{P}}=F^{\rm{P}}-vP'^{\rm{P}}=\frac{1}{3\pi^{2}\gamma v} \int_{0}^{\infty} d\omega\, \omega^{4} \Im\alpha_{\rm{P}}(\omega)\int_{y_{-}}^{y_{+}} dy \,(y-\gamma)f^{\rm{P}}(y)\left(\frac{1}{e^{\beta\omega y}-1}-\frac{1}{e^{\beta'\omega}-1}\right).
\end{equation}
In Eq.~\eqref{5.7}, the term involving $ \beta' $ clearly does not contribute to the integral on account of the oddness of the $ y $ integrand around $ y=\gamma $. Physically, this reflects the fact that the emitted dipole radiation does not have a momentum bias, and therefore does not contribute to the frictional force in frame $ \mathcal{P} $. See Ref.~\cite{Xin:eqf1} for a detailed explaination. As a result, $ F'^{\rm{P}} $ is independent of the temperature of the particle so long as $ \Im\alpha $ does not depend on temperature. In fact, $ F'^{\rm{P}} $ in Eq.~\eqref{5.7} for an out-of-NESS particle precisely equals the NESS quantum vacuum friction $ \tilde{F}^{\rm{P}} $ shown in Eq.~\eqref{eq6-2}, $ F'^{\rm{P}}=\tilde{F}^{\rm{P}} $. Therefore, like $ \tilde{F}^{\rm{P}} $, $ F'^{\rm{P}} $ is also negative definite.

The quantum vacuum frictional force in frame $ \mathcal{R} $, $ F $, consists of two terms, i.e., $ F=F'+vP' $. The first term, $ F' $, is the steady state contribution $ F'=\tilde{F} $. The second term, $ vP' $, is therefore the nonsteady part, which is due to the rest mass change of the particle.\footnote{This is purely an inertial effect. It even comes into play in the electrodynamics of moving classical dipoles already discussed in Ref.~\cite{Kim:dipole} where the friction $ F $ simply equals $ vP' $. There, the friction on the moving classical dipole in its rest frame, $ F' $, is zero.} This term does not have a definite sign because the particle could either gain or lose rest mass.
%Introducing $ I(\xi) $ in Eq.~\eqref{eq5-21}, $ F' $ can be written as 
%\begin{equation}\label{5.8}
%F'=\frac{1}{2\pi^{2}\gamma^{2}v^{2}}\int_{y_{-}}^{y_{+}} dy \, (y-\gamma) I(\beta y).
%\end{equation}
%Because $ I(\xi) $ is monotonically decreasing in its argument, $ F' $ is negative definite. As a result, $ F_{\rm{ext}}=-F' $  must be in the same direction as $ v $. 
%This simply proves that the neutral particle can not be used as a perpetual motion machine to extract the vacuum energy and convert it into useful work. 

As a result, $ F $ is no longer negative definite out of NESS. That is to say, the quantum vacuum ``friction'' in $ \mathcal{R} $ could become a pushing force on the particle by adjusting the temperature of the particle. Let us now attempt to give another physical interpretation why such a positive friction should occur. If we consider the combined system of particle and radiation, the only net force on the system is $ F_{\rm{ext}} $ provided by some external agent. This external force will cause the total momentum of the combined system to change. The rate of change of the momentum of the radiation equals $ -F $, which is the reaction force to the electromagnetic force (quantum frictional force) exerted on the radiation fields by the particle. The rate of change of the momentum of the particle due to its varying mass is given by $ vP' $. Applying Newton's second law to the combined system of particle and radiation leads to $ F_{\rm{ext}} =-F+vP'$, just as given in Eq.~\eqref{5.5}. Here, $ F_{\rm{ext}} $ is positive definite while $ vP' $ could be positive or negative depending on the radiation temperature, $ T $, and the actual temperature of the particle, $ T' $. As a result, $ F $ could be of either sign. In particular, if the particle gains mass in a rate greater than that provided by the external force, $ vP'>F_{\rm{ext}} $, the electromagnetic force exerted on the particle by the radiation fields will be positive.

%And now we understand physically why such a positive friction could occur from Eq.~\eqref{5.5}. Out of NESS, the particle has freedom to gain mass or increase its internal energy, which produces an effective drag force, $ F_{\rm{eff}}=-v\, dm/dt'=-vP'<0 $, on the particle. If this effective drag is to overcome the positive external force, $ F_{\rm{eff}}+F_{\rm{ext}}<0 $, a positive frictional force, $ F $, is required. 

It is also interesting to study the temperature of the particle, $ T_{0} $, at which the quantum vacuum friction in frame $ \mathcal{R} $ becomes zero. This may be found by solving the equation, $ F=0 $, where $ F $ is the quantum vacuum friction in Eq.~\eqref{FISO}. For the temperature-independent model of the gold nanosphere described in Sec.~\ref{nano1}, the equation   to be solved is
\begin{equation}\label{5.81}
\int_{0}^{\infty} du\frac{u^{5}}{(1-u^{2})^{2}+u^{2}\epsilon^{2}}\int_{y_{-}}^{y_{+}}dy \left(y-\frac{1}{\gamma}\right)\frac{1}{e^{2x_{1} u y}-1}=\int_{0}^{\infty} du\frac{u^{5}}{(1-u^{2})^{2}+u^{2}\epsilon^{2}}\int_{y_{-}}^{y_{+}}dy \left(y-\frac{1}{\gamma}\right)\frac{1}{e^{2x_{1}u/r_{0}}-1}.
\end{equation}
Here, the definitions for dimensionless variables in Eq.~\eqref{dimensionless} still apply, except that we use $ r_{0}=T_{0}/T $ to denote the temperature ratio for $ F=0 $, to be distinguished from the temperature ratio in NESS, $ \tilde{r}=\tilde{T}/T $. In the low-temperature regime ($x_{1}\gg 1  $ or $ T\ll 30\,000 \,\rm{K} $), $ \Im\alpha(\omega) $ reduces to the $ n=1 $ monomial model, for which we already know that $ F $ is negative definite [see footnote \ref{ft4} for a detailed explanation] and therefore no solution could be found for Eq.~\eqref{5.81}. In the high-temperature regime ($ x_{1}\ll 1 $ or $ T\gg 30\,000\,\rm{K} $), $ \Im\alpha(\omega) $ reduces to the $ n=-3 $ monomial model and we therefore obtain the high-temperature limit of the ratio,
\begin{equation}\label{5.82}
r_{0}=\sqrt{\frac{1}{\gamma^{2}v^{2}}\left(\frac{1}{2v}\ln\frac{1+v}{1-v}-1\right)}.
\end{equation}
For intermediate temperatures around the resonance ($x_{1}\sim 1 $ or $ T \sim 30\,000 \rm{K} $), we may set $ u=1 $ and find the approximate ratio,
\begin{equation}\label{5.83}
r_{0}=2x_{1}\left(\ln\left\lbrace 1+2\gamma^{2}v^{3}\left[\int_{y_{-}}^{y_{+}} dy \left(y-\frac{1}{\gamma}\right)\frac{1}{e^{2x_{1}y}-1}\right]^{-1}\right\rbrace\right)^{-1},
\end{equation}
where the $ y $ integral in Eq.~\eqref{5.83} can be expressed in terms of dilogarithms.

\begin{figure*}
\subfloat[]{\label{figr0a}%
\includegraphics[width=0.48\linewidth]{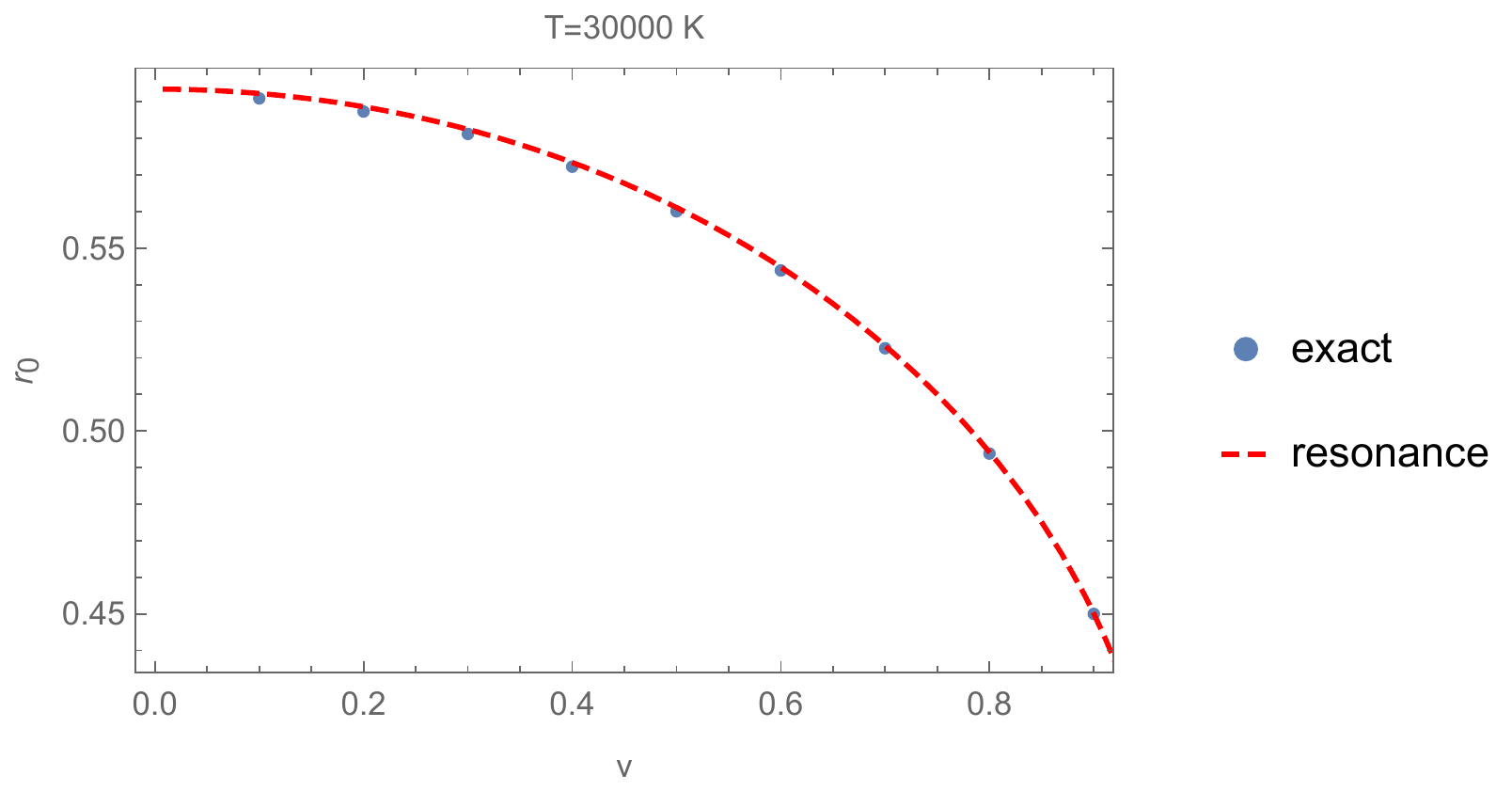}%
}
\subfloat[]{\label{figr0b}%
\includegraphics[width=0.48\linewidth]{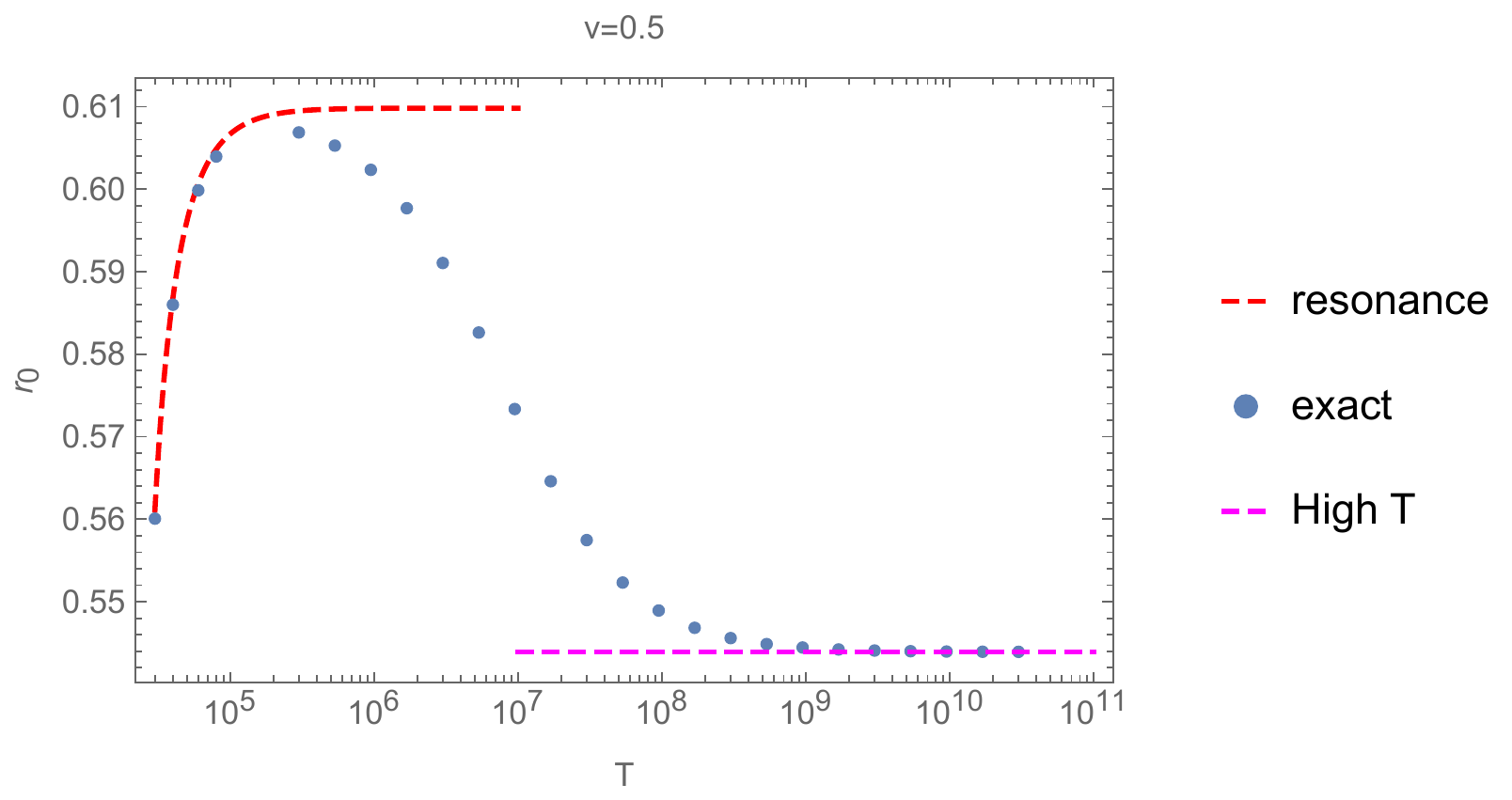}%
}
\caption{For the constant damping model of the gold nanosphere, the ratio $r_{0}= T_{0}/T $ is solved numerically from Eq.~\eqref{5.81} and illustrated. Here, $ T_{0} $ is the temperature of the particle at which the quantum vacuum frictional force on it is zero in the rest frame of radiation. (a) The velocity dependence of $ r_{0} $ at $ T=30\,000 \,\rm{K}$ $(x_{1}=1.01) $ is illustrated. The blue dots display the numerical solution obtained by solving Eq.~\eqref{5.81} directly. The dashed red curve shows $ r_{0} $ calculated from the resonance approximation Eq.~\eqref{5.83}. (b) The temperature dependence of $ r_{0} $ for fixed velocity $ v=0.5 $ is illustrated. The exact numerical data are shown by the blue dots. The dashed red curve plots the resonance approximation. The dashed magenta curve  shows the value of the high-temperature limit of the ratio according to Eq.~\eqref{5.82}, $ r_{0}=\sqrt{3(\ln3-1)}=0.544 $. } 
\label{figr0}
\end{figure*}

It is seen in Fig.~\ref{figr0a} that, at $ T=30\,000\, \rm{K} $, the prediction given by the resonance formula Eq.~\eqref{5.83} agrees very well with the exact numerical data for $ r_{0} $. In Fig.~\ref{figr0b}, we see that the ratio $ r_{0} $, for fixed velocity $ v=0.5 $, is described well by the resonance approximation only up to about $ 10^{5}\,\rm{K} $. Then the ratio goes through a transition region ($ 10^{5}-10^{9} \,\rm{K} $) and eventually decays to its high-temperature limit given by Eq.~\eqref{5.82} at around $ 10^{9}\,\rm{K} $.
For the Bloch-Gr\"{u}neisen model, we also give a high-temperature prediction of $ r_{0} $ in Eq.~\eqref{C10} of Appendix \ref{ApC}.

From Eq.~\eqref{5.5}, we see that the sign of the frictional power, $ P' $, decides both the sign of the total force on the particle and the sign of its mass change. Using the definition for $ I(\xi) $ in Eq.~\eqref{eq5-21}, $ P'^{\rm{P}} $ in Eq.~\eqref{eqPprime} can be rewritten as
\begin{equation}\label{eq5-11-2}
P'^{\rm{P}}=\frac{1}{3\pi^{2}}\int_{y_{-}}^{y_{+}} f^{\rm{P}}(y)\left[I(\beta y)-I(\beta')\right],
\end{equation}
where the first term corresponds to the absorbed power due to the field fluctuations and the second term corresponds to the emitted power due to the dipole fluctuations.
We can replace the first term using the NESS condition Eq.~\eqref{eq5-22} 
\begin{equation}\label{eq5.10}
\int_{y_{-}}^{y_{+}} dy \, f^{\rm{P}}(y)I(\tilde{{\beta}})=\int_{y_{-}}^{y_{+}} dy \,f^{\rm{P}}(y)I(\beta y).
\end{equation}
As a result, the frictional power now reads simply
\begin{equation}\label{5.11}
P'=\frac{1}{3\pi^{2}}\left[I(\tilde{\beta})-I(\beta')\right]\int_{y_{-}}^{y_{+}} dy \, f^{\rm{P}}(y),
\end{equation}
where the integral on $ y $ becomes $ 3 $ for the isotropic polarization and $ 1 $ for other diagonal polarizations.
Because $ I(\xi) $ is a decreasing function, the sign of $ P'^{\rm{P}} $ is determined by the deviation of the temperature of the particle from its NESS temperature as below,
\begin{equation}\label{5.12}
\addtolength{\arraycolsep}{-3pt}
\left\{%
 \begin{array}{lcrcl}
T'<\tilde{T} \,\Rightarrow\,  P'^{\rm{P}}>0, \,\\\\
T'=\tilde{T} \,\Rightarrow\,  P'^{\rm{P}}=0, \,\\\\
T'>\tilde{T} \,\Rightarrow\,  P'^{\rm{P}}<0. \,\\
 \end{array}
 \right.
\end{equation}

By now, we have come to an understanding of the issue of the stability of NESS. Imagine the particle slightly deviates from NESS with a temperature lower than the NESS temperature, $ T'<\tilde{T} $. According to Eq.~\eqref{5.12}, the particle will absorb net energy thereafter, which will in turn raise the temperature of the particle $ T' $ \footnote{Here, we assume the change of the particle's temperature is adiabatic so that it still has a well-defined temperature while it is away from NESS, even though the particle is in neither an equilibrium state nor a steady state. } so that it becomes closer to the NESS temperature $ \tilde{T} $. Therefore, the particle would tend to return to NESS after deviating from it. This seems to provide theoretical support for experimentally measuring the NESS temperature of the particle as a feasible signature for the quantum vacuum frictional effects.

One might imagine that a positive quantum vacuum frictional force qualifies the particle to function as a perpetual motion machine, which could extract vacuum energy from the radiation fields and turn it into useful mechanical work endlessly. We argue that this is not possible. First, the external agent keeps providing energy to the particle-radiation system and the law of conservation of energy is never violated. Therefore, the particle is not a perpetual motion machine of the first kind. Second, the positive ``friction" only occurs if the temperature of the particle is lower than the NESS temperature and the particle gains internal energy. Heat transfers to a relatively colder reservoir.\footnote{The particle is colder in the sense that its actual temperature is lower than the temperature needed to reach the steady state where there is no net heat transfer.} In this process, the entropy should increase and the second law of thermodynamics is respected. Therefore, the particle is not a perpetual motion machine of the second kind either.   

\section{conclusions}\label{conclusions}
In this paper, we studied the quantum vacuum frictional phenomenon associated with an intrinsically dissipative particle.
We not only derived conditions for the particle to be in the nonequilibrium steady state (NESS) and found expressions for quantum vacuum friction in NESS, but also extended the analysis to the out-of-NESS scenario.

Independent of the particle's polarizability, the NESS temperature, $ \tilde{T} $, has a general lower bound, $ T/\gamma $, the Planck-Einstein transformed temperature of the blackbody radiation. Furthermore, the nonrelativistic limit of the particle's NESS temperature always equals the radiation temperature.
Under the NESS conditions, the quantum vacuum friction on the dissipative particle is shown to be negative definite and to be related to the friction for a nondissipative particle, investigated in our previous paper \cite{Xin:eqf1}. We also show that the NESS quantum vacuum friction reduces to the well-known Einstein-Hopf drag in the nonrelativistic limit.

The NESS temperature ratio and the NESS quantum vacuum friction are first calculated explicitly for the resonance model and the monomial models and then for a gold nanosphere modelled either with a constant damping (constant damping model) or a temperature-dependent damping (Bloch-Gr\"{u}neisen model). The NESS temperature ratio and friction for the resonance model and the monomial models can be worked out analytically, providing insight to the numerical results of the more realistic models used for the gold nanosphere. The monomial models are good approximations for the gold nanosphere with constant damping in the two extreme temperature limits. The resonance model approximates the behavior in the intermediate temperature region. In addition, the nonmonotonic behavior (in velocity) of the NESS friction on the gold  nanosphere in the ultra-relativistic region is mainly associated with the resonance contributions. For both models (constant damping model and the Bloch-Gr\"{u}neisen model) of the gold nanosphere, the NESS temperature ratio is found to be maximized around $ T=3000\, \rm{K} $ and the deviation of the NESS temperature from the radiation temperature is quite noticeable around this maximum. The magnitude of the NESS quantum vacuum friction increases with temperature and reaches the order of picoNewtons if the background temperature could be raised to $ 10^{4} \,\rm{K} $. Comparing the Bloch-Gr\"{u}neisen model with the constant damping model of the gold nanosphere, the NESS friction is enhanced for higher temperatures but suppressed for lower temperatures, and the high-temperature limit of the NESS temperature ratio is raised to above $ 1 $.

The energetics out of NESS is different from that in NESS because the particle can absorb or emit net energy, thereby changing its rest mass. The change of the particle's rest mass produces a change in its momentum in the rest frame of radiation ($\mathcal{R}$). As a result, the quantum vacuum friction, $ F $, in frame $ \mathcal{R}$, no longer balances the external force, $ F_{\rm{ext}} $, needed to keep the particle moving at constant velocity, and it is allowed to take either sign or even zero. Nonetheless, the quantum vacuum friction, $ F' $, in frame $ \mathcal{P}$, is still negative definite. And we find that $ F_{\rm{ext}} $ must be opposite to $ F' $ and positive definite, which excludes the possibility of converting the vacuum energy into useful mechanical work using the moving particle as a perpetual motion machine. It is precisely the difference between the particle's actual temperature, $ T' $, and its NESS temperature, $ \tilde{T} $, that determines whether the particle absorbs or emits energy. Only if $ T'=\tilde{T} $ is the particle's internal energy conserved. The particle will absorb energy if $ T'<\tilde{T} $ and emit energy if $ T'>\tilde{T} $. To sum up, the particle tends to return to NESS after deviating from it. If the temperature of a nanosphere moving at constant velocity could be measured, it is then expected to be the NESS temperature, which would show some deviation from the temperature of the lab environment. Such an experiment could in principle be used to identify the quantum vacuum frictional effects discussed in this paper.

%acknowledgements
\begin{acknowledgments}
We thank the US National Science Foundation, grants Nos. 1707511, 2008417, for partial support of this work. We thank S. Fulling, P. Parashar and J. J. Marchetta for insightful comments. We thank G. V. Dedkov for pointing us to their papers so that we are able to confirm the agreement with their results. This paper reflects solely the authors' personal opinions and does not represent the opinions of the authors' employers, present and past, in any way.
\end{acknowledgments}

%appendix
\appendix
\section{THE EQUIVALENCE OF THE LORENTZ FORCE LAW AND THE PRINCIPLE OF VIRTUAL WORK}\label{apC}
In this appendix, we apply the Lorentz force law to a moving electric dipole and illustrate that the Lorentz force can be equivalently calculated through differentiating a free energy including both a electric dipole contribution and a motion-induced magnetic dipole contribution. 

The Lorentz force density reads
\begin{equation}\label{A1}
\vb{f}(t,\vb{r})=\rho(t,\vb{r})\vb{E}(t,\vb{r})+\vb{j}(t,\vb{r})\cross\vb{B}(t,\vb{r}).
\end{equation}
The charge and current densities corresponding to a moving electric dipole with constant velocity $ \vb{v} $ is\footnote{If one only considers the requirement of current continuity, $ \partial_{t} \rho+\div \vb{j}=0 $, there might be another choice of the current, $ \vb{j}_{2}(t, \vb{r})= -\vb{d}(t) \div{\vb{v}}\delta(\vb{r}-\vb{v}t)+\dot{\vb{d}}(t)\delta(\vb{r}-\vb{v}t)$, which differs from $ \vb{j}(t,\vb{r}) $ in Eq.~\eqref{A2b} by a total curl. In fact, the current we use is associated with the convection current, which includes both the polarization current $ \vb{j}_{2}(t, \vb{r}) $ and a motion-induced magnetization current $ \vb{j}_{m}(t, \vb{r})=\curl [\vb{d}(t)\cross\vb{v}\delta(\vb{r}-\vb{v}t)] $. See Ref.~\cite{Hnizdo} for a discussion.  Therefore, if we use instead $ \vb{j}_{2}(t, \vb{r}) $ in the derivation, we will not see the motion-induced magnetic dipole term in Eq.~\eqref{A8}. Such ambiguity of current for a moving dipole can be avoided by deriving it from the electric and magnetic polarization fields $ \vb{P} $ and $ \vb{M} $ in Ref.~\cite{Kim:dipole}.}
\begin{subequations}\label{A2}
\begin{equation}\label{A2a}
\rho(t,\vb{r})=-\div \vb{d}(t)\delta(\vb{r}-\vb{v}t),
\end{equation}
\begin{equation}\label{A2b}
\vb{j}(t,\vb{r})=-\vb{v} \div{\vb{d}}(t)\delta(\vb{r}-\vb{v}t)+\dot{\vb{d}}(t)\delta(\vb{r}-\vb{v}t).
\end{equation}
\end{subequations}
After integrating the Lorentz force density $ \vb{f} $ over all space, we find the Lorentz force on the dipole to be
\begin{equation}\label{A3}
\vb{F}(t)=\vb{d}(t)\cdot\grad \vb{E}(t,\vb{v}t)+\vb{d}(t)\cdot\grad\left[\vb{v}\cross\vb{B}(t,\vb{v}t)\right]+\dot{\vb{d}}(t)\cross\vb{B}(t,\vb{v}t).
\end{equation}
To prove the principle of virtual work, we are to write the right side of Eq.~\eqref{A3} in the form of a total derivative on the spatial arguments. 

The second term in Eq.~\eqref{A3} can be written as
\begin{align}\label{A4}
\vb{d}(t)\cdot\grad\left[\vb{v}\cross\vb{B}(t,\vb{v}t)\right]&=-\vb{d}(t)\cross\left[\grad\cross\left(\vb{v}\cross\vb{B}(t,\vb{v}t)\right)\right]+\grad\left[\vb{d}(t)\cdot\left(\vb{v}\cross\vb{B}(t,\vb{v}t)\right)\right]\nonumber\\
&=\vb{d}(t)\cross\left[\vb{v}\cdot\grad\vb{B}(t,\vb{v}t)\right]+\grad\left[\vb{d}(t)\cdot\left(\vb{v}\cross\vb{B}(t,\vb{v}t)\right)\right],
\end{align}
where we have used $ \div \vb{B}=0 $ in the second equality.
The third term in Eq.~\eqref{A3} can be written as 
\begin{equation}\label{A5}
\dot{\vb{d}}(t)\cross\vb{B}(t,\vb{v}t)=\frac{d}{dt}\left[\vb{d}(t)\cross\vb{B}(t,\vb{v}t)\right]-\vb{d}(t)\cross\frac{\partial}{•\partial t}\vb{B}(t,\vb{v}t)-\vb{d}(t)\cross\left[\vb{v}\cdot\grad\vb{B}(t,\vb{v}t)\right],
\end{equation}
where the middle term can be broken into two pieces with the use of Faraday's law,
\begin{equation}\label{A6}
-\vb{d}(t)\cross\frac{\partial}{•\partial t}\vb{B}(t,\vb{v}t)=\vb{d}(t)\cross\left[\grad\cross\vb{E}(t,\vb{v}t)\right]=\grad\left[\vb{d}(t)\cdot\vb{E}(t,\vb{v}t)\right]-\vb{d}(t)\cdot\grad \vb{E}(t,\vb{v}t).
\end{equation}

When we put Eqs.~\eqref{A4}--\eqref{A6} back into Eq.~\eqref{A3}, several terms cancel and only three terms remain in the expression for the Lorentz force:
\begin{equation}\label{A7}
\vb{F}(t)
=\grad\left[\vb{d}(t)\cdot\vb{E}(t,\vb{v}t)\right]+\grad\left[\left(\vb{d}(t)\cross\vb{v}\right)\cdot\vb{B}(t,\vb{v}t)\right]+\frac{d}{dt}\left[\vb{d}(t)\cross\vb{B}(t,\vb{v}t)\right].
\end{equation}
A more general  derivation of the Lorentz force on neutral particles including the contribution of an intrinsic magnetic dipole moment can be found in Ref.~\cite{Schwinger:ce}. Since the neutral particle considered in the paper does not possess a magnetic dipole polarizability or magnetic dipole fluctuations, Eq.~\eqref{A7} is already sufficient for our purpose. In addition, the last term in Eq.~\eqref{A7}, being a total derivative with respect to time, would not contribute to the quantum frictional force, which is really the average time rate of change in the frictional impulse over a large time period. 

As a result, we are able to express the relevant Lorentz force $ F $ as the gradient on the same free energy $ \mathcal{F} $ used in the power expression \eqref{eq2.8} and therefore proves the principle of virtual work,
\begin{equation}\label{A8}
\vb{F}=-\grad \mathcal{F}, \quad \mathcal{F}=-\vb{d}(t)\cdot\vb{E}(t,\vb{v}t)-\bm{\mu}_{v}(t)\cdot\vb{B}(t,\vb{v}t),
\end{equation}
where we have identified $ \vb{d}(t)\cross\vb{v}=\bm{\mu}_{v}(t) $ as the magnetic dipole moment induced by the movement of the electric dipole. The analysis above is general and Eq.~\eqref{A8} is applicable in any frame of reference. In particular, in the rest frame of the particle $ \mathcal{P} $, the magnetic term in the free energy vanishes and the principle of virtual work reads
\begin{equation}\label{A9}
\vb{F}'=-\grad' \mathcal{F'}, \quad \mathcal{F'}=-\vb{d}'(t')\cdot\vb{E}'(t',\vb{0}).
\end{equation}

\section{THE BLOCH-GR\"UNEISEN MODEL IN THE HIGH-TEMPERATURE LIMIT}\label{ApC}
%Consider a nanosphere of radius $a$ made of dielectric material with permittivity modelled by a Lorentzian oscillator:
%\begin{equation}
%\varepsilon(\omega)=1+\frac{\omega_p^2}{\omega_0^2-\omega^2-i\omega\nu},
%\end{equation}
%where $\omega_p$ is the plasma frequency, $\omega_0$ is the resonance frequency, and $\nu$ is the damping (resistivity) parameter. The imaginary part of the polarizability of the nanosphere is then
%\begin{equation}
%\operatorname{Im} \alpha(\omega)=\frac{V \omega_p^2 \,\omega \nu}{\left(\omega^2-\omega_1^2\right)^2+\omega^2\nu^2},
%\end{equation}
%where $\omega_1\equiv \sqrt{\omega_0^2+\frac{\omega_p^2}{3}}$ is the shifted resonance frequency and $V=\frac{4\pi a^3}{3}$ is the volume of the nanosphere. 
In this appendix, we obtain high-temperature asymptotic expressions for the power and the frictional force in the case that the damping parameter has the linear high-temperature dependence of the Bloch-Gr\"uneisen model~\cite{Gruineisen, Bloch, Ziman} as shown in Eq.~\eqref{BG2}, 
\begin{equation}\label{C1}
\nu=\frac{2\pi \eta}{\beta'}=2\pi \eta T',
\end{equation}
for some constant $\eta>0$.\footnote{In terms of the parameters in Eq.~\eqref{BG1}, $ \eta=\nu_{0}/8\pi k_{B}\theta  $ and it evaluates to be $ \eta=0.219 $ for gold.}

The power in the rest frame of an isotropic particle is given by
\begin{equation}\label{C2}
\begin{aligned}
P'&=\frac{1}{2\pi^2\gamma v}\int_0^{\infty}d\omega\, \omega^4\operatorname{Im} \alpha(\omega) \int_{y_{-}}^{y_{+}} d y\left(\frac{1}{e^{\beta\omega  y}-1}-\frac{1}{e^{\beta'\omega}-1}\right)\\
&=\frac{V\omega_p^2\,\nu}{2\pi^2\gamma v}\int_0^{\infty}d\omega\, \frac{\omega^5}{\left(\omega^2-\omega_1^2\right)^2+\omega^2\nu^2} \int_{y_{-}}^{y_{+}} d y\left(\frac{1}{e^{\beta\omega y}-1}-\frac{1}{e^{\beta'\omega}-1}\right)\\
&=\frac{V\omega_p^2\,\eta}{\pi\gamma v \beta'^3}\int_{y_-}^{y_+}d y\int_0^{\infty}dz\,\left(\frac{z^5}{r^2 y^2\left[\left(z^2-\omega_1^2\beta'^2r^2 y^2\right)^2+4\pi^2 \eta^2 r^2 y^2 z^2\right]}-\frac{z^5}{\left(z^2-\omega_1^2\beta'^2\right)^2+4\pi^2 \eta^2 z^2}\right)\frac{1}{e^z-1},
\end{aligned}
\end{equation}
where $r \equiv \frac{\beta}{\beta'}=\frac{T'}{T}$. 
In the high-temperature limit, $\beta, \beta' \to 0$, and $ r=\beta/\beta' $ tends to a finite value, the dependence on the shifted resonance frequency is suppressed. Retaining only the leading term in this limit, we obtain
\begin{equation}\label{C3}
\begin{aligned}
P'&=\frac{V\omega_p^2\,\eta}{\pi\gamma v \beta'^3}\int_{y_-}^{y_+}d y\int_0^{\infty}dz\,\left(\frac{z^3}{r^2 y^2\left(z^2+4\pi^2 \eta^2 r^2 y^2\right)}-\frac{z^3}{z^2+4\pi^2 \eta^2}\right)\frac{1}{e^z-1}\\
&=\frac{V\omega_p^2\, \pi \eta}{\gamma v \beta'^3} \int_{y_-}^{y_+}dy\,\left\{\frac{1}{r^2y^2}\left[\frac16-2\eta^2r^2y^2
\left(\ln(\eta ry)-\psi(\eta ry)-\frac{1}{2\eta ry}\right)\right]-\left[\frac16-2\eta^2\left(\ln \eta -\psi(\eta) -\frac{1}{2\eta}\right)\right]\right\}\\
&=\frac{V\omega_p^2 \,\pi \eta r^{3}}{3\beta^3}\left[\frac{1}{r^2}+\frac{6\eta}{\gamma v r}\ln\left(\frac{y_+\Gamma(\eta ry_+)}{\Gamma(\eta ry_-)}\right)-1-6\eta-12\eta^2\left(\ln r +\psi(\eta)+\frac{\ln y_+}{v}-1\right)\right],
\end{aligned}
\end{equation}
where we have employed the integral representation of the digamma function,
\begin{equation}\label{C4}
\psi(s)\equiv \frac{d}{ds}\ln \Gamma(s)=\ln s -\frac{1}{2s} -2 \int_0^{\infty}dt\,\frac{t}{(t^2+s^2)(e^{2\pi t}-1)}, \quad \operatorname{Re} s>0,
\end{equation}
which follows immediately on differentiation of Binet's second integral formula for the (log) gamma function \cite{Whittaker, Erdelyi, Gradshteyn}, to deduce that 
\begin{equation}\label{C5}
\int_0^{\infty}dz\, \frac{z^3}{(z^2+u^2)(e^z-1)}=\frac{\pi^2}{6}-\frac{u^2}{2}\left[\ln\left(\frac{u}{2\pi}\right)-\psi\left(\frac{u}{2\pi}\right)-\frac{\pi}{u}\right], \quad \operatorname{Re} u>0.
\end{equation}
When the NESS condition, $P'=0$, holds, it is clear from Eq.~\eqref{C3} that the NESS temperature ratio $\tilde{r}$ satisfies the implicit equation
\begin{equation}\label{C6}
\tilde{r}^2\left[1+6\eta+12\eta^2\left(\ln \tilde{r} +\psi(\eta) +\frac{\ln y_+}{v}-1\right)\right] -\frac{6\eta \tilde{r}}{\gamma v}\ln\left(\frac{y_+\Gamma(\eta \tilde{r}y_+)}{\Gamma(\eta \tilde{r}y_-)}\right)=1
\end{equation}
in the high-temperature limit. 

Similarly, the NESS frictional force in Eq.~\eqref{eq6-2}, which is equivalent to the frictional force in frame $ \mathcal{P} $ in Eq.~\eqref{5.7}, is given by
\begin{equation}\label{C7}
\begin{aligned}
\tilde{F}=F'&=\frac{1}{2\pi^2\gamma^2 v^2}\int_0^{\infty}d\omega\, \omega^4\operatorname{Im} \alpha(\omega) \int_{y_{-}}^{y_{+}} d y\,(y-\gamma)\,\frac{1}{e^{\beta\omega  y}-1}\\
&=\frac{V\omega_p^2\,\nu}{2\pi^2\gamma^2 v^2}\int_0^{\infty}d\omega\, \frac{\omega^5}{\left(\omega^2-\omega_1^2\right)^2+\omega^2\nu^2} \int_{y_{-}}^{y_{+}} d y\,(y-\gamma)\,\frac{1}{e^{\beta\omega y}-1}\\
&=\frac{V\omega_p^2\,\eta}{\pi\gamma^2 v^2 \beta'^3}\int_{y_-}^{y_+}d y\,\frac{(y-\gamma)}{\tilde{r}^2y^2}\int_0^{\infty}dz\,\frac{z^5}{\left(z^2-\omega_1^2\beta'^2\tilde{r}^2 y^2\right)^2+4\pi^2 \eta^2 \tilde{r}^2 y^2 z^2}\,\frac{1}{e^z-1},
\end{aligned}
\end{equation}
which becomes, in the high-temperature limit, 
\begin{equation}\label{C8}
\begin{aligned}
\tilde{F}=F'&=\frac{V\omega_p^2\,\eta}{\pi\gamma^2 v^2 \beta'^3}\int_{y_-}^{y_+}d y\,\frac{(y-\gamma)}{\tilde{r}^2y^2}\int_0^{\infty}dz\,\frac{z^3}{z^2+4\pi^2 \eta^2 \tilde{r}^2 y^2}\frac{1}{e^z-1}\\
&=\frac{V\omega_p^2\, \pi \eta}{\gamma^2 v^2 \beta'^3} \int_{y_-}^{y_+}dy\,\frac{(y-\gamma)}{\tilde{r}^2y^2}\left[\frac16-2\eta^2\tilde{r}^2y^2\left(\ln(\eta \tilde{r}y)-\psi(\eta \tilde{r}y)-\frac{1}{2\eta \tilde{r}y}\right)\right]\\
&=\frac{2V\omega_p^2 \,\pi \eta \tilde{r}^{3}}{\gamma^2v^2\beta^3}\left\{\frac{1}{\tilde{r}^2}\left[\frac{\ln y_+-\gamma^2v}{6}+\ln\left(\frac{G(1+\eta \tilde{r}y_+)}{G(1+\eta \tilde{r}y_-)}\right)\right]-\frac{\gamma \eta}{\tilde{r}} \left[v\ln(2\pi)+\ln\left(\frac{y_+\Gamma(\eta \tilde{r}y_+)}{\Gamma(\eta \tilde{r}y_-)}\right)\right]+\eta^2\left(\gamma^2v+\ln y_+\right)\right\},
\end{aligned}
\end{equation}
where $G$ is the Barnes (double gamma) $G$-function \cite{Barnes, NIST, Voros, Adamchik, CPChen}. 

It follows from Eq.~\eqref{C3} and Eq.~\eqref{C8} that, in the high-temperature limit, the force in the rest frame of the blackbody radiation, $F=F'+vP'$, is given by 
\begin{equation}\label{C9}
\begin{aligned}
F=\frac{2V\omega_p^2 \,\pi \eta r^{3}}{\gamma^2v^2\beta^3}&\left\{\frac{1}{r^2}\left[\frac{\ln y_+-v}{6}+\ln\left(\frac{G(1+\eta ry_+)}{G(1+\eta ry_-)}\right)\right]-\frac{\gamma \eta}{r} \left[v\ln(2\pi)+\left(1-v^2\right)\ln\left(\frac{y_+\Gamma(\eta ry_+)}{\Gamma(\eta ry_-)}\right)\right]\right.\\
&-\left. \gamma^2v^3\left(\frac16+\eta+2\eta^2\left(\ln r+\psi(\eta)\right)\right)+\eta^2\left(\left(3\gamma^2-2\right)v+\left(3-2\gamma^2\right)\ln y_+\right)\right\}.
\end{aligned}
\end{equation}
In particular, when the frictional force becomes zero in frame $ \mathcal{R} $, $ F=0 $, the corresponding temperature ratio $ r_{0} $ satisfies the implicit equation in the high temperature limit
\begin{equation}\label{C10}
\begin{aligned}
&r_{0}^2\left[\gamma^2v^3\left(\frac16+\eta+2\eta^2\left(\ln r_{0} +\psi(\eta)\right)\right)+\eta^2\left(\left(2-3\gamma^2\right)v+\left(2\gamma^2-3\right)\ln y_+\right)\right]\\
&\quad+\gamma \eta r_{0} \left[v\ln(2\pi) +\left(1-v^2\right) \ln\left(\frac{y_+\Gamma(\eta r_{0}y_+)}{\Gamma(\eta r_{0}y_-)}\right)\right]=
\frac{\ln y_+-v}{6}+\ln\left(\frac{G(1+\eta r_{0}y_+)}{G(1+\eta r_{0}y_-)}\right).
\end{aligned}
\end{equation}

The exact analytical expressions in Eq.~\eqref{C6} and Eq.~\eqref{C10} enable exploration of the variation of $r$ with $v$ across the entire $v$ domain, in the high-temperature limit. They therefore provide a useful check of some of the numerical results obtained by other means.

\section{SOLUTION FOR $ r_{0} $ IN THE LOW-TEMPERATURE, HIGH-VELOCITY REGIME}\label{apB}
In Sec.~\ref{out}, we commented that quantum vacuum friction $ F $ in frame $ \mathcal{R} $ can be made zero only for high temperatures, for example, $ 30\,000 \,\rm{K} $ for the gold nanosphere. In fact, this is only true in the moderate velocity regime. It is possible to find solutions to Eq.~\eqref{5.81} for lower temperatures in the ultrarelativistic regime.\footnote{Of course, ultrarelativistic speed is hard to achieve experimentally for neutral particles.} This appendix explores that possibility.

Let us define the left hand side of Eq.~\eqref{5.81} as $ f_{L}(\gamma, x_{1}, \epsilon) $ and the right hand side as $ f_{R}(\gamma, x_{1},\epsilon, r_{0}) $. In the low-temperature regime, $ x_{1}\gg 1 $, for finite $ r_{0} $, it is clear that $ f_{R}$ is dominated by low frequency-contributions $ u\ll 1 $. The contribution around the resonance $ u\sim 1 $ is suppressed because of the exponential factor so that $ f_{R} $ reduces to
\begin{equation}\label{B1}
 f_{R}(\gamma, x_{1},\epsilon, r_{0})\sim 2\Gamma(6)\zeta(6)\gamma^{2}\left(\frac{r_{0}}{2x_{1}}\right)^{6},
\end{equation}
where we have also applied the high velocity limit, $ \gamma\gg 1 $. For $ f_{L} $, we note the exponential factor contains $ y $. In the ultrarelativistic region, the integration on $ y $ is essentially taken over the interval $ [1/2\gamma,2\gamma] $. The lower values of $y$ allow the resonance around $ u\sim 1 $ to contribute significantly to the $ u $ integration while the higher values of $ y $ are suppressed because $ x_{1} $ is already large. As a result, the $ u $ integration in $ f_{L} $ collapses, leaving
\begin{equation}\label{B2}
f_{L}(\gamma, x_{1}, \epsilon)\sim\frac{\pi}{2\epsilon}\int_{1/2\gamma}^{2\gamma} dy \left(y-\frac{1}{\gamma}\right) \frac{1}{e^{2x_{1}y}-1}.
\end{equation}
Equating $ f_{R} $ in Eq.~\eqref{B1} and $ f_{L} $ in Eq.~\eqref{B2} gives an approximation for $ r_{0} $ in the low-temperature, high-velocity regime,
\begin{equation}\label{B3}
r_{0}=x_{1} \left[\frac{126}{\pi^{5}\gamma^{2}\epsilon}\int_{1/2\gamma}^{1} dy \left(y-\frac{1}{\gamma}\right) \frac{1}{e^{2x_{1}y}-1}\right]^{1/6},
\end{equation}
where we delibrately cut off  the $ y $ integration on the upper limit because this will not hurt the accuracy in the large $ x_{1} $ limit but improve the efficiency of the numerical evaluation of the integral.

Let us note the zero of the integral inside the bracket of Eq.~\eqref{B3} gives precisely the lower bound of the velocity for which Eq.~\eqref{5.81} still has a solution. At room temperature $T=300\,\rm{K}$ ($ x_{1}=101 $), this lower bound is found to be $ \gamma=115 $.
\begin{figure}[h!]
 \includegraphics[width=0.7\linewidth]{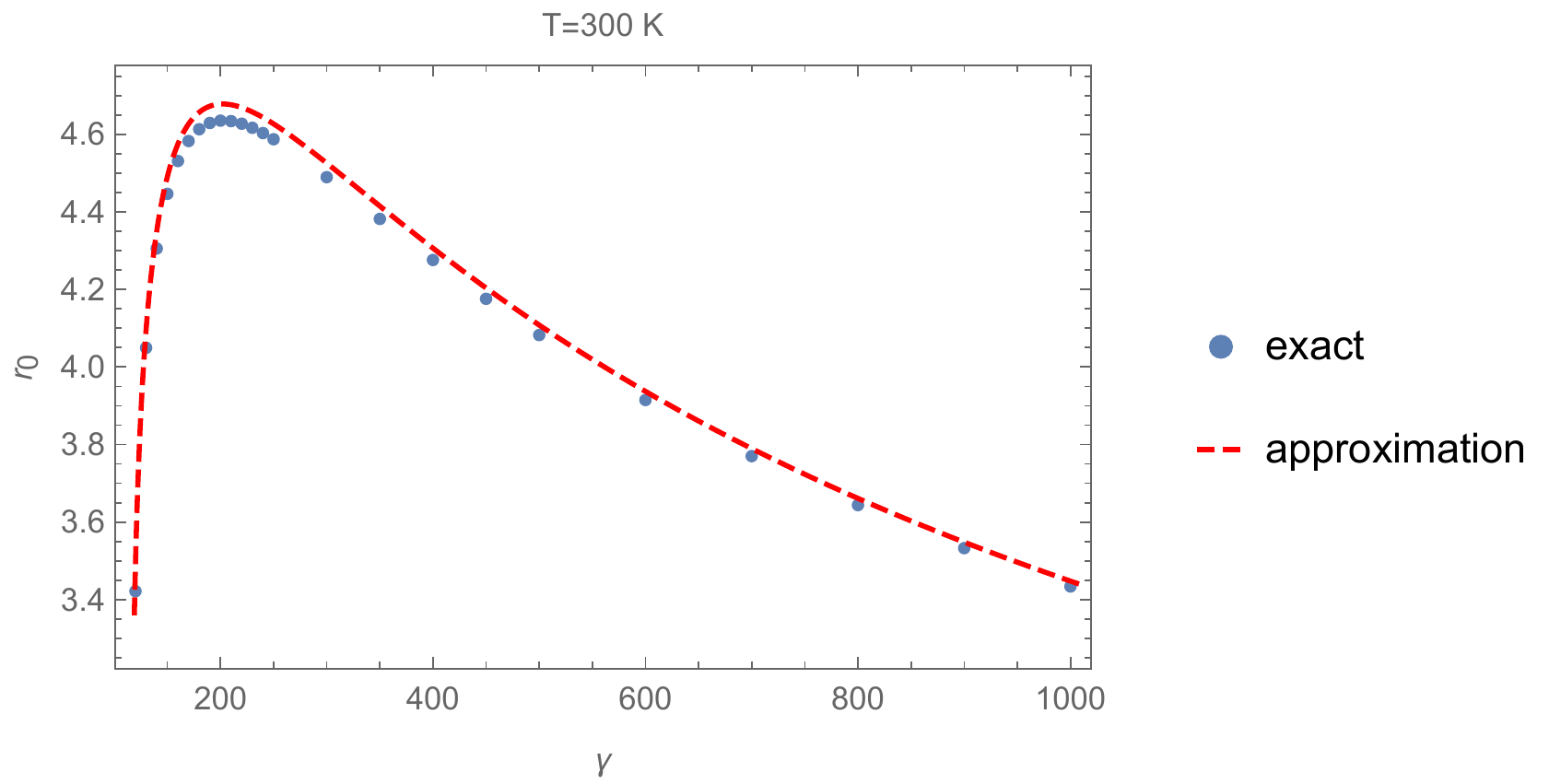}
 \caption{The solution to Eq.~\eqref{5.81} at $ T=300 \,\rm{K} $ is illustrated for ultrarelativistic velocities. The exact numerical solution is shown by the blue dots. The dashed red curve shows the approximate formula \eqref{B3}. \label{figr0hiv}}
 \end{figure} 
In Fig.~\ref{figr0hiv}, we show both the exact numerical solution and the approximation of $ r_{0} $ for ultrarelativistic velocites $ \gamma\in [120,1000] $ at room temperature. The approximate formula Eq.~\eqref{B3} works reasonably well in the velocity region covered. It is seen that $ r_{0} $ can become substantially greater than $ 1 $ in such a high velocity regime.

%references
\bibliography{eqfcite}

\end{document}